\documentclass[sn-mathphys]{sn-jnl}
\usepackage{graphicx}
\graphicspath{{./pics/}}
\newcommand{\isis}{IS$\odot$IS }
\newcommand{\m}[1]{\mathbf{#1}}
\newcommand{\vkh}{von K\'arm\'an Howarth~}

\jyear{2022}
\begin{document}
\title[PSP Turbulent Energy Transfer]{Observations of Cross Scale Energy Transfer in the Inner Heliosphere by Parker Solar Probe}

\author*[1,2]{\fnm{Tulasi N} \sur{Parashar}}\email{tulasi.parashar@vuw.ac.nz} 
\author[2]{\fnm{William H} \sur{Matthaeus}}\email{whm@udel.edu}   

\affil*[1]{\orgdiv{School of Chemical and Physical Sciences}, \orgname{Victoria University of Wellington}, \orgaddress{\street{Gate 7, Kelburn Parade, Kelburn}, \city{Wellington}, \postcode{6012}, \country{New Zealand}}}

\affil[2]{\orgdiv{Department of Physics and Astronomy}, \orgname{University of Delaware}, \orgaddress{\street{Sharp Laboratory}, \city{Newark}, \state{Delaware}, \postcode{19711}, \country{USA}}}

\abstract{
The solar wind, a continuous flow of plasma from the sun, not only shapes the near Earth space environment but also serves as a natural laboratory to study plasma turbulence in conditions that are not achievable in the lab. Starting with the Mariners, for more than five decades, multiple space missions have enabled in-depth studies of solar wind turbulence. Parker Solar Probe (PSP) was launched to explore the origins and evolution of the solar wind. With its state-of-the-art instrumentation and unprecedented close approaches to the sun, PSP is starting a new era of inner heliospheric exploration. In this review we discuss observations of turbulent energy flow across scales in the inner heliosphere as observed by PSP. After providing a quick theoretical overview and a quick recap of turbulence before PSP, we discuss in detail the observations of energy at various scales on its journey from the largest scales to the internal degrees of freedom of the plasma. We conclude with some open ended questions, many of which we hope that PSP will help answer.
}
\keywords{keyword1, Keyword2, Keyword3, Keyword4}


\maketitle
\section{Introduction}
\label{intro}
Since the advent of the space race, the importance of space weather and space environment in general has increased in our lives. The solar wind that shapes this space environment has been a subject of intensive study ever since its prediction and first measurements \cite{ParkerApJ58a,NeugebauerJGR66}. Million degree hot corona is responsible for the acceleration of the solar wind and its eventual escape from the sun into the interplanetary medium. The solar wind, like most naturally occuring as well as man made plasmas, is turbulent in nature \cite{krommes2002fundamental, tsytovich2016introduction, yamada2008anatomy, ColemanApJ68, MatthaeusJGR82, BrunoLRSP13}. The solar wind, being easily accessible through many space missions, serves as a natural laboratory for studying plasma turbulence in situ. These measurements have allowed testing and refinement of plasma turbulence theories, which are relevant for many other astrophysical systems such as the interstellar medium \cite{cordes1985small, falceta2014turbulence, armstrong1995electron}, accretion disks \cite{balbus1998instability, abramowicz2013foundations}, and the intracluster medium \cite{SchueckerAA04, churazov2012x, mohapatra2020turbulence}.

Turbulence is believed to be an important player in heating the solar corona \cite{hendrix1996magnetohydrodynamic, MattEA99-ch, cranmer2007self}. The subsonic coronal plasma accelerates to become supersonic at a few solar radii and eventually super Alfv\'enic at around $\sim 10-20 R_\odot$. Beyond this Alfv\'en critical region, the strong coronal magnetic field loses control of the plasma. The shears introduce Kelvin Helmholtz like dynamics, creating large scale roll ups in the solar wind \cite{DeforestApJ16, TelloniApJ22}. These roll-ups push the solar wind towards isotropization at the largest scales introducing an important step in its turbulent evolution \cite{RuffoloApJ20}. As the solar wind evolves, the turbulence is believed to keep the wind hotter than what is expected from a simple adiabatic expansion \cite{RichardsonEA95}. The outer scale of turbulence, the Alfv\'enicity, and the amplitude of turbulent fluctuations all play an important role in the evolution of the solar wind. Accurate understanding of turbulent processes and their evolution is also critical for improving our global heliospheric models \cite{UsmanovApJ11, GamayunovApJ12, SokolovApJ13, oran2014coronal, ChhiberApJS17, vanderHolstApJ22}.
%

Iconic missions such as Mariner, Voyager, Helios, and Ulysses ushered an era of exploration of macroscopic as well as turbulent properties of the solar wind. We refer the reader to excellent reviews by Tu \& Marsch \cite{TuSSR95}, Bruno \& Carbone \cite{BrunoLRSP13}, and Verscharen et. al. \cite{VerscharenLRSP19} for a comprehensive view of solar wind turbulence in the heliosphere before PSP. 
The basic macroscopic properties of the solar wind are well described 
in the seminal book by Hundhausen \cite{Hundhausen}.
Parker Solar Probe \cite{FoxSSR16}, with its state-of-the-art instrumentation and unprecedented close approaches to the sun is enabling hitherto impossible studies of plasma turbulence 
close to the sun. In this review paper we discuss how PSP has enhanced our understanding of the turbulent transfer of energy across scales in the inner heliosphere.

We start by describing a phenomenology of scale to scale spectral transfer from energy containing to kinetic scales in section \ref{spectral_transfer}. In section \ref{oldies} we discuss the findings from prior missions such as Voyager, Helios, and Ulysses on the evolution of turbulence in the heliosphere. In section \ref{psp_stuff} we discuss the new findings enabled by the Parker Solar Probe mission before concluding with a summary of the findings and potential future directions in section \ref{conclusions}.

\section{Turbulent scale-to-scale transfer of energy}\label{spectral_transfer}
Fully developed turbulence is ideally 
characterized \cite{BatchelorTHT} 
by an input of energy at some large scales, which is then conservatively transferred to progressively smaller scales in the inertial range. The cascaded energy is eventually converted to internal energy at the dissipative scales \cite{TennekesLumley}. The large scale dynamics of plasmas are well described by magnetohydrodynamic description down to fairly small scales \cite{WuPRL13, KarimabadiPP13, WanPP16}. However, in kinetic plasmas, the hydrodynamic notion of a single dissipative scale is replaced by a multitude of smaller scales including the inertial lengths and gyro-radii of protons and electrons, the Debye length and other hybrid scales. The nature of the cascade modifies at some of these scales, rendering the hydrodynamic cascade picture to be relatively simpler in comparison. More sophisticated theories of turbulence are needed to describe the nature of turbulence at kinetic scales \cite{SchekochihinApJS09, BoldyrevApJ13, EyinkPRX18}. Even with the lack of a well accepted kinetic plasma turbulence theory, we can study the transfer of energy across scales in a quantitative way, down to kinetic scales and smaller with appropriate methods. We now describe a phenomenology of such a transfer from the largest scales to kinetic scales and into the internal degrees of freedom.

{\bf \em Energy at large scales:} The energy at the largest scales is input by direct sources of energy or by large scale instabilities and is subsequently cascaded down to smaller scales \cite{BiskampBook03}. In hydrodynamics the turbulent cascade adjusts in such a way as to balance the energy input at the largest scales by dissipation at the small scales \cite{KarmanPRSLA38}. This \vkh description of the decay when generalized to MHD can be written as \cite{HossainPFL95,WanJFM12, BandyopadhyayPRX18}
\begin{equation}
    \epsilon_\pm = -C_\pm \frac{Z_\pm^2 Z_\mp}{\lambda_\pm}
    \label{eqn:Yaglom}
\end{equation}
where $\epsilon_\pm = \frac{d Z_\pm^2}{dt}$ is the decay rate for $Z_\pm =\langle \lvert\mathbf{z}_\pm\rvert^2\rangle$ with the Els\"asser variables defined as $\m{z}_\pm = \m{v} \pm \m{b}/\sqrt{\mu_0 \rho}$ with $\m{v}$, $\m{b}$, and $\rho$ being the fluctuating velocity, magnetic field, and density, and $\lambda_\pm$ the energy containing scales. The proportionality constant $C_\pm$ can depend on a lot of conditions such as the Reynolds number $Re$, cross helicity $\sigma_c$, and the ratio of thermal to magnetic pressures $\beta=8\pi\mu_0nk_BT/B^2$ \cite{MccombPRE15, MatthaeusApJL16, BandyopadhyayPRX18}.

The \vkh similarity, although a simplified large scale description, describes the balance of large scale energy input and dissipation really well not only in hydrodynamics but also in kinetic plasmas down to very small scales (of the order of a few ion inertial scales $d_i$) \cite{WuPRL13, ParasharApJ15}. In the solar wind the \vkh similarity has been recently shown to be applicable to the magnetic field fluctuations \cite{RoyApJL21}. As discussed in the section \ref{psp_stuff}, PSP has allowed an examination of the balance between energy input and heating rate.

{\bf \em Energy in the inertial range:} Assuming isotropy, homogeneity, constancy of scale-to-scale energy transfer rate $\epsilon$, and locality of transfer in the scale space, Kolmogorov \cite{Kolmogorov41a} identified the power spectrum of hydrodynamic turbulence to be $E(k) = C \epsilon^{2/3}k^{-5/3}$ (K41), where $E(k)$ is the energy density in wavenumber $k$ and $\epsilon$ is assumed to be constant across scales and uniform in space. The K41 phenomenology when extended to MHD predicts spectral slopes varying between $k^{-5/3}$ and $k^{-3/2}$ \cite{KraichnanPFL65, GoldreichApJ95, VermaPP99, ZhouRMP04}. This scaling has been observed in hydrodynamic turbulence behind a grid \cite{ChampagneJFM78}, Earth's magnetosheath \cite{ParasharPRL18}, solar wind \cite{KiyaniPTRSA15}, interstellar medium \cite{FraternaleApJ19}, and the intracluster medium \cite{SchueckerAA04}. If one considers non-uniform dissipation, intermittency emerges while minimally affecting the isotropic form of the spectral law \cite{KolmogorovJFM62, PolitanoPRE95, VermaPR04}.

The cascade of energy from large to small scales in the inertial range is quantitatively described, under the assumptions of homogeneity, time stationarity, and isotropy, by the so called third order law. For the conservative part of the cascade, in hydrodynamics, the third order structure function is related to decay rate by $\langle \delta u_\ell^3 \rangle = - \frac{4}{5}\epsilon \ell$, where $\ell$ is the lag and $\delta u$ is the magnitude of a velocity increment computed at lag $\ell$ \cite{PopeBook}. This third order law was generalized to incompressible MHD by Politano and Pouquet in 1998 \cite{PolitanoPRE98}.

\begin{equation}
    Y^\pm(\ell) = \langle \delta\m{z}^\mp \lvert \delta\m{z}^\pm \rvert^2\rangle = \frac{4}{3} \epsilon^\pm \ell
    \label{eqn:thirdorder}
\end{equation}
where $\ell$ is the lag at which the increment $\delta \m{z}^\pm(\m{x},\ell) = \m{z}^\pm(\m{x}+\ell)-\m{z}^\pm(\m{x})$ is computed, and $\langle \ldots \rangle$ denote appropriate averaging. The incompressible third order law can further be generalized to include more physics in the form of anisotropy \cite{PodestaJFM08}, compressibility \cite{AndresPRE17}, shears \cite{WanPP09}, Hall physics \cite{GaltierPRE08}, some combination of such effects \cite{FerrandJPP21}, or be generalized to electron MHD  \cite{GaltierJGR08}. This \vkh Yaglom Politano Pouquet (KHYPP) law or Politano-Pouquet (PP) law and many of its extensions to Hall/compressible MHD have been used to measure and test the cascade rates in simulations, solar wind, and Earth's magnetosheath \cite{Sorisso-ValvoPRL07, MarinoApJL08, OsmanEA11-3rd, VerdiniApJ15, HellingerApJL18, BandyopadhyayPRL20-PP}. The cascade rates have also been compared to the expected rates of plasma heating in the solar wind to test if turbulence can account for heating of the solar wind (see discussion in sections \ref{oldies} and \ref{psp_stuff}).

The cascade of energy is actually even more complicated in models as simple as compressible MHD where a compressive cascade proceeds in parallel to a magnetic cascade \cite{AluiePRL11}. There exists a small scale `decoupled range' where the magnetic energy and kinetic energy cascades proceed conservatively with the same cascade rate \cite{BianPRL19}. In this picture the exchange between kinetic and magnetic fluctuations happens at relatively large scales in the inertial range. There are also suggestions that magnetic reconnection, and in some situations large scale instabilities, could potentially bypass the cascade and transfer energy directly into the kinetic range and into internal degrees of freedom \cite{SquirePRL17, FranciApJL17, KunzJPP20}.

The third order law approach becomes cumbersome with the addition of more physics. Moreover the accuracy of this approach depends on the terms retained in the fluid model. An alternative approach to studying scale-to-scale transfer of energy in the fully kinetic limit is to apply scale filtering techniques to the Vlasov equation \cite{YangPP17, EyinkPRX18, CamporealePRL18, CerriPP20}. Starting with the Vlasov equation, applying scale filtering techniques, one arrives at \cite{YangPP17, YangApJ22}
\begin{eqnarray}
    \partial_t \Tilde{\mathcal{E}}_\alpha^f + \nabla\cdot \m{J}_\alpha^u & = & - \Pi_\alpha^{uu}-\Phi_\alpha^{uT}-\Lambda_\alpha^{ub} \nonumber \\
    \partial_t \Bar{\mathcal{E}}^m + \nabla\cdot\m{J}^b & = & - \sum_\alpha \Pi_\alpha^{bb} + \sum_\alpha \Lambda_\alpha^{ub}
    \label{scale_filter}
\end{eqnarray}
\begin{figure}[!ht]
\centering
\includegraphics[width=0.5\textwidth]{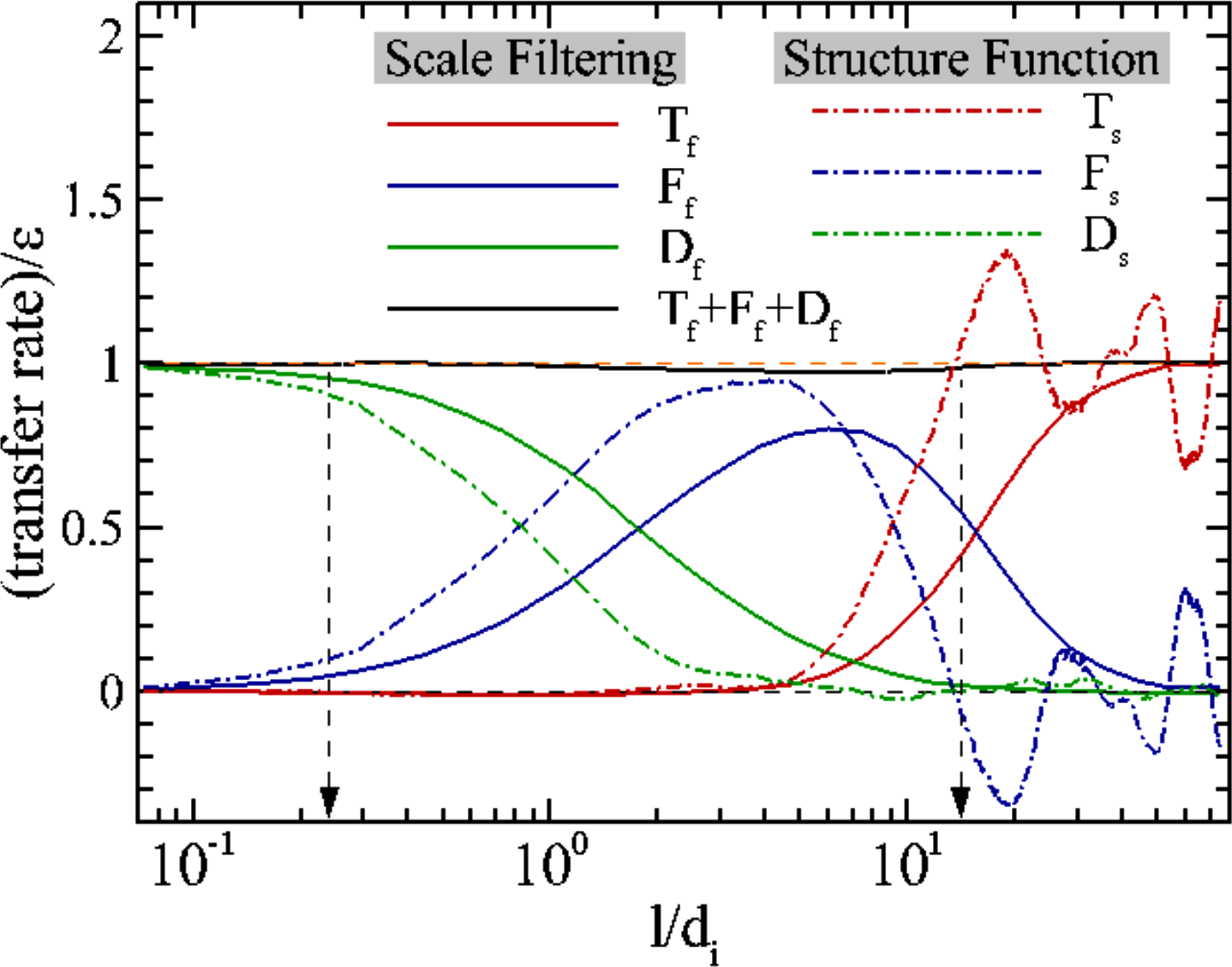}
\caption{Scale filtered energy fluxes (solid lines) compared to transfer terms from \vkh equations generalized to incompressible Hall MHD for a fully kinetic 2.5D simulation of plasma turbulence. The scale filtered energy equations show good numerical energy conservation at all scales while energy conservation has to be imposed on the \vkh equations given the lack of kinetic and compressive physics. (Reproduced with permission from \cite{YangApJ22}).}
\label{YanApJ22}       
\end{figure}
where $\Bar{~}$ represents a scale filtered quantity and $\Tilde{~}$ represents a density weighted filtered quantity. $\Tilde{\mathcal{E}}_\alpha^f$ is the filtered fluid flow energy, $\Bar{\mathcal{E}}^m$ is the filtered electromagnetic energy, $\m{J}_\alpha^u$ and $\m{J}^b$ are spatial transport terms, $\Pi_\alpha^{uu}$, and $\Pi_\alpha^{bb}$ are the subgrid scale fluxes, $\Phi_\alpha^{uT}$ is the rate of flow energy conversion to internal energy through pressure strain interactions (see the discussion of energy at kinetic scales below), and $\Lambda_\alpha^{uT}$ is the rate of energy conversion from electromagnetic fluctuations into fluid flow through filtered $\m{j}_\alpha\cdot\m{E}$. For details of these equations see \cite{YangApJ22}. 

The equations \ref{scale_filter} can be combined to write:
\begin{equation}
\underbrace{\partial_t \left<\sum_\alpha \widetilde{E}^{f}_{\alpha} + \overline{E}^m\right>}_{T_f-\epsilon}
=-\underbrace{\left<\sum_\alpha \left( {\Pi}^{uu}_{\alpha} + {\Pi}^{bb}_{\alpha}\right)\right>}_{F_f}- \underbrace{\left<\sum_\alpha {\Phi}^{uT}_{\alpha}\right>}_{D_f} \label{eq:filtered-Ef+Em}.
\end{equation}
with $T_f-\epsilon$, $F_f$, and $D_f$ representing the decay of energy, the inertial range fluxes, and ``dissipation'' respectively. These quantities can be directly compared to the generalized \vkh equations as shown in Fig. \ref{YanApJ22}. The solid lines represent scale filtered quantities, and the dashed lines correspond to equivalent terms in the \vkh equations. The structure function approach seems to achieve a range reminiscent of an inertial range where the inertial range flux is comparble to the decay rate $\epsilon$, while the scale filtered flux remains short of $\epsilon$ even at its peak. However, the scale filtered equations show energy conservation, within numerical error, across scales, giving a quantitative handle on flow of energy across scales in an accurate manner. The \vkh equations contain varied limits of physics based on the model for which they are written \cite{HellingerApJL18} and hence energy conservation has been imposed to estimate the surrogate dissipation (shown as a green dot-dashed line). The situation is even more dramatic when the bandwidth available for the cascade is small (e.g. in 3D fully kinetic simulations or the Earth's magnetosheath). See Yang et. al. \cite{YangApJ22} for more detailed discussion of these issues. In a recent preprint Hellinger et. al. \cite{HellingerArXiv22} have included pressure strain interactions to modify the von K\'arm\'an equations. This modification extends the range of validity of these equations down to sub-proton scales implying the important role played by pressure-strain interactions in the kinetic range energy transfer; see also \cite{YangApJ22}.

{\bf \em Energy transfer at kinetic scales:} At the ion kinetic scales, some of the energy is removed into heating the ions, and the rest of it cascades down to smaller scales, eventually dissipating at electron scales. Heating of plasma\footnote{We use the terms `{\em dissipation}' and `{\em heating}' in a relaxed way to mean a transfer of turbulent fluctuation energy into the internal degrees of freedom of the plasma regardless of the (ir)reversible nature of the processes involved.} can potentially happen in many ways including wave-particle interactions such as Landau damping \cite{HollwegPRL71, ChenNature19}, cyclotron resonances \cite{HollwegJGR02,KasperPRL13}, magnetic pumping \cite{DawsonNF65, LichkoApJL17}, and stochastic heating \cite{ChandranApJ10-1, XiaApJ13, MalletJPP19, CerriApJ21, MartinovicApJ21}. In the stochastic heating picture, particles experiencing large electric fluctuation changes at their gyro scales can get stochastic kicks perpendicular to the mean magnetic field changing their magnetic moment. This effect, that depends on turbulent fluctuation amplitude at the proton gyro scale gets enhanced near intermittent structures such as current sheets \cite{ChandranApJ10-1, XiaApJ13, MalletJPP19}. Moreover, landau damping has also been shown to occur in or near current sheets \cite{TenbargeApJ13}. 

Many factors such as plasma $\beta$, turbulence amplitude, proton-electron temperature ratio, Alfv\'enicity etc. can potentially regulate the fraction of energy going into heating the ions \cite{WuPRL13, HughesGRL14, MatthaeusApJL16}. A simplified view \cite{MatthaeusApJL16} proposes that the ratio of the local nonlinear time at the ion scales to the cyclotron time is an important factor in deciding the partitioning of energy between ions and electrons. If the kinetic scale nonlinear time is comparable to or smaller than the proton cyclotron time, significant nonlinear evolution of turbulent magnetic fluctuations happens within a gyroperiod and hence the protons can get significant stochastic kicks, leaving a smaller amount of energy to cascade down to electron scales and eventually heat them. Such dependence of relative proton-electron heating has been shown to hold in simulations as well as recently in MMS data \cite{MatthaeusApJL16, RoyApJL22}.

\begin{figure}[!ht]
\centering
\includegraphics[width=0.5\textwidth]{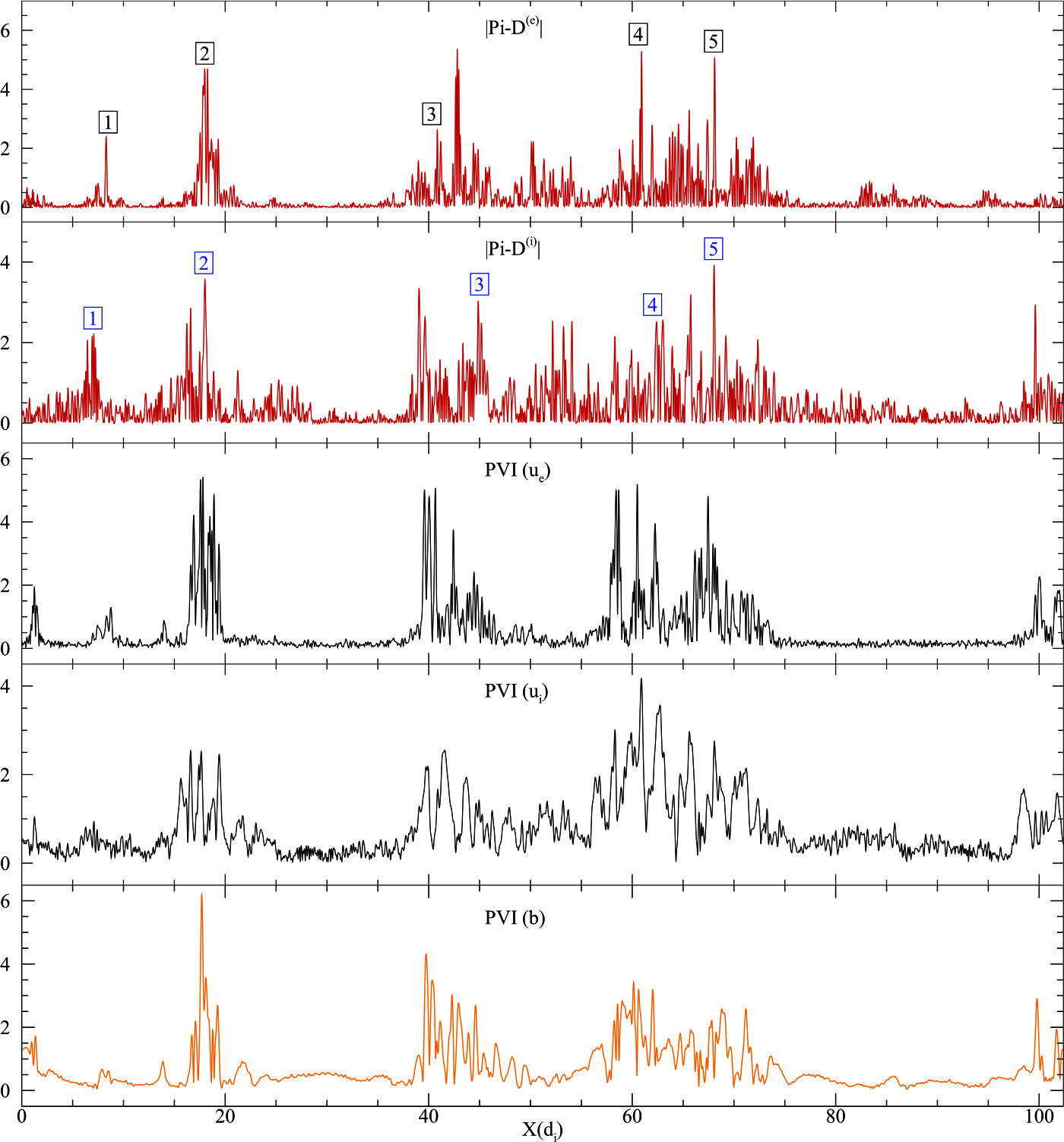}
\caption{Pi-D, the pressure strain interaction, is active near current sheets and is spatially more correlated with velocity strains than current. (Reproduced with permission from \cite{YangPP17}.)}
\label{YangPP17-PVI}
\end{figure}

Although the exact processes responsible for heating the ions can vary from one scenario to the other, the mathematical terms responsible for the transfer are fairly straightforward to understand. The transfer of energy from electromangnetic fields and bulk flow energy into internal degrees of freedom happens via a collisionless generalization of viscosity. The equations for time evolution of electromagnetic energy $E^m$, fluid flow kinetic energy $E^f_\alpha$, and internal energy $E^{th}_\alpha$ can be written in a straightforward manner from the Vlasov Maxwelll set of equations as \cite{YangPRE17}

\begin{eqnarray}
\partial_t E^{f}_\alpha + \nabla \cdot \left( E^{f}_\alpha \boldsymbol{u}_\alpha + \boldsymbol{P}_\alpha \cdot \boldsymbol{u}_\alpha \right) &=&
\left( \boldsymbol{P}_\alpha \cdot \nabla \right) \cdot {\boldsymbol{u}}_\alpha+
\boldsymbol{j}_\alpha  \cdot  \boldsymbol{E}, \nonumber \\
\partial_t E^{th}_\alpha + \nabla \cdot \left( E^{th}_\alpha \boldsymbol{u}_\alpha + \boldsymbol{h}_\alpha \right) &=& -\left( \boldsymbol{P}_\alpha \cdot \nabla \right) \cdot \boldsymbol{u}_\alpha, \nonumber \\
\partial_t E^{m} + {\frac{c}{4\pi}} \nabla \cdot \left( \boldsymbol{E} \times \boldsymbol{B} \right) &=& -\boldsymbol{j} \cdot \boldsymbol{E},
\label{yangequations}
\end{eqnarray}
where the subscript $\alpha=e, i$ represents the species, $\boldsymbol{P}_\alpha$ is the pressure tensor, $\boldsymbol{h}_\alpha$ is the heat flux vector, $\boldsymbol{j}=\sum_{\alpha} \boldsymbol{j}_\alpha$ is the total electric current density, and $\boldsymbol{j}_\alpha=n_\alpha q_\alpha \boldsymbol{u}_\alpha$ is the electric current density of species $\alpha$. The divergence terms simply transport energy in its current form and are not responsible for conversion from one form into another. The $\boldsymbol{j} \cdot \boldsymbol{E}$ term is responsible for transfer of energy from electromagnetic fields into bulk flow and the $-\left( \boldsymbol{P}_\alpha \cdot \nabla \right) \cdot \boldsymbol{u}_\alpha = P^{(\alpha)}_{ij} \nabla_i\, u^{(\alpha)}_j$ term (called PS for short) is responsible for transferring energy from bulk fluid motions into internal degrees of freedom \cite{DelSartoPRE16, YangPRE17}. 

The pressure tensor $P^{(\alpha)}_{ij}$ can be separated into a trace and a traceless part by defining $P^{(\alpha)}_{ij} = p_\alpha \delta_{ij}  + \Pi^{(\alpha)}_{ij}$ where $p_\alpha = \frac13 P^{(\alpha)}_{jj}$ and, $\Pi_{ij} = P_{ij} - p\delta_{ij}$. The stress tensor $S^{(\alpha)}_{ij}=\nabla_i u^{(\alpha)}_j$ can be similarly decomposed $S^{(\alpha)}_{ij}  = \frac13\theta_\alpha\delta_{ij} + D^{(\alpha)}_{ij} + \Omega^{(\alpha)}_{ij}$ where $\theta=\nabla \cdot {\mathbf u}$, $D^{(\alpha)}_{ij} = \frac12\left ( \nabla_i u^{(\alpha)} _j + \nabla_ju^{(\alpha)}_i\right ) $, and $\Omega^{(\alpha)}_{ij} = \frac12\left ( \nabla_i u^{(\alpha)} _j - \nabla_ju^{(\alpha)}_i \right ) $. With these decompositioins, the pressure stress interaction separates as $\left( \bf{P}_\alpha \cdot \nabla \right) \cdot {\bf u}_\alpha = p^{(\alpha)} \theta^{(\alpha)} + \Pi^{(\alpha)}_{ij}D^{(\alpha)}_{ij}$. The first part is typically abbreviated as $p\theta$ and the second part as Pi-D. $p\theta$ is the familiar heating/cooling due to compressions rarefactions term. The pressure tensor is symmetric and hence only the symmetric stresses of bulk velocity interact with the traceless part of the pressure tensor to achieve the conversion into internal energy. In the highly collisional limit, the Pi-D term reduces to the familiar viscous heating term \cite{HuangBook}.

Kinetic activity, including the heating of the ions via the Pi-D channels takes place intermittently {\em near} strong current sheets \cite{ServidioPRL12, ServidioApJL14, FranciAIP15, DelSartoPRE16, ParasharApJ16b}. Sheared magnetic fields produce strong current sheets, which in turn develop vortex quadrupoles near them \cite{MatthaeusGRL82, ParasharApJ16b}. Although the vorticity is the antisymmetric part of the velocity strain tensor, the vortices are stretched into sheet like structures in the large Reynolds number limit, creating symmetric parts of the velocity strain tensor. This symmetric part contracts with the traceless pressure tensor to transfer energy from bulk turbulent motions into internal degrees of freedom \cite{DelSartoPRE16, YangPRE17}. The pressure strain interaction has been shown to be an effective description of plasma heating in simulations as well as in magnetosheath data \cite{YangPP17, SitnovGRL18, MatthaeusApJ20, BandyopadhyayPRL20}.

The intermittent sites near which the dissipation occurs can be identified using partial variance of increments (PVI), defined as $\mathcal{I}=\lvert\Delta \m{b}(t,\Delta t)\rvert/\sqrt{\langle\lvert \m{b}(t,\Delta t)\rvert^2\rangle}$ where $\Delta \m{b}(t,\Delta t) = \m{b}(t+\Delta t)-\m{b}(t)$ \cite{GrecoSSR18}, or via the unaveraged kernel of the third order law (Eqn. \ref{eqn:thirdorder}), also called Local Energy Transfer rate (LET) \cite{Sorisso-ValvoSP18}. Fig \ref{YangPP17-PVI} shows cuts of Pi-D for ions and electrons from a 2.5D fully kinetic simulation along with PVI computed from velocities of ions and electrons and magnetic field \cite{YangPP17}. The locations of enhanced dissipation, identified by spikes in Pi-D, are clustered near large PVI values. Large PVI values have been shown to be strongly correlated with hotter ions in the solar wind \cite{OsmanApJL11},
and with higher fluxes of energetic particles \cite{TesseinApJL13}. More on this below.
    
An imbalanced cascade, with different powers in the $\m{z}^+$ and $\m{z}^-$ fluctuations and relevant for example for inner heliospheric conditions, could modify the cascade at ion kinetic scales affecting the ion heating rates and resulting cascade to smaller scales. The `helicity barrier' inhibits the cascade of energy to scales smaller than proton kinetic scales, resulting in a build-up of energy at the proton scales. This build-up of proton scale kinetic energy can result in generation of cyclotron waves, which can heat the protons perpendicularly \cite{SquireNatureAstron22}. This enhanced energy dissipation at ion kinetic scales can result in very steep spectra just below ion kinetic scales (approaching $k^{-4}$), eventually returning to the more familiar $k^{-8/3}$ close to electron scales.
  
\begin{figure}[!ht]
\centering
\includegraphics{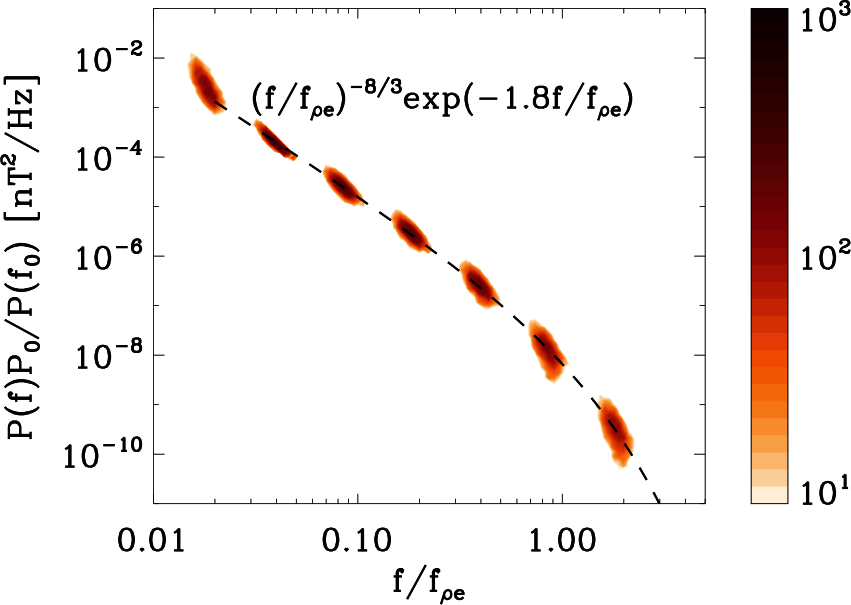}
\caption{Magnetic energy spectra in the inner heliosphere computed using SCM data from the Helios spacecraft. The spectra show a powerlaw superposed with exponential decay of magnetic fluctuations at electron scales. (Reproduced with permission from \cite{AlexandrovaPRE21}).}
\label{AlexandrovaPRE21}       
\end{figure}

Close to electron scales, the magnetic energy spectra are exponentially damped. Fig. \ref{AlexandrovaPRE21} shows magnetic energy spectra computed using magnetic field measurements from the search coil magnetometer (SCM) on Helios. The 3344 individual spectra have been rescaled by their amplitude at roughly 20 electron gyroperiods. The colours represent 2D histograms with darker colours representing more points. The spectra show a power-law behaviour with $f^{-8/3}$ superposed with exponential decay at electron scales indicating strong damping of magnetic fluctuations \cite{AlexandrovaApJ12, TenbargeApJ13, ArroArXiv21}.

\begin{figure}[!htb]
\includegraphics[width=0.45\textwidth]{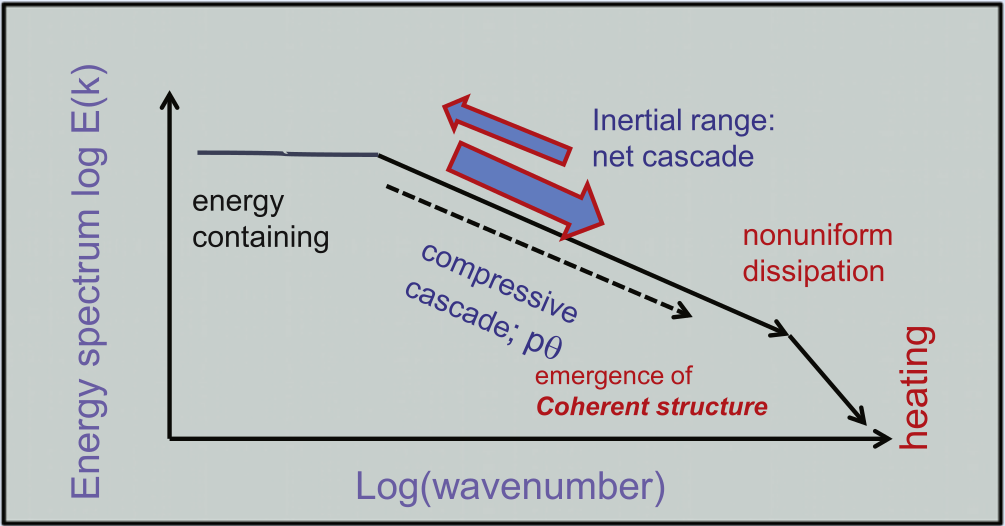}
\includegraphics[width=0.54\textwidth]{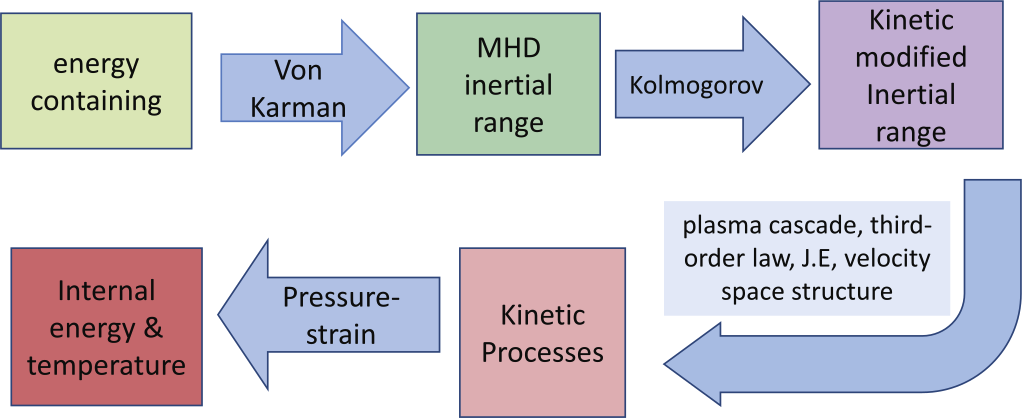}
\caption{A schematic representation of the flow of energy from large to kinetic scales in kinetic plasma turbulence.}
\label{MatthaeusApJ20}       
\end{figure}

Based on the above considerations, an overall view of the cascade of energy from large to small scales emerges to be as follows (see Fig. \ref{MatthaeusApJ20} and \cite{MatthaeusApJ20} for an in-depth discussion): The energy containing scales input energy at the von K\'arm\'an rate into the inertial range. In the inertial range a cascade of energy to smaller scales via incompressible as well as compressive cascades transfers energy conservatively to smaller scales. The cascade also generates intermittent structures where pressure strain interactions transfer energy into internal degrees of freedom. The remaining part of the energy is transferred
down to smaller scales where kinetic effects become dominant
and dissipation ends the cascade.

\section{Turbulence in the heliosphere before PSP}
\label{oldies}
The dynamics of turbulence and its effects on the evolution of the solar wind have been studied in-depth since the late 60s. The observations from Mariners, Voyagers, Helios, and Ulysses revealed how the turbulent power, Alfv\'enicity, power spectra, spectral anisotropy, turbulent cascade, and intermittency evolve with heliocentric distance.  For comprehensive view of solar wind turbulence we refer the reader to Tu \& Marsch, Bruno \& Carbone, and Verscharen et. al. \cite{TuSSR95, BrunoLRSP13, VerscharenLRSP19}.

Among the pioneering and landmark early studies of turbulence in the interplanetary medium, an important example is the work of Coleman \cite{ColemanApJ68}. This study synthesized analysis of Mariner 2 data, taking in to account earlier observations \citep{HolzerJGR66, Coleman66} as well as the important suggestion \citep{SturrockPRL66} that energy in waves or turbulence may be responsible for heating the corona. 
Coleman \cite{ColemanApJ68}
developed this idea by postulating that the 
physical processes leading this heating would be governed approximately by ideas from classical 
hydrodynamic turbulence theory 
\cite{Chandra49-apj}
adapted to plasma in the approach of
Kraichnan \cite{KraichnanPFL65}. This lead to a heating rate due to the cascade that was found to be reasonably in accord with observed temperatures at 1au. 
Several decades of research have elaborated on these ideas.

\begin{figure}[!ht]
\centering
\includegraphics[width=0.7\textwidth]{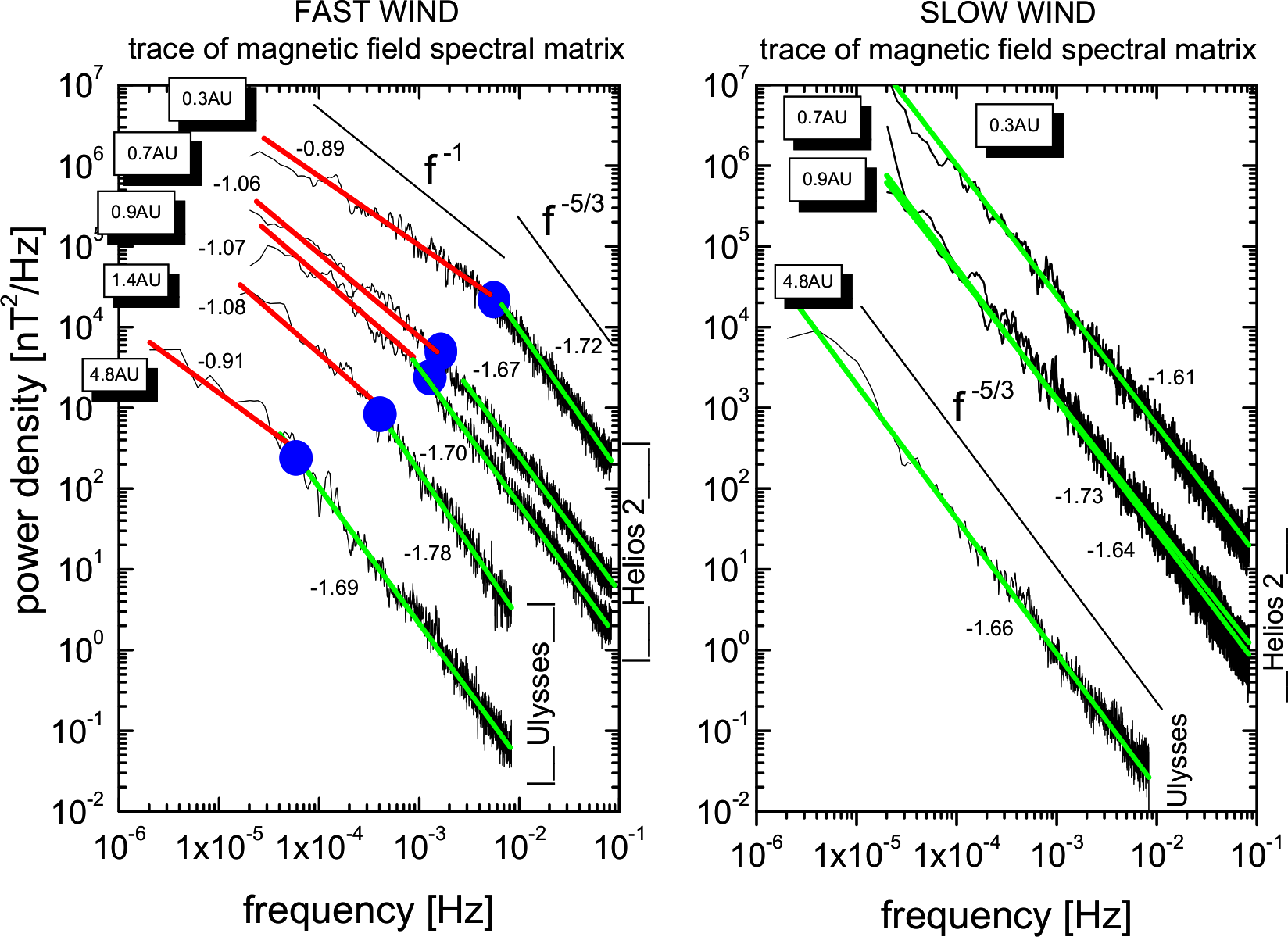}
\caption{Magnetic field spectra for various heliocentric distances computed from Helios and Ulysses data. Left panel shows fast wind cases and the right panel shows slow wind cases. Spectral power decreases with increasing heliocentric distance regardless of the speed of the wind. The fast wind spectra show a transition from Kolmogorov like $f^{-5/3}$ to $f^{-1}$ at large scales. The break frequency moves to larger scales with increasing heliocentric distance. The slow wind does not show such transition. (Reproduced with permission from \cite{BrunoLRSP13})}
\label{BrunoEMP09}       
\end{figure}

As the solar wind expands, the turbulent power decreases with heliocentric distance \cite{BelcherJGR74}. From earlier inner heliospheric observations \citep{RobertsEA90},
one may argue that this decrease is consistent with WKB theory
\cite{VermaRoberts93}. However it is also 
consistent with a driven, dissipative expanding MHD system \cite{ZankJGR96} which gives very similar radial profiles for appropriate parameter choices.
Interestingly the solutions are {\it not} 
consistent with an undriven dissipative turbulence system, which would provide power at 1 au that is less than what is observed. 
The observed decreasing total power reflects 
as well in the reduced spectral densities with increasing heliocentric distance \cite{BavassanoJGR82, HorburyJGR01, BrunoEMP09}. 
Describing this evolution of 
the spectrum and understanding 
the somewhat subtle physical effects that enter this description has been the subject of intensive study, even until the time of this writing.

Fig. \ref{BrunoEMP09} shows power spectra for magnetic field computed from Helios and Ulysses data \cite{BrunoEMP09}. The left panel shows trace magnetic power spectra for fast wind intervals and the right panel shows the same for slow wind intervals at various heliocentric distances. Two important features are clearly identified. The fast wind spectra show a break at large scales and transition from $f^{-5/3}$ to $f^{-1}$ at some large scale. The location of this spectral break shifts to lower frequencies or larger scales as the solar wind expands in the heliosphere. The break frequency follows a power law decrease of $R^{-1.5}$ with heliocentric distance. There is some variability in the powerlaw values of this variation. This will be discussed below as well as in section \ref{psp_stuff}. Secondly, the slow wind spectra do not show a transition to an $f^{-1}$ regime, 
potentially owing to the more advanced state of 
evolution of the observed 
slow wind. 

The first theoretical studies to attempt 
description of the radial evolution of heating \cite{Hollweg86, HollwegJohnson88} and the radial evolution of the spectral shapes
\cite{TuJGR84,TuJGR88} made major steps
towards merging the ideas of turbulence theory with spatial transport modeling of radial evolution of solar wind properties, a theory classically exemplified 
by WKB theory \cite{Hollweg74}.
It was soon recognized that 
refinements of these approaches were required for greater veracity, including the crucial development of 
{\it non-WKB transport} theory \cite{MarschTu89, ZhouMatt89}
and transport theory to describe the 
turbulence fluctuations at 
{\it energy containing scales}
\cite{MatthaeusJGR94} that feed energy into the inertial range

The energy containing scale, identified by the correlation length $\Lambda = \tau_{corr}V_{sw}$, increases with increasing heliocentric distance \cite{SmithJGR01, RuizSP14}. Measurements from Voyager and Ulysses show the increase to follow $R^{0.45}$, in contrast to the expectation of break frequency variation of $R^{-1.5}$.  Along with the highly nonadiabatic behavior of the proton temperature \cite{RichardsonEA95,SmithJGR01}, the observed variation of correlation
scale is a strong indication of the macroscopic influence of active turbulence evolution in the interplanetary medium.

\begin{figure}[!ht]
\centering
\includegraphics[width=0.5\textwidth]{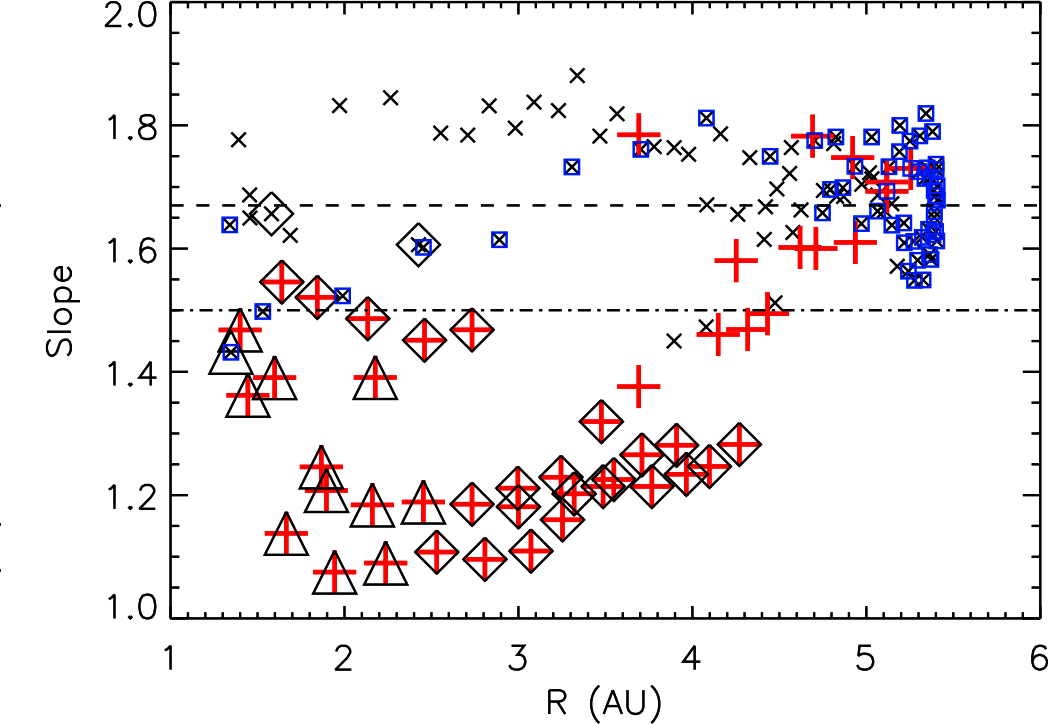}
\caption{Slopes of velocity spectra from Ulysses spacecraft as a function of heliocentric distance. Red plus signs represent intervals with wind speed greater than 675 km/s, all other points are black crosses. Intervals with signed Alfv\'enicity between 0.33 and 0.5 are enclosed in diamonds and the ones with high Alfv\'enicity are enclosed in triangles. Points within 20$^\circ$ of the ecliptic are enclosed in blue squares. (Reproduced with permission from \cite{RobertsJGR10}).}
\label{RobertsJGR10}       
\end{figure}

The characteristics of the plasma 
velocity fluctuations are also indicative of turbulence and of turbulence evolution as the solar wind ages. 
The velocity spectra evolve with heliocentric distance towards a state that is closer to a Kolmogorov-like spectral index. 
Fig. \ref{RobertsJGR10} shows slopes of velocity wavenumber
spectra from Ulysses data as a function of heliocentric distance \cite{RobertsJGR10}. The slopes were computed using power law fits to the velocity spectra in the $10^{-5}$ Hz to $10^{-4}$ Hz range. Red pluses represent very fast wind with speed $> 675$ km/s and black crosses represent rest of the intervals. These intervals show a significant scatter in spectral slopes between 1-2AU but show a gradual rise of the slope from $\sim -1.1$ to $\sim -5/3$ in the outer heliosphere. A reason behind this could potentially be that the $10^{-5} - 10^{-4}$ Hz fitting range might be in the energy containing range in the inner heliosphere and in the inertial range as the wind expands and the break point moves to lower frequencies (see for example Fig \ref{BrunoEMP09}). The points enclosed in blue squares are within 20$^\circ$ of the ecliptic and hence directly comparable with results from other missions such as Helios and PSP. The slopes near the ecliptic vary from $-3/2$ to $-5/3$ as heliocentric distance increases. 
As interesting as this may be, one must recall that the standard Kolmogorov theory applied to MHD does not make a specific universal prediction about 
the velocity spectrum itself, but rather, in the usual sense,
for the total incompressive energy spectrum.

\begin{figure}[!ht]
\centering
\includegraphics[width=0.8\textwidth]{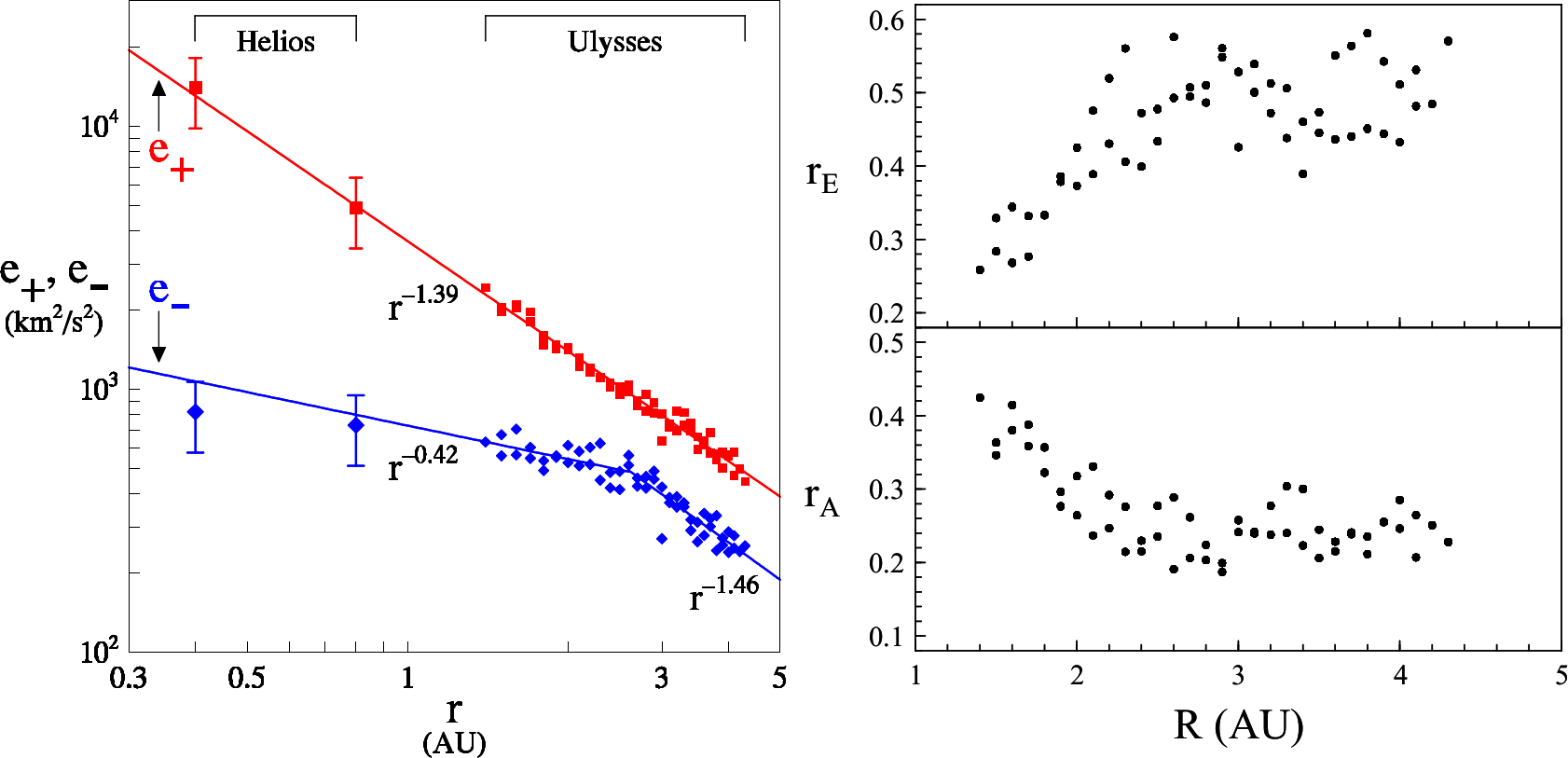}
\caption{Radial variation of Els\"asser variances from Helios and Ulysses (left panel), Els\"asser ratio (top right panel) and Alfv\'en ratio (bottom right panel). The outward Els\"asser variable decays faster than the inward Els\"asser variable until roughly 2.5AU after which the outward and inward Els\"asser variables show roughly the same energy while the Alfv\'en ratio fluctuates near 0.2. (Reproduced with permission from \cite{BrunoLRSP13})}
\label{BavassanoJGR00}       
\end{figure}

The turbulence is in an ``imbalanced'' high cross helicity 
state in the inner heliosphere. It is highly Alfv\'enic with the outward propagating Els\"asser field ($\m{z}^+$) dominating the energy budget. 
This is usually argued to be a consequence of the launching of outward-propagating waves in the lower corona and photosphere, along with the potential filtering effect that may occur at the Alfv\'en critical surface
(Only outward waves propagate away.) 
However it is well known 
that an admixture of inward-type cross helicity is required to drive an incompressive MHD cascade \cite{KraichnanPFL65}. A widely accepted explanation for how the cascade is enabled is that interactions with 
Alfv\'en speed gradients, i.e., ``reflections,'' \cite{MattEA99-ch}
or, equally well, interaction with shears \cite{MattEA99-swh} 
that tap velocity field energy and produced {\it both} senses of Elas\"asser propagation, can produce the required flux of inward fluctuations. 

High latitude solar wind turbulence shows a smaller inertial range than what is observed in the ecliptic because of the lack of shears \cite{GoldsteinGRL95}, or perhaps weaker shears
\cite{BreechEA08}. 
As the solar wind expands, the cross helicity and the Alfv\'enicity of the solar wind fluctuations decrease \cite{RobertsJGR87, GoldsteinGRL95, MatthaeusGRL04, BavassanoJGR00}. Fig. \ref{BavassanoJGR00} shows the radial evolution of Els\"asser variances (left panel), the ratio of energies contained in the Els\"asser variables, called Els\"asser ratio in 
the top right panel, and the Alfv\'en ratio in the bottom right panel. The outward Els\"asser energy decreases significantly faster than the inward Els\"asser energy out to roughly 2.5AU beyond which both decrease in a similar fashion. The ratio of the two Els\"asser energies increases gradually 
as the turbulence becomes less Alfv\'enic; this ratio
fluctuates around 0.5 beyond 2.5AU. 

Apart from the variation of cross helicity, there is also systematic
radial evolution of the  Alfv\'en ratio. This 
quantity is defined as the ratio of energy density in the velocity fluctuations to energy density in magnetic fluctuations, i.e., 
$r_A = \langle {\bf v}^2 \rangle /\langle {\bf b}^2 \rangle$.   
In the inner heliosphere, the inertial range 
$r_A$ decreases and stabilizes at values around $r_A \approx 1/2$.  Like cross helicity, $r_A$ is influenced by both expansion and shear. 
An equivalent quantity is the {\it residual energy}
defined as $\sigma_r = (\langle {\bf v}^2 \rangle - \langle {\bf b}^2 \rangle)/(\langle {\bf v}^2 \rangle + \langle {\bf b}^2\rangle )$,
a quantity that is not related to ideal invariants and so is not associated with a conserved spectral flux. Even if it cannot 
``cascade'' in the usual sense, $\sigma_r$ 
exhibits distinctive properties, such as attaining moderately negative values $\sigma_r \sim -1/3$
in the inertial range of MHD turbulence and in the solar wind 
over a fairly wide range of parameters. (For observations, see, e.g.,
\cite{MatthaeusJGR82}.)
There have been numerous phenomenological theories 
developed to describe 
the behavior of residual energy, e.g., 
\cite{StriblingMatt91,MullerGrappin04,BoldyrevEA11,GrappinEA16}.

\begin{figure}[!ht]
\centering
\includegraphics[width=0.5\textwidth]{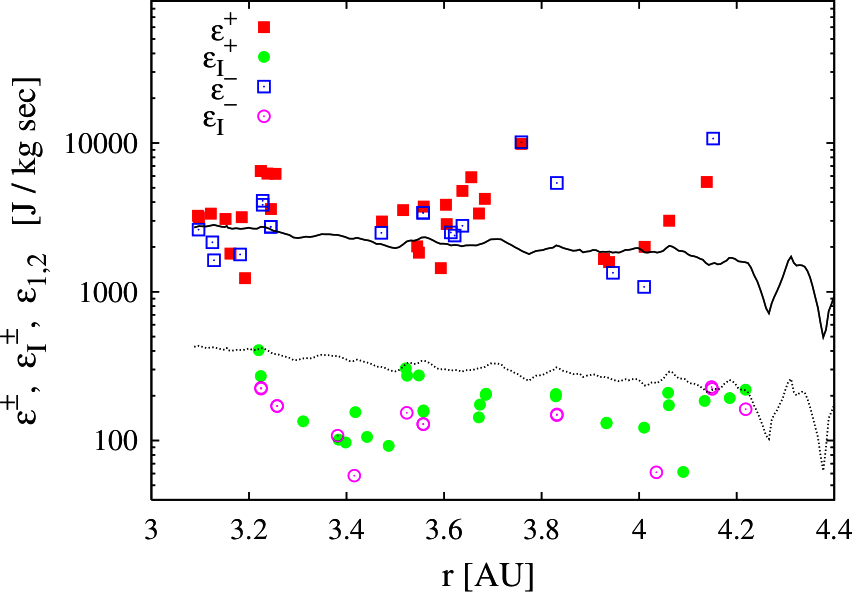}
\caption{Incompressible as well as compressible cascade rates (defined using density weighted Els\"asser variables), computed from Ulysses data, as a function of heliocentric distance. The lines show estimated heating rates for solar wind obtained from temperature profiles. (Reproduced with permission from \cite{CarbonePRL09}) }
\label{CarbonePRL09}       
\end{figure}

The temperature of the solar wind drops slower than expected as it expands until the pickup ions introduce a new significant source of energy in its evolution \cite{MarschJGR82, WangJGR01, MatteiniGRL07, HellingerJGR13}. The heating rate computed from large scale properties in the solar wind at 1AU is sufficiently larger than the required heating rate \cite{VasquezJGR07}. The nonlinear cascade of energy to smaller scales, measured by the third order law (Eqn. \ref{eqn:Yaglom}), has also been established in the solar wind \cite{Sorisso-ValvoPRL07, MacBrideEA08,CarbonePRL09,MacBrideEA-sw11, MarinoApJ12}. The incompressible cascade rate, studied in the polar wind using Ulysses data, can provide a significant fraction of energy required to heat the solar wind \cite{MarinoApJL08}. When generalized phenomenologically to include compressibility effects via density weighted Els\"asser fields, the cascade rate increases significantly. Fig. \ref{CarbonePRL09} shows cascade rates computed for both incompressible Els\"asser variables as well as density weighted Els\"asser variables. The compressive estimates are about an order of magnitude higher and for both estimates follow the heating rate required for heating the solar wind protons \cite{VermaJGR95, VasquezJGR07, MarinoApJL08}.

Near kinetic scales the cascaded energy is transferred partially into proton internal energy and partially cascaded down to electron scales. Data from Helios missions have been used to compute the expected stochastic heating rates in the inner heliosphere \cite{BourouaineApJ13, MartinovicApJ19}. The stochastic heating rate appears to be sufficient to heat the solar wind. However, the radial dependence of the SH rate is very steep ($r^{-2.5}$) and it decreases more rapidly than the expected heating rate. This disparity is larger for the fast solar wind streams. The stochastic heating process is enhanced in the presence of the intermittent structures \cite{ChandranApJ10-1, XiaApJ13, MalletJPP19}.

In simulations, and in observations at 1au, the heating of plasmas has been shown to happen intermittently \cite{ParasharPP11, OsmanApJL11, OsmanPRL12a, TenbargeApJ13, YangPP17}. 
Until recently there have not been many studies
that investigated intermittent heating behaviour in the inner heliosphere;
this is discussed more in a later section below. 
However, the radial evolution of intermittency has been studied in some detail \cite{BrunoJGR03, ParasharApJL19, CuestaApJS22}. The kurtosis at a given time-scale seems to increase with increasing heliocentric distance \cite{BrunoJGR03}. When plotted as a function of plasma scales (e.g. multiples of proton inertial length $d_p$), the kurtosis drops with increasing heliocentric distance \cite{ParasharApJL19, CuestaApJS22}. The proton inertial length $d_p$ increases faster than the outer scale of the turbulence in expanding wind. This reduces the bandwidth available for the inertial range cascade, and hence the level of intermittency as identified by scale dependent kurtosis at a given plasma scale.


\section{Parker Solar Probe Observations}\label{psp_stuff}
Parker Solar Probe was launched in 2018 to study the origins and evolution of the solar wind \cite{FoxSSR16}. The science objectives of the probe are to ``{\em trace the flow of energy that heats the corona and accelerates the solar wind}'', to ``{\em determine the structure and dynamics of the magnetic fields at the sources of solar wind}'', and to ``{\em explore the mechanisms that accelerate and transport energetic particles}'' \cite{FoxSSR16}. The mission carries four instruments: Electromagnetic Fields Investigation (FIELDS) \cite{BaleSSR16}, Integrated Science Investigation of the Sun (\isis) \cite{McComasSSR16}, Solar Wind Electrons Alphas and Protons (SWEAP) \cite{KasperSSR16}, and Wide-field Imager for Solar Probe (WISPR) \cite{VourlidasSSR16}. The data from FIELDS, SWEAP, and \isis have extensively been used to study the origins and evolution of solar wind turbulence and its role in energetic particle dynamics. In this section we discuss PSP's contributions to our understanding of how the turbulent energy flows from large to small scales in the inner heliosphere. After a brief overview of the turbulence properties (see \cite{RaouafiEA22SSR} 
for an in-depth review), we provide an in-depth discussion of the turbulent transfer of energy from large scales to small scales.

As one approaches closer to the sun, the turbulence becomes more structured. Intermittent patches of reversals in radial magnetic field are enbedded in `{\em smoother and less turbulent flow with near radial magnetic field}' \cite{BaleNature19}. These routinely observed feature of the solar wind, also called `magnetic switchbacks' typically show a sharp reversal in the sign of the radial component of the magnetic field \cite{BaloghGRL99, MatteiniGRL14, BorovskyJGR16, deWitApJS20}. The origins of switchbacks are debated. Proposed mechanisms involve interchange reconnection, in-situ generation by expanding turbulence, and velocity shears \cite{FiskApJL20, SquireApJL20, RuffoloApJ20}. In the shear driven picture, large velocity shears across magnetic flux tubes are dominated by the strong magnetic field in the corona. After the Alfv\'en critical zone, the magnetic field is not strong and velocity shears can produce ``flocculated'' roll ups and switchbacks in-situ \cite{DeforestApJ16, ChhiberApJL18, RuffoloApJ20}. A consequence of such shear driven in-situ generation of switchbacks is that their number should drop in the sub-Alfv\'enic wind inside magnetically dominated corona. Early hints of that are already being observed in the PSP data \cite{BandyopadhyayApJL22, KasperPRL21}. A significant amount of literature stemming from PSP observations has focused on comparing and contrasting the nature of turbulence in and outside switchbacks (see e.g. \cite{deWitApJS20, BourouaineApJL20, MartinovicApJS20, TeneraniApJS20, McManusApJS20, MartinovicApJ21, BandyopadhyayApJL22} and references therein for a sampling of topics studied in the context of switchbacks by PSP).

\begin{figure}[!ht]
\centering
\includegraphics[width=0.51\textwidth]{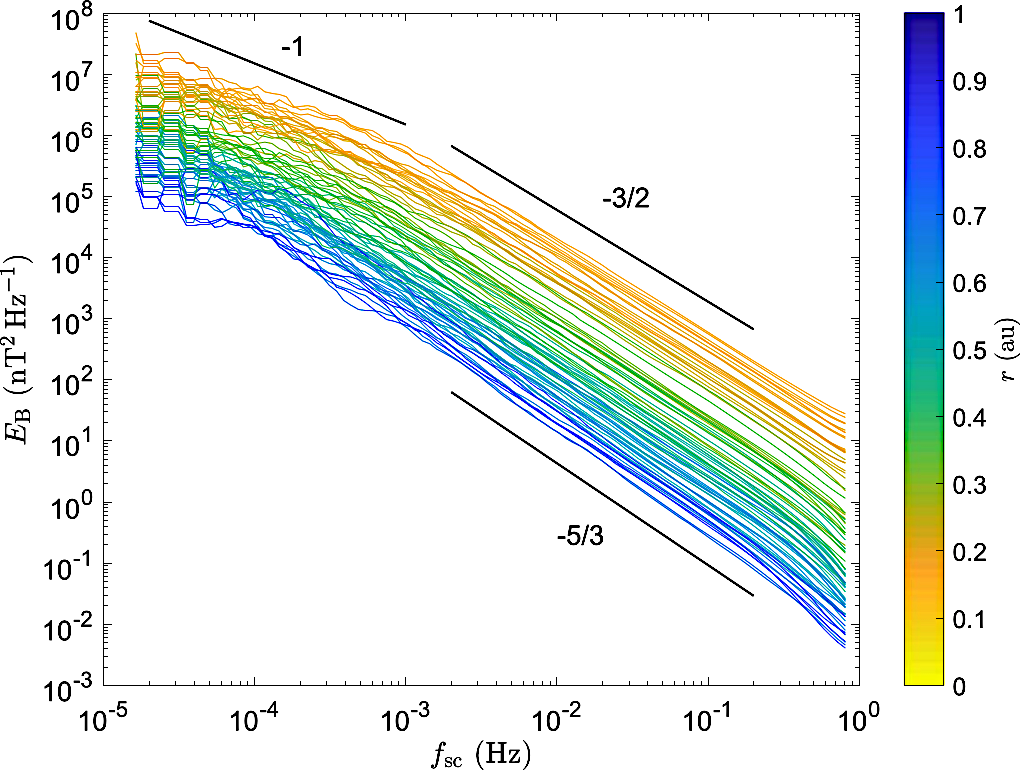}
\includegraphics[width=0.48\textwidth]{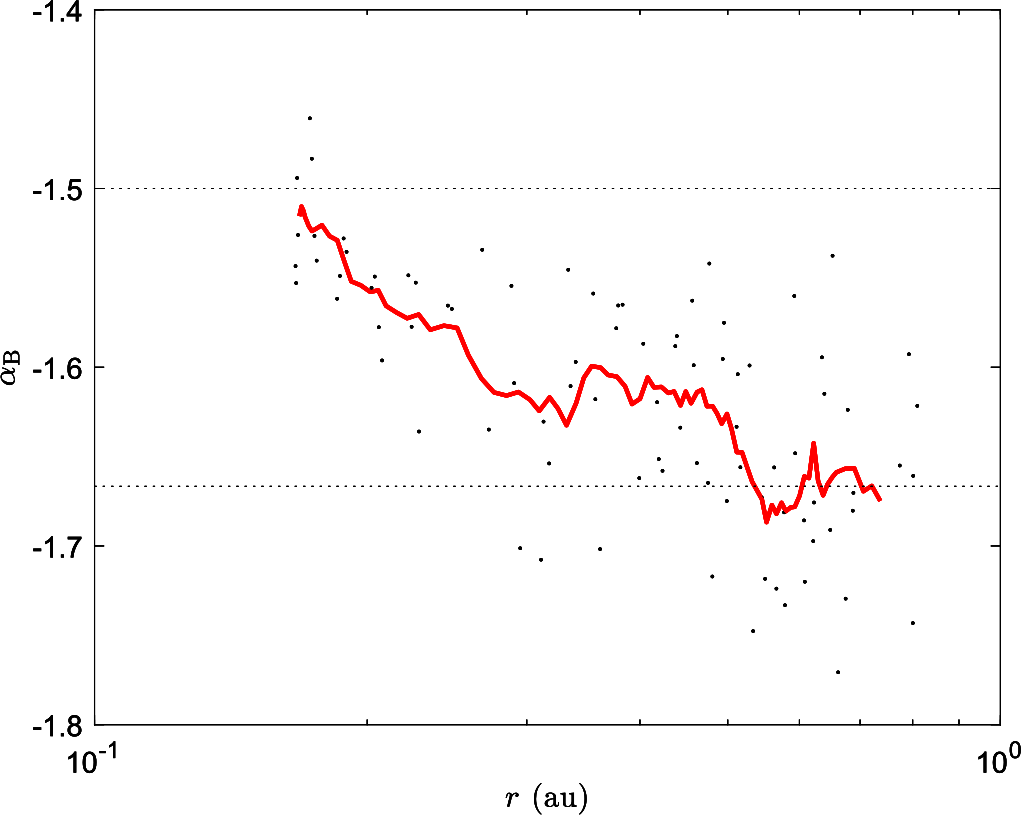}
\caption{(Left) Magnetic field spectra in the energy containing and MHD inertial range computed from PSP magnetic field data, (Right) Spectral slopes in the inertial range. The spectral slopes show a clear transition from Kolmogorov like to IK like as the radial distance decreases. (Reproduced with persmission from \cite{ChenApJS20}.)}
\label{ChenSlopes}
\end{figure}

The turbulent power, at the energy containing and inertial range scales, increases sunwards, consistent with earlier findings \cite{BelcherJGR74, BavassanoJGR82, HorburyJGR01, BrunoEMP09, ChenApJS20}.The inertial range slopes of the magnetic field power spectra gradually shift from Kolmogorov-like $k^{-5/3}$ \cite{Kolmogorov41a} to Irochnikov-Kraichnan like $k^{-3/2}$ \cite{KraichnanPFL65} with decreasing radial distance \cite{ChenApJS20}. Fig. \ref{ChenSlopes} shows magnetic field spectra from the first two orbits of PSP in the left panel and the computed slopes in the right panel. Net power in the magnetic fluctuations is seen to rise sunwards. Most of the spectra show a $1/f$ range at the largest scales and the inertial range slopes transition from $-3/2$ in the inner heliosphere to $-5/3$ at 1au. Another analysis, using Hilbert Huang Transform (HHT), computed multifractal scalings for magnetic field fluctuations observed by PSP \cite{AlbertiApJ20}. The spectral slopes are confirmed to transition from IK like -3/2 to Kolmogorov like -5/3 at 0.4au. The turbulence also shows a transition at this radial distance from monofractal to multi-fractal nature.

The outer scale of the turbulence, as characterized by the correlation time, changes from $\sim 500 s$ at around 0.17AU to a couple of hours at around 1AU \cite{ParasharApJS20, ChenApJS20}. The spectral break in Fig. \ref{ChenSlopes} between $f^{-1}$ and $f^{-5/3}$ regimes is seen to shift to lower frequencies with increasing heliocentric distance, consistent with earlier studies \cite{BavassanoJGR82, BrunoEMP09, BrunoApJL14}. The large-scale break frequency has been shown to follow a power-law variation with heliocentric distance \cite{BrunoLRSP13, ChenApJS20, WuApJL20}. The power-law exponent was found to be $\sim 1.1$ by \cite{ChenApJS20}. Although \cite{WuApJL20} did not perform a fit to their data, the qualitative behaviour shown by them is similar to that shown by \cite{BrunoLRSP13} with an exponent of $\sim 1.5$.  Interestingly, these variations are seen in previous studies as well \cite{HorburyAA96, KleinSW92, BrunoLRSP13, RuizSP14}. The slope found by \cite{ChenApJS20} is shallower than what was found by \cite{BrunoLRSP13} and steeper than those found by \cite{RuizSP14}. Recently, Cuesta et. al. \cite{CuestaApJS22} studied the heliocentric variation of the outer scale using data from Helios, Voyager, and three intervals from PSP. The Voyager values are consistent with \cite{RuizSP14}, Helios data show a slightly steeper rise, and the three PSP intervals are placed to indicate a very steel slope in the inner heliosphere. The turbulence evolution models \cite{ZankJGR96, MatthaeusJPP96} predict slopes shallower than -1 and are consistent with \cite{RuizSP14}. A comprehensive study is needed to understand the origins of such differences. At the kinetic scales, the break frequency between the inertial range and sub-proton range also shifts to lower frequencies as $r^{-1.1}$ \cite{DuanApJS20}. The kinetic scale break frequency is observed to be closely correlated with the cyclotron frequency along with other parameters such as plasma $\beta$.

Consistent with earlier observations, the Alfv\'enicity of the fluctuations decreases with heliocentric distance. The cross helicity as well as the energy budget of sunward and outward Els\"asser variables, decreases as the wind expands \cite{ShiAA21}. The dominant outward Els\"asser variable decays faster than the sunward Els\"asser variable (see Fig \ref{BavassanoJGR00}, reproduced from \cite{BavassanoJGR00, BrunoLRSP13}, and equivalent figures in \cite{ChenApJS20}). Some intervals in the inner heliosphere show a decrease in cross helicity across scales, indicating a possibility of strong velocity shears destroying the cross helicity in the inertial range \cite{ParasharApJS20}. MHD simulations with shears as well as analysis of Helios intervals with shear show a similar reduction in cross helicity at shear sites \cite{RobertsJGR92}.

\begin{figure}[!ht]
\centering
\includegraphics[width=0.7\textwidth]{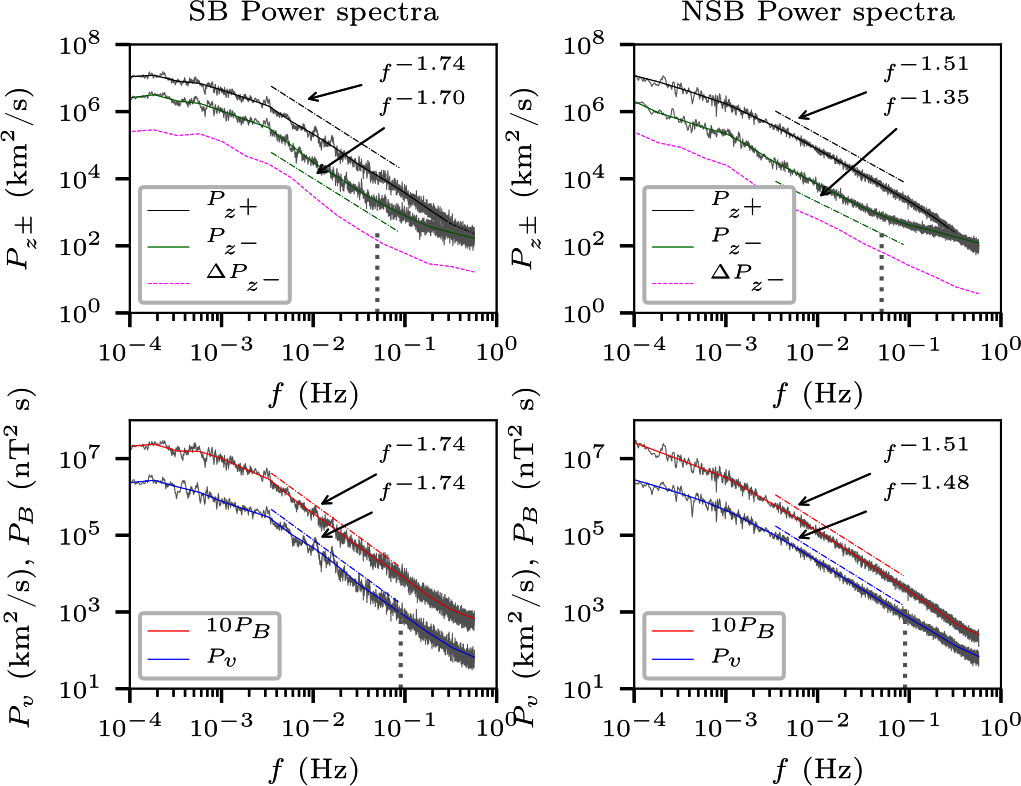}
\caption{Power spectra for the Els\"asser variables, velocity as well as the magnetic field in switchback (SB, left panels) and non switchback (NSB, right panels) intervals. The spectra show Kolmogorov like spectral slopes close to $-5/3$ inside SBs and IK like $-3/2$ in NSB intervals.}
\label{BourouaineApJ20}
\end{figure}

While many quantities important for turbulence, e.g. the magnitudes of wind speed, temperature, magnetic field and density, remain comparable between SBs and nearby `quiet' regions, some properties of turbulence such as spectral signatures, decay rates, and intermittency properties vary significantly between switchbacks and non-switchback intervals \cite{BourouaineApJL20, MartinovicApJ21}. Fig. \ref{BourouaineApJ20} shows power spectra in the spacecraft reference frame for intervals with and without switchbacks from PSP's first encounter in November 2018. Left panels show Els\"asser, velocity, and magnetic field spectra for the switchback interval and the right panels show the same spectra for non-switchback intervals. The spectra for all variables in the SB intervals show Kolmogorov like $f^{-5/3}$ spectra, while the NSB intervals show IK like $f^{-3/2}$ spectra. A potential reason behind this could be intense driving of turbulence by velocity shears that are likely responsible for switchbacks \cite{RuffoloApJ20}.  The more evolved turbulence in the SB regions produces larger PVI events with higher probability \cite{MartinovicApJ21}.

\begin{figure}[!ht]
\centering
\includegraphics[width=0.7\textwidth]{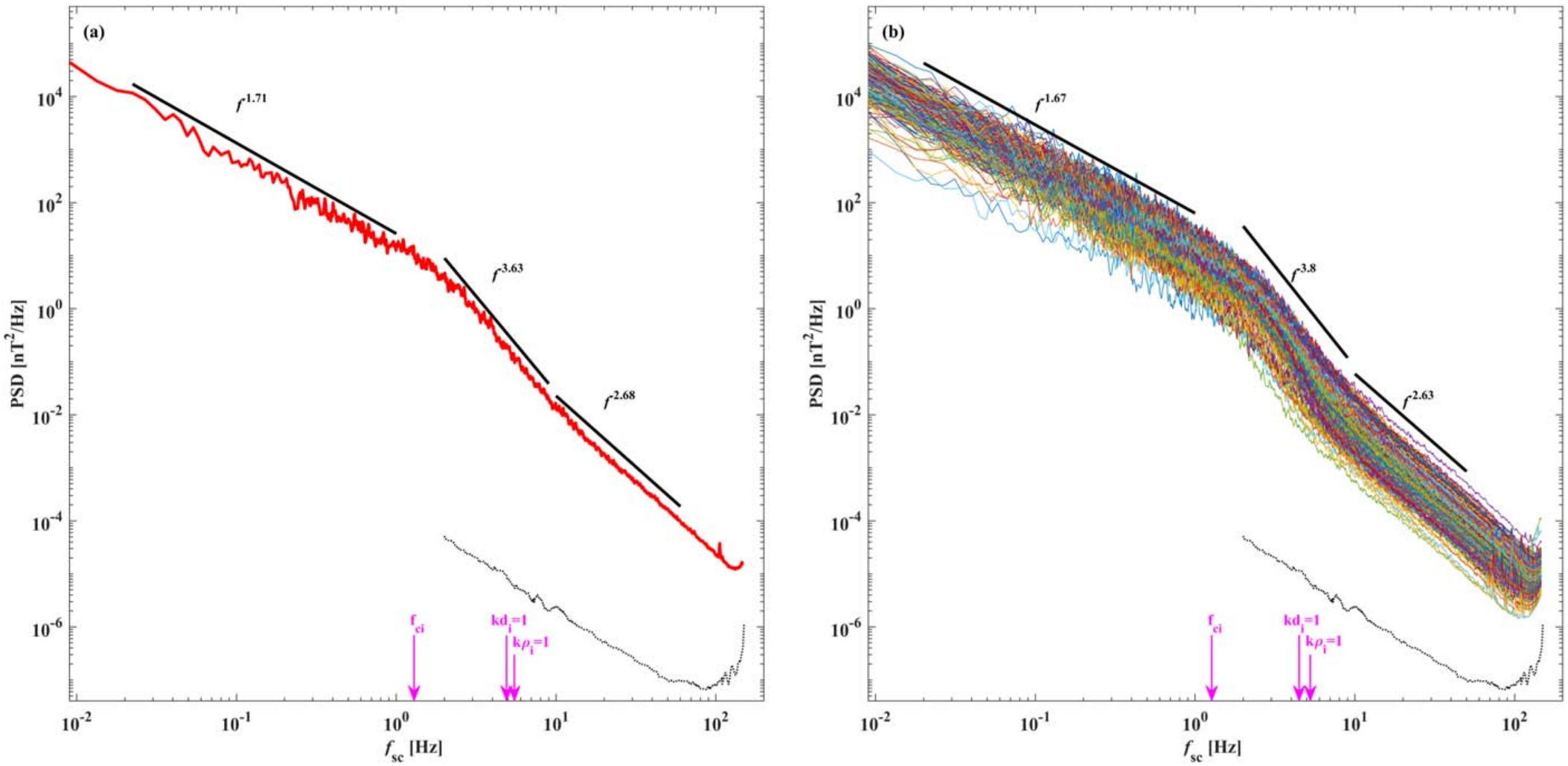}
\caption{The magnetic field spectra show a steep transition range at the ion kinetic scales. The transition range typically shows a spectral slope of $\sim -4$ before turning to the familiar spectral slope of $\sim -8/3$ at the higher frequencies. (Reproduced with permission from \cite{HuangApJL21}).}
\label{HuangApJL21}
\end{figure}

Another commonly observed feature of turbulence in the inner heliosphere is the enhanced steepening of magnetic field power spectra near the ion kinetic scales. Fig \ref{HuangApJL21} shows magnetic field spectra computed using the merged Fluxgate magnetometer (FGM) and Search Coil Magnetometer (SCM) data \cite{BowenJGR20, HuangApJL21}. The spectra shown are from encounter 1 of PSP, specifically from November 4-7, 2018. A transition range near proton kinetic scales is seen with a slope of $f^{-4}$. This sharp transition indicates a modified cascade or enhanced dissipation at kinetic scales or a combination thereof. The spectra return to $f^{-8/3}$ approaching electron kinetic scales. Such transition range has also been observed at 1AU in Wind observations \cite{DenskatJGR84, LeamonJGR99}. One of the possible explanations proposed for this transition is the ``helicity barrier'', which emerges in imbalanced turbulence and reduces electron dissipation of kinetic Alfv\'en waves (KAWs) \cite{MeyrandJPP21, SquireNatureAstron22}. 

\begin{figure}[!ht]
\centering
\includegraphics[width=\textwidth]{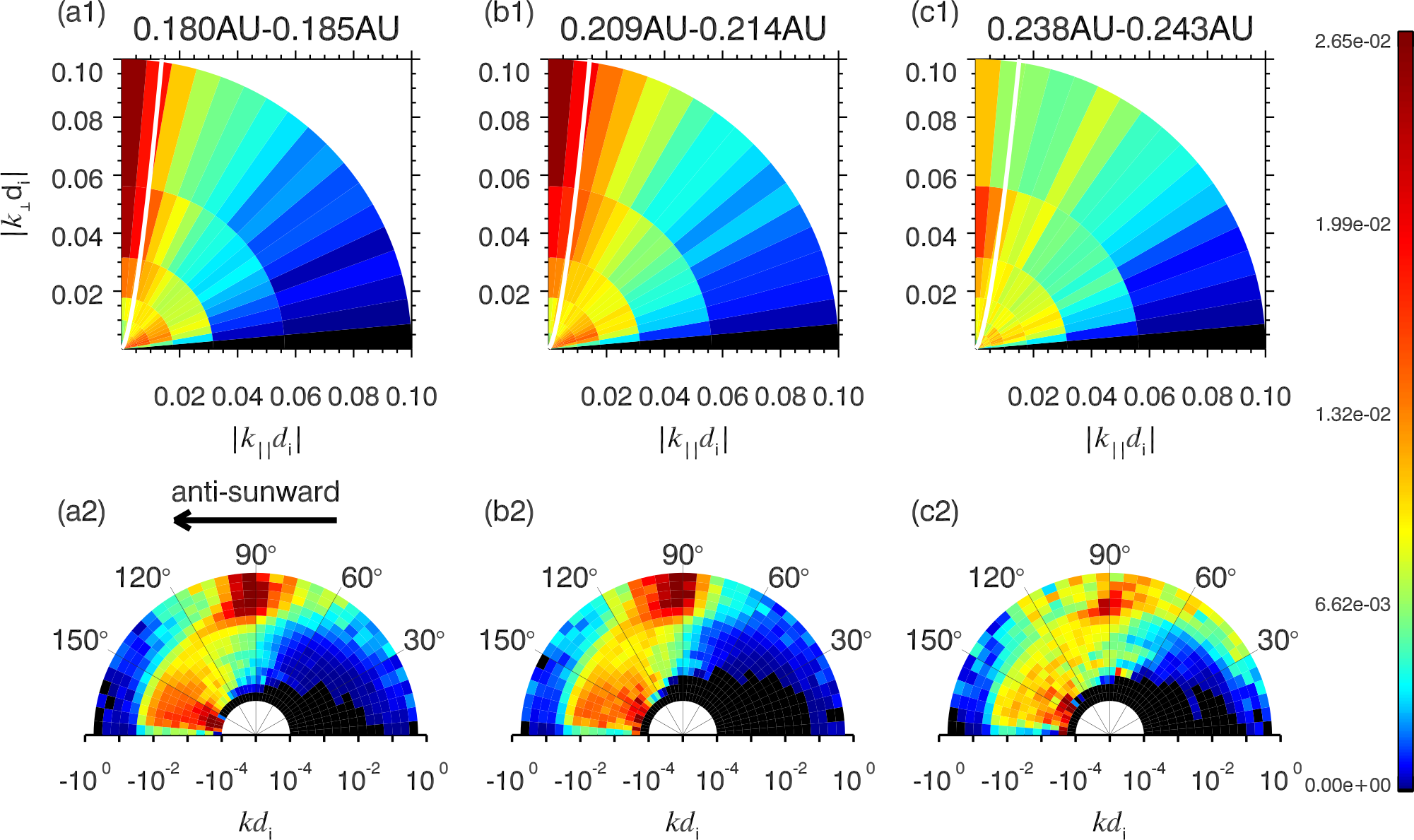}
\caption{PDFs of wavevectors identified in PSP data during the first encounter of PSP. Top row shows the PDFs in $k_\perp-k_\parallel$ plane and the bottom row shows the data in $k-\theta$ plane. Dominance of predominantly parallel wavevectors for large scales and predominantly perpendicular wavevectors for smaller wavenumbers is evident. (Reproduced with permission from \cite{ZhuApJL20})}
\label{ZhuApJL20}       
\end{figure}

As expected for MHD turbulence, the fluctuations are highly anisotropic in nature. Fig \ref{ZhuApJL20} shows an example from the first encounter of PSP \cite{ZhuApJL20}. The top row shows probability distribution functions (PDFs) of identified wave vectors in the $k_\perp - k_\parallel$ plane, and the bottom row shows the PDFs in $k-\theta$ plane. The wavenumbers studied were in the MHD regime. In the inertial range, for $kd_i < 0.02$ the fluctuations are predominantly parallel. Towards the tail of the inertial range, for $kd_i > 0.02$, the fluctuations are predominantly perpendicular. Also the outward component of the Alfv\'enic fluctuations dominates. The fluctuations are still highly anisotropic in the transition and kinetic ranges, with power in perpendicular fluctuations being roughly an order of magnitude higher than the parallel fluctuations \cite{DuanApJL21,ZhangApJL22}.

In summary, the turbulence observed by PSP shows a lot of structure with interspersed switchbacks and ``quiet'' Alfv\'enic intervals. The SBs generally have more evolved turbulence compared to NSB intervals. The power is distributed highly anisotropically in the wavenumber space. The Alfv\'enicity of the wind decreases with increasing heliocentric distance, and the imbalance of turbulence is likely responsible for a steep transition range near proton kinetic scales. We now discuss the PSP observations of energy transfer across scales.

\subsection{Energy at large scales}

\begin{figure}[!ht]
\centering
\includegraphics[width=\textwidth]{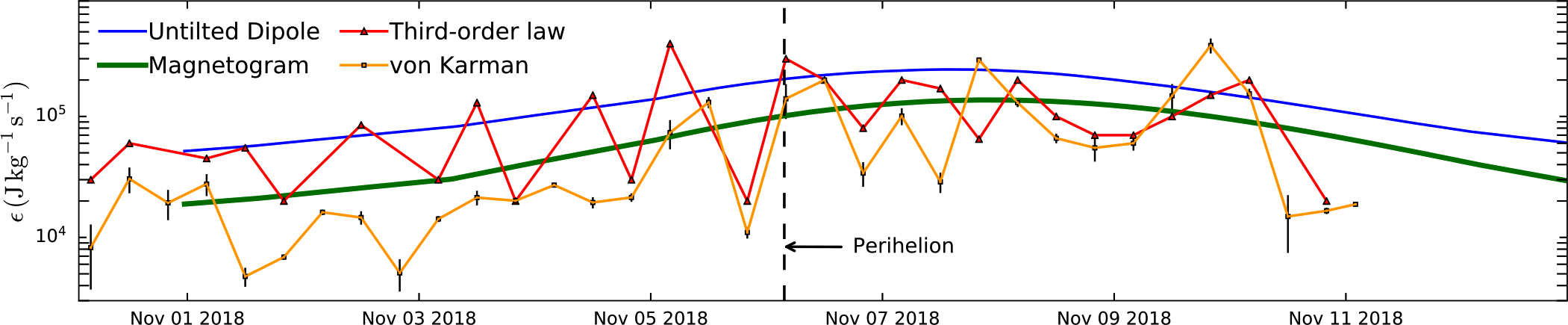}
\caption{Decay rates from PSP's first encounter, computed in two ways: squares with orange line show the large scale von K\'arm\'an estimate and triangles with red line show the third order law estimates. The thin blue line and thick green line show two global simulations with untilted dipole and the relevant solar magnetogram as inputs for solar magnetic field. (Reproduced with permission from \cite{BandyopadhyayApJS20b})}
\label{Bandyopadhyay-radial}
\end{figure}
The plasma driven at large scales by various solar inputs, in-situ velocity shears, and instabilities receives energy at a rate quantified by the von K\'arm\'an decay rate. Fig \ref{Bandyopadhyay-radial} shows the decay rates computed from encounter 1 of PSP \cite{BandyopadhyayApJS20b}. The squares with orange line show the von K\'arm\'an decay rate based on large scale parameters and the triangles with red line show the energy transfer rates estimated by fitting the third order structure functions following the Politano Pouquet law (see next subsection for more details). The two smooth solid lines show two decay rates computed from two global simulations \cite{ChhiberApJS19}. The thin blue line is a simulation with sun's magnetic field represented as an untilted dipole, and the green line is a simulation with the magnetogram from the time of encounter 1. The decay rates increase with decreasing heliocentric distance, changing by more than an order of magnitude within a few solar radii in which the encounter data were collected. The overall decay rates obtained near the perihelion are a couple of orders of magnitude larger than $1000$ J Kg$^{-1}$s$^{-1}$ decay rate that is observed at 1au \cite{CoburnPTRSA15}. The von K\'arm\'an decay rates and the third order law estimates match better in the outgoing part of E1, presumably because that was highly Alfv\'enic slow wind, which might have a higher density of intermittent structures. The global simulations estimate the decay rates reasonably well with the magnetogram simulation overlapping really well with the two estimates from PSP. 

The heating rates in the inner heliosphere have also recently been computed from angular broadening studies of Crab nebula \cite{RajaApJ21}. The angular broadening observations were used to estimate the average density perturbations, which were then converted to velocity fluctuations assumig kinetic Alfv\'en wave like properties. The decay rates computed from these velocity fluctuation levels are in the same ballpark as the numbers quoted in \cite{BandyopadhyayApJS20b}.


\begin{figure}[!ht]
\centering
\includegraphics[width=0.5\textwidth]{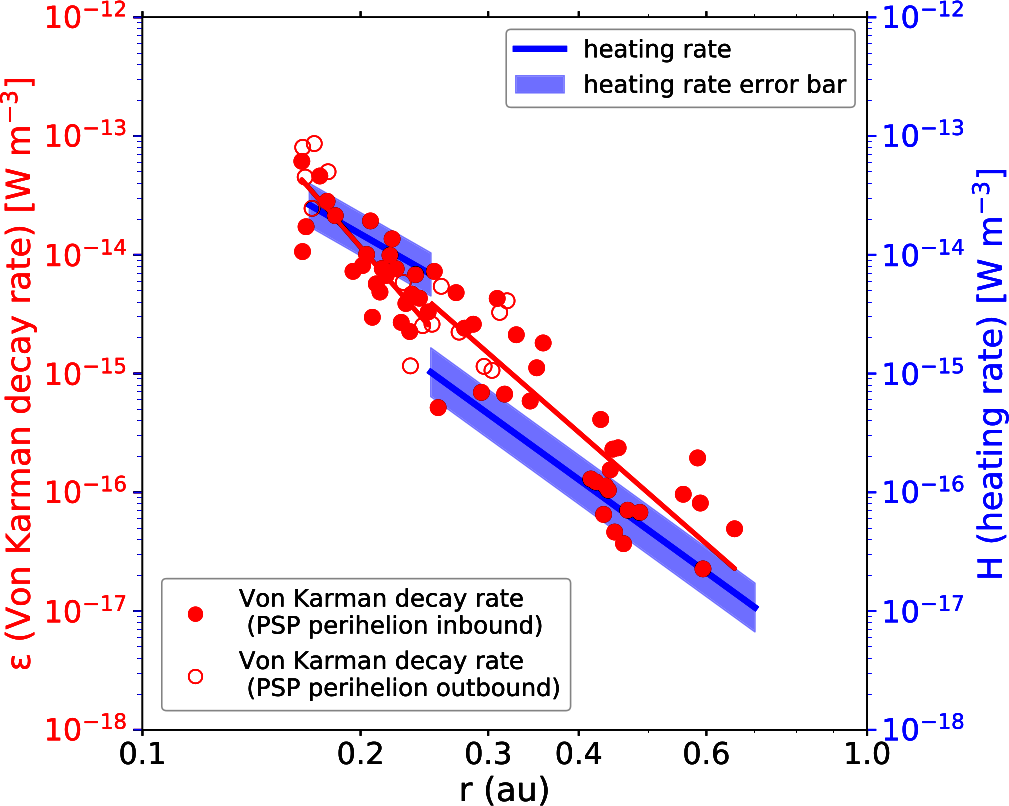}
\caption{von K\'arm\'an decay rates compared to solar wind heating rates estimated from first three encounters of PSP. The red symbols represent decay rates and the lines and shaded region represent proton heating rates. (Reproduced with permission from \cite{WuApJ22a})}
\label{WuApJ22}       
\end{figure}

As seen in simulations \cite{WuPRL13, ParasharApJ15, MatthaeusApJL16, ShayPP18} and in previous observations \cite{VasquezJGR07, StawarzApJ09, CoburnPTRSA15}, the von K\'arm\'an decay rate determines the rate of dissipation even in kinetic plasmas. PSP has already been used to test this balance in the inner heliosphere. Fig \ref{WuApJ22} shows the von K\'arm\'an decay rates and proton heating rates esitmated using first three encounters worth of data \cite{WuApJ22a}. The red dots represent the von K\'arm\'an decay rate and the lines+shaded regions represent estimates of solar wind heating rates of protons \cite{TuSSR95, WuApJL20, MartinovicApJS20}. The von K\'arm\'an decay rate is not only consistent with the energy supply rate estimated from the evolution of the large scale break frequency (see fig. 3 of \cite{WuApJ22a}) but is also consistent with proton heating rate. This balance between the energy input rate and dissipation implies that the energy cascades in the inertial range down to kinetic scales in a conservative fashion.

\subsection{Energy in the inertial range}
\begin{figure}[!ht]
\centering
\includegraphics[width=0.4\textwidth]{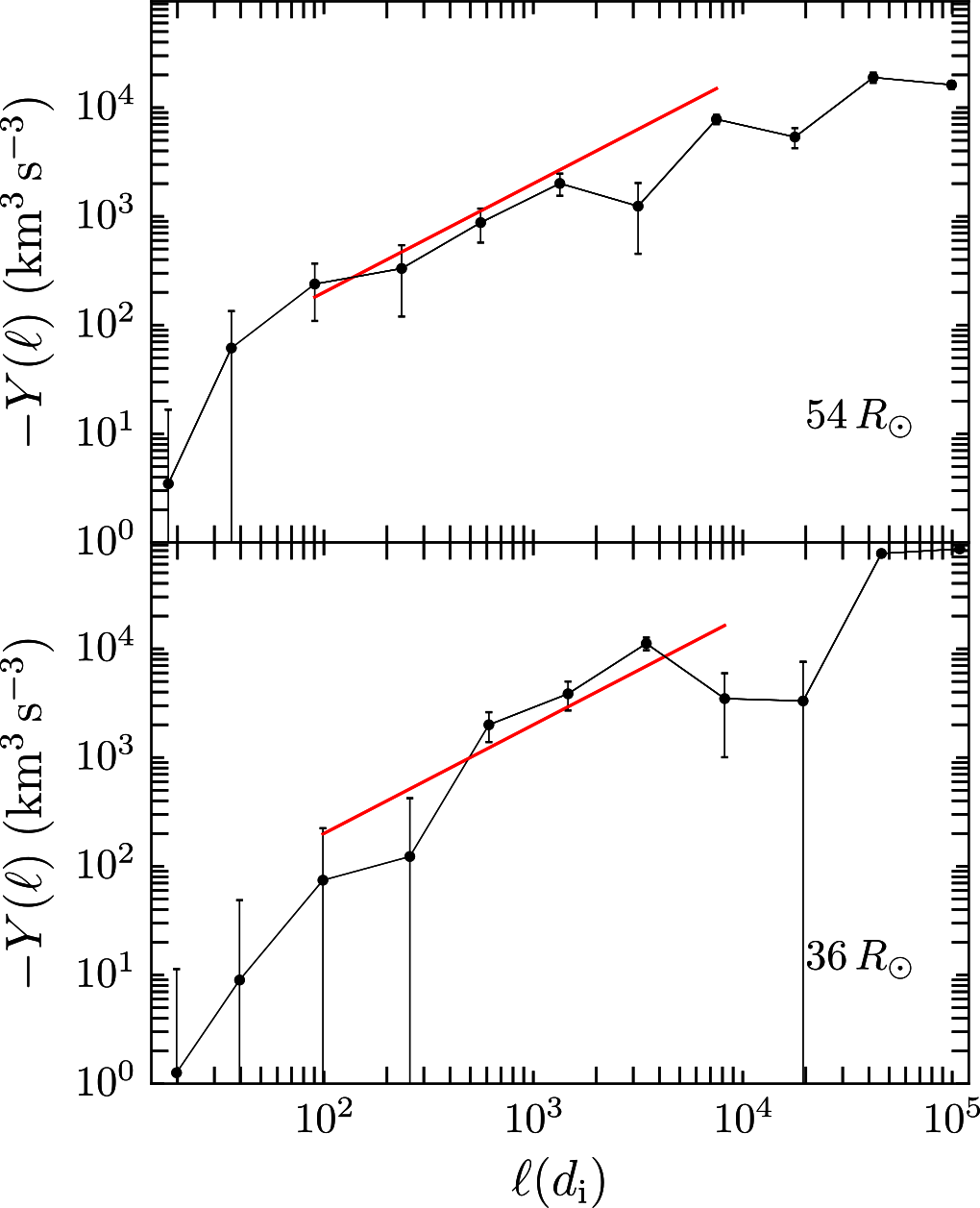}
\caption{Incompressible Yaglom fluxes computed using PSP data from the first encounter \cite{BandyopadhyayApJS20b}. The red lines represent fits used to obtain the cascade rate. (Reproduced with permission from \cite{BandyopadhyayApJS20b}).}
\label{Bandyopadhyay-yaglom}
\end{figure}

The energy input at the large scales cascades down to kinetic scales through the inertial range. This cascade rate can be quantified using the Politano Pouquet MHD generalization of Yaglom's 3rd order law (see e.g. Eqn. \ref{eqn:thirdorder}). Fig \ref{Bandyopadhyay-yaglom} shows two examples of the third order fluxes computed at two different radial distances using the incompressible version of the PP law. The red lines show linear scaling laws. These linear scalings, when identified, can be used to estimate the cascade rate in the inertial range \cite{BandyopadhyayApJS20b}. The cascade rates identified this way are compared with the von K\'arm\'an decay rates in Fig \ref{Bandyopadhyay-radial} as squares with an orange line. The third order law decay rates are comparable to the von K\'arm\'an decay rates and are also comparable to simulation findings. A recent study compared the cascade rates in sub-Alfv\'enic wind with super-Alfv\'enic wind \cite{ZhaoApJL22}. The cascade rate was computed to be higher in the sub-Alfv\'enic interval compared to the super-Alfv\'enic interval although longer sub-Alfv\'enic intervals would be needed to get statistically significant results. The cascade rates were also dominated by compressive terms compared to incompressive terms.

It is possible to estimate the decay rates from the isotropic and anisotropic versions of the PP law. Both isotropic and anisotropic cascade rates increase with decreasing heliocentric distance. This increase correlates well with the well established increase in fluctuation amplitude, Alfv\'enicity, and temperature with decreasing heliocentric distance \cite{AndresArXiv21}. The compressible cascade rates computed from PSP's first encounter have been compared to compressible cascade rates at 1AU (THEMIS data) and 1.6AU (near Mars using MAVEN data). The compressible cascade rates show a drop by five orders of magnitude \cite{AndresApJ21} between 0.2au (PSP) and 1.6au (MAVEN). The density fluctuations are larger in the inner heliosphere. As the wind expands, it approaches a nearly incompresible state \cite{MatthaeusJGR90, AdhikariApJ17}. This change in the nature of solar wind turbulence could be related to the decrease in compressible decay rates.

\begin{figure}[!ht]
\centering
\includegraphics[width=\textwidth]{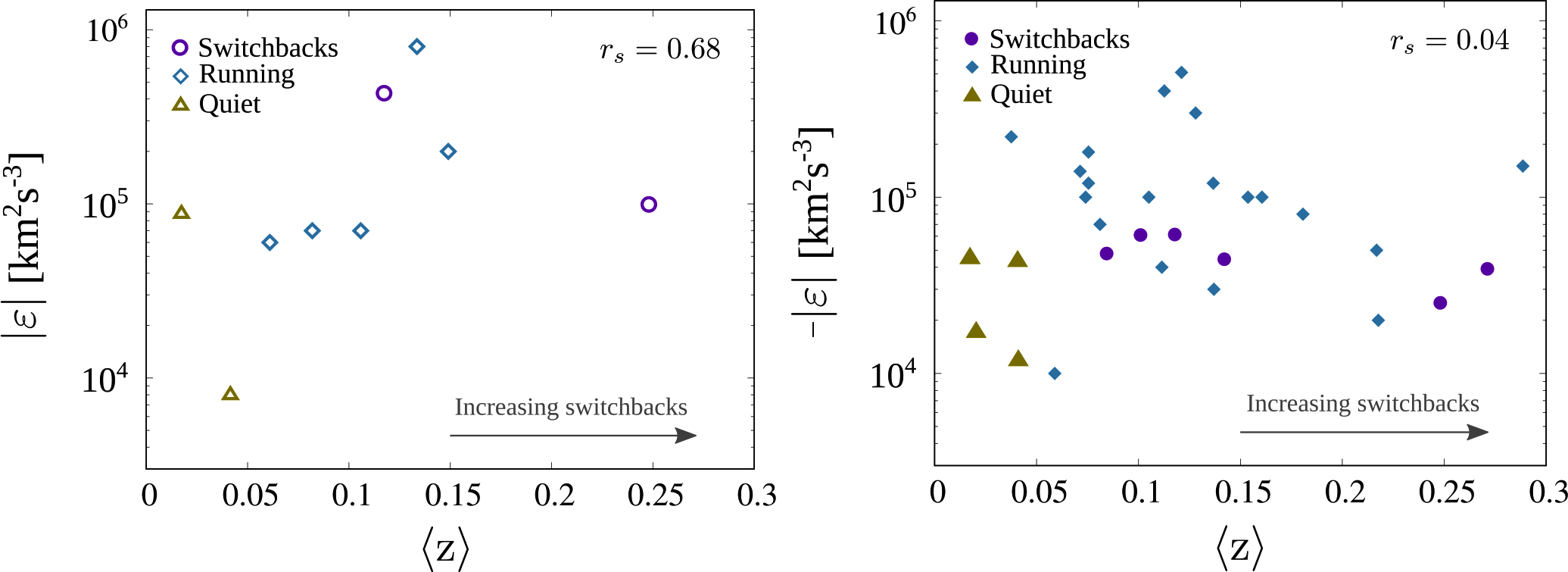}
\caption{Cascade rates computed using Yaglom law inside and outside switchbacks \cite{HernandezApJL21}. (Reproduced with permission from \cite{HernandezApJL21})}
\label{HernandezApJL21}
\end{figure}

The presence of switchbacks can affect the cascade rates as well. Fig. \ref{HernandezApJL21} shows the cascade rates computed in switchbacks identified during the first encounter of PSP \cite{HernandezApJL21}. Some switchback intervals show positive decay rate and some show negative decay rates. Although the interpretation of a negative cascade rate is unclear, it could imply a local transfer of energy to larger scales. The intervals that show a net positive cascade rate in the inertial rage show a visibly identifiable correlation with the switchback parameter $Z$ \cite{HernandezApJL21} \cite[see][for the definition of $Z$ and detailed analysis of switchback properties]{deWitApJS20}. No such correlation is observed in cases when a negative cascade rate is identified in an interval \cite{HernandezApJL21}. The finding that the cascade rate is enhanced by the presence of switchbacks is consistent with the notion of more evolved turbulence in switchbacks (see e.g. Fig. \ref{BourouaineApJ20}). A local weak formulation of PP98 law gives a local in space and time dissipation function $\mathcal{D}$ \cite{DavidApJ22}. This local dissipation function also shows different behavior in and outside the switchbacks. It is predicted to follow $\sigma^0$ scaling in the inertial range, $\sigma^2$ in the viscous/dissipative range, and $\sigma^{-1}$ at the discontinuities where $\sigma$ is the scale at which the local dissipation measure is computed. The dissipation measure $\mathcal{D}$, when averaged over the interval, shows behaviour similar to the third order law but locally it shows an unexplained scaling of $\sigma^{-3/4}$ at the locations of switchbacks. 

\subsection{Energy at kinetic scales}
The cascaded energy is partially dissipated at the proton kinetic scales and part of it is cascaded down to electron scales. Various dissipative mechanisms and pathways have been proposed to explain dissipation in kinetic plasmas. These include wave-particle interactions such as cyclotron resonance \cite{HollwegJGR02} and Landau damping \cite{TenbargeApJ13}, stochastic heating \cite{ChandranApJ10-1, MartinovicApJ21}, and heating at intermittent locations \cite{ParasharPP11, TenbargeApJ13, WanPP16}. Heating at intermittent locations can happen because of processes such as reconnection \cite{ShayPP18}, and possibly because of Landau damping as well \cite{TenbargeApJ13}. A mechanism agnostic approach focuses on the action of pressure strain interaction term, which also happens intermittently \cite{YangPP17}  (see section 2 for a detailed discussion).

PSP's state-of-the-art very high time cadence measurements in the inner heliosphere have enabled identification of various wave modes including circularly polarized waves and electrostatic waves \cite{BowenApJS20, BowenApJ20, VernieroApJS20, MalaspinaApJS20, VechAA21, ZhaoApJ21, CattellApJL21}. The waves are found to last anywhere from a few seconds \cite{BowenApJS20} to ``wave storms'' a few hours long \cite{VernieroApJ22}. Coincident with the ion scale waves is a broadening of the beam of proton velocity distribution function (VDF) that is being termed as ``hammerhead'' of the proton VDF. It was also shown by \cite{VernieroApJ22} that these features are consistent with the expectations of quasilinear diffusion in velocity space in the presence of waves. A recent paper studies proton VDFs in the presence of cyclotron waves \cite{BowenArXiv21}. The VDFs show signatures consistent with qualisilinear diffusion in the velocity space, indicating a possibility of cyclotron heating. This quasilinear diffusion is accompanied by steepening of magnetic power spectra to $\sim f^{-4}$ near proton scales. A potential explanation for such a combination of observations was recently proposed \cite{SquireNatureAstron22}. It is suggested that a helicity barrier inhibits a cascade of energy to smaller scales \cite{MeyrandJPP21}. This results in a proton scale build up of energy and a steep transition range just below proton scales. This proton scale build up of energy in-turn generates ion-cyclotron waves, which can participate in preferential resonant heating of ions \cite{SquireNatureAstron22}. This steepening of magnetic spectra at proton scales has also been used to estimate a dissipative removal of energy at proton scales. It was estimated that the power removed from the magnetic fluctuations at proton scales is sufficient to heat the solar wind \cite{BowenPRL20}.

\begin{figure}[!ht]
\centering
\includegraphics[width=0.49\textwidth]{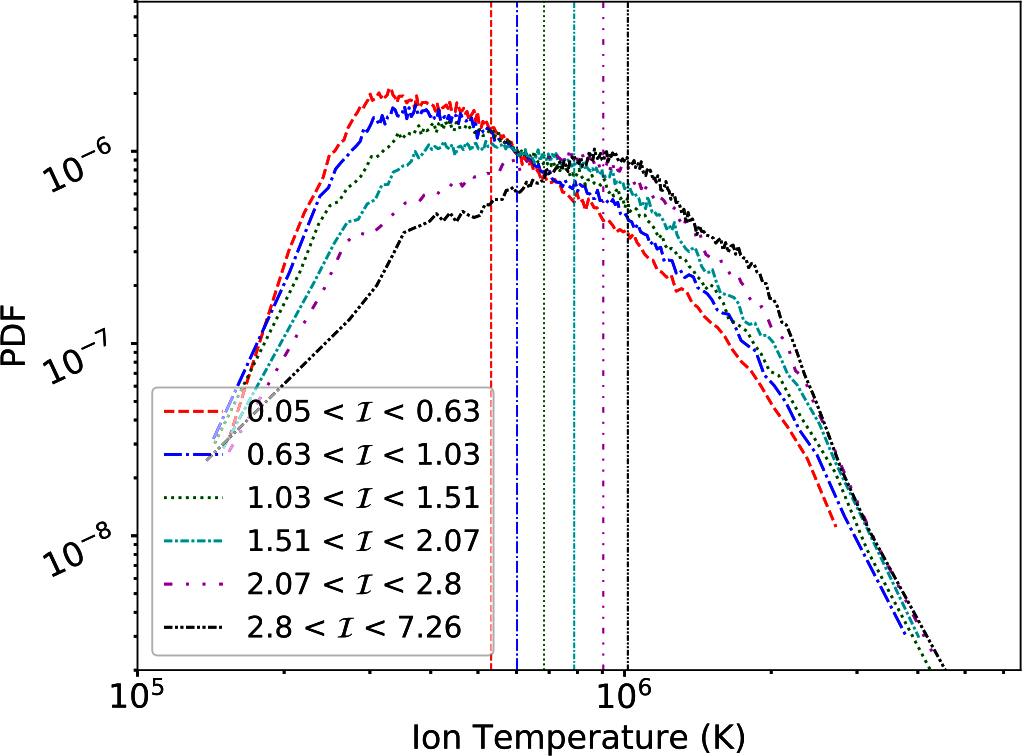}
\caption{PDFs of temperature conditioned on PVI value. Verticle lines show the median temeprature for each subset of the data. A clear rise of the median temperature with PVI is seen. (Reproduced with permission from \cite{QudsiApJ20}).}
\label{Qudsi-TPVI}
\end{figure}

Another pathway to dissipation is stochastic heating \cite{ChandranApJ10-1, CerriApJ21} where particles experience stochastic kicks by the changing electric potential during their orbit (see section 2 for a discussion). The stochastic heating rate increases with decreasing heliocentric distance as $Q_\perp \propto r^{-2.5}$ \cite{MartinovicApJ21}. It is larger in the fast wind compared to slow wind. Moreover, stochastic heating is enhanced inside the switchback regions. As mentioned before, stochastic heating enhances in the presence of intermittent structures. SBs show enhanced probability of finding large PVI values, and hence the turbulence in SBs is likely more intermittent. This could potentially enhance the stochastic heating rate. There is ample evidence for intermittent heating at 1au as well as in the inner heliosphere.


At 1AU, the protons are observed to be hotter near strong PVI sites \cite{OsmanApJL11}. Similar analysis in the inner heliosphere using PSP yields no new surprises. Fig \ref{Qudsi-TPVI} shows the probability distribution functions of temperature conditioned on PVI values. The probability of finding a higher temperature, say greater than $10^6$K, is higher for data conditioned on higher PVI values and the opposite is true for lower temperatures, say a few $10^5$K. This is also quantified by the increasing median value of temperature with increasing PVI value. Vertical lines in Fig \ref{Qudsi-TPVI} show this systematic trend in the median temperature rise, indicating hotter populations for larger PVI values.

\begin{figure}[!ht]
\centering
\includegraphics[width=0.5\textwidth]{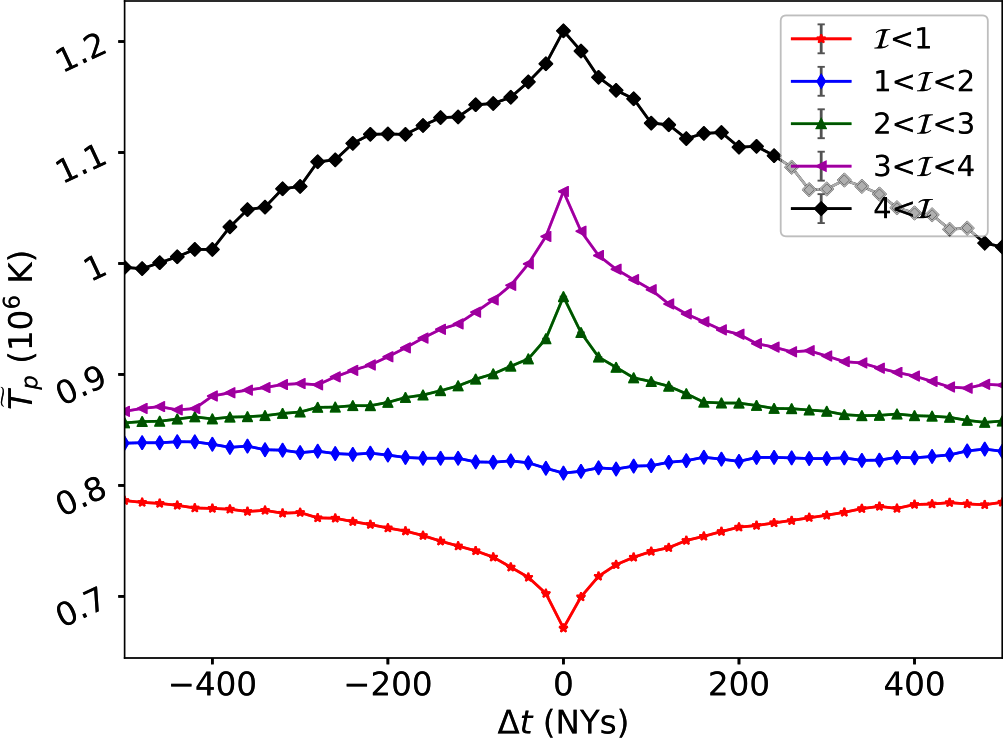}
\includegraphics[width=0.49\textwidth]{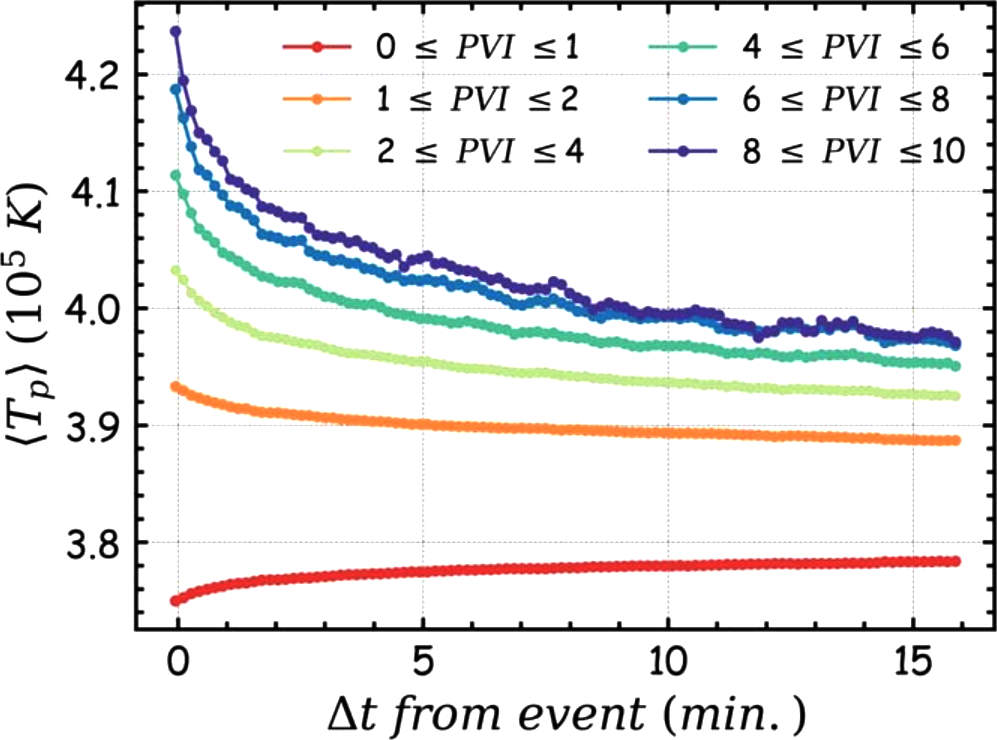}
\caption{Conditioned temperature of protons near PVI structures (left: from second half of PSP's encounter 1, right: from first six encounters' worth of data). See text for details. Mean temperatures are higher near higher PVI values. (Reproduced with permission from \cite{QudsiApJS20, Sioulas_2022})}.
\label{Qudsi-Tdt}       
\end{figure}

The proximity of dissipation and intermittent structures can also be quantified by computing the average temperature in the vicinity of PVI structures. This technique has been used to not only identify heating at intermittent structures \cite{OsmanPRL12a}, but also to show enhanced energetic particle fluxes near such intermittent structures \cite{TesseinApJL13}. The average is computed as $\Tilde{T}_p(\Delta t, \theta_1, \theta_2) =  \langle T_p(t_\mathcal{I}+\Delta t) \lvert \theta_1 \lt \mathcal{I} \lt \theta_2 \rangle $ where $\Tilde{T}_p$ is the conditional average temperature for all events, $\Delta t$ is the lag relative to the PVI position, $t_\mathcal{I}$ is the location of PVI event, and $\theta_1$ and $\theta_2$ are the thresholds for seleccting particular PVI values. Fig \ref{Qudsi-Tdt} shows conditional average temperatures for various PVI thresholds from the second half of PSP's encounter 1 in the left panel and from the first six encounters in the right panel. Evidently, the temperature in the vicinity of larger PVI values is larger, even out to about a correlation length away, when compared to smaller PVI values. The fact that PVI values appear to be clustered in the inner heliosphere \cite{ChhiberApJS20} could be one of the potential reasons behind consistently higher temperatures this far away from the intermittent PVI events.

Another possibility to study intermittent dissipation is to use LET, the local kernel of the third order law (Eqn. \ref{eqn:thirdorder}). Although LET varies significantly from one point to the other, its average at a given scale quantifies the net flow of energy into/out of that scale \cite{SorrisoJPP18}. In a similar analysis to Fig. \ref{Qudsi-Tdt} using Helios data, the LET was shown to correlate better than PVI with the mean temperature even though there was shown to be a strong correlation between LET and PVI \cite{Sorisso-ValvoSP18}. A potential reason behind the apparent lack of correlation between PVI and temperature in Helios data could be an artefact of the coarse resolution of Helios data which would not allow resolution of large small scale increments and hence the most intense PVI structures. PSP, with its fast measurements, allows resolution of PVIs at much finer scales and recovers the behaviour shown in Fig. \ref{Qudsi-Tdt}. 

\begin{figure}[!ht]
\centering
\includegraphics[width=0.49\textwidth]{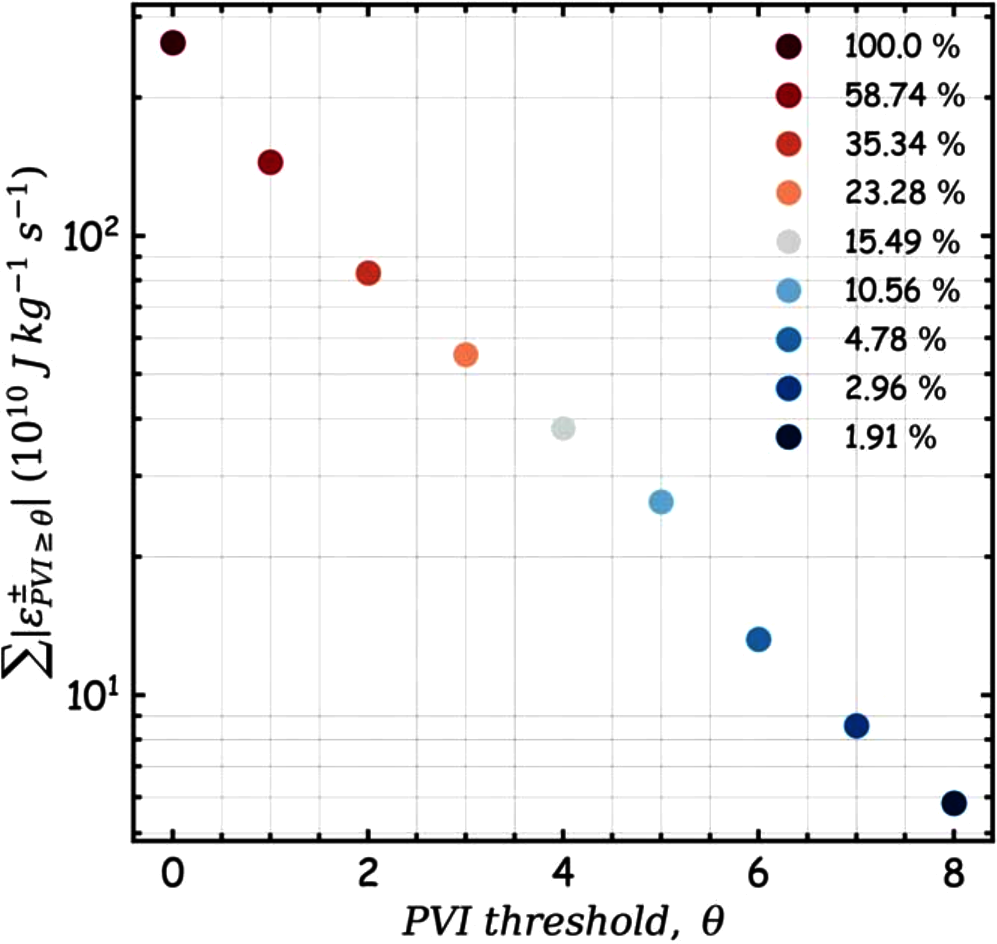}
\caption{Dissipation measure contributions from PVI structures. The sum of the absolute value of the LET measure, conditioned over PVI decreases with increasing PVI threshold. PVI $\gt 5$ contribute $\sim 11\%$ to the net dissipation \cite{OsmanPRL12a, Sioulas_2022} (Reproduced with permission from \cite{Sioulas_2022})}
\label{SioulasApJ22}       
\end{figure}

Larger PVI events occupy smaller fraction of the total volume but contribute significantly more to the budget of internal energy $U = C n k_B T$, where $C$ is the specific heat capacity, $k_B$ is the Boltzmann constant, $n$ the density, and $T$ the temperature. At 1AU, using ACE data, it has been shown that PVI $> 2.4$ occupy only 19\% of the volume but contribute 50\% to the internal energy budget, while PVI $> 5$ occupy only 2\% of the volume but contribute $\sim 11\%$ to the internal energy budget \cite{OsmanPRL12a}. Fig. \ref{SioulasApJ22} shows the total absolute value of LET conditioned on PVI threshold from first six encounters of PSP \cite{Sioulas_2022}. There are some similarities and differences in these findings when compared to Osman et. al. \cite{OsmanPRL12a}. Here PVI $>2$ contribute $\sim 35\%$ to the absolute LET measure, significantly less than what was estimated by \cite{OsmanPRL12a}. On the other hand, PVI $>5$ contribute $\sim 11\%$ which is consistent with the findings of \cite{OsmanPRL12a}. Evidently LET is a very different measure of dissipation compared to internal energy. The differences in the inferences of \cite{Sioulas_2022} and \cite{OsmanPRL12a} could potentially stem from the different nature of these measures, or from other considerations such as enhanced possibility of wave damping \cite{BowenArXiv21, SquireNatureAstron22}, or from yet largely unexplored clustering properties of the PVI events \cite{ChhiberApJS20} or the transition of turbulence to monofractal behaviour at sub-proton scales potentially owing to scale invariance of current sheets in the kinetic range \cite{ChhiberApJL21}. Moreover, LET and PVI are scale dependent quantities. Similar studies with varied LET computation scale are required to explore such connections further. 

\section{Summary and Conclusions} \label{conclusions}

The origins and evolution of the solar wind have been a mystery ever since its prediction and discovery \cite{ParkerApJ58a, NeugebauerJGR66, TuSSR95, MarschLRSP06, BrunoLRSP13, VerscharenLRSP19}. Many iconic missions have enhanced our understanding of the processes that contribute to origins and evolution of the solar wind. However, the solar coronal dynamics were largely studied using remote observations until very recently. Parker Solar Probe is allowing the exploration of hitherto unexplored regions approaching and 
inside the Alfv\'en critical surface. Recently PSP entered the magnetically dominated solar corona  between 19.7 and 18.4 solar radii $R_\odot$ \cite{KasperPRL21}. As PSP approaches closer to the sun, eventually reaching 9.8 $R_\odot$, it will allow exploration of various processes active in the solar corona, including the turbulence that potentially plays an important role in the heating of the Solar Corona. 

In this paper we have focused primarily on the transfer of energy across scales
in the inner heliosphere as observed by the Parker Solar Probe in the last four years. 
Rather than adopting a more detailed and mathematical {\it scale to scale} transfer approach
\cite{Verma2019} that may not be amenable to spacecraft
analysis, we adopt a more empirical approach based on the taxonomy of 
energy-containing scales, inertial range scales and kinetic scales. 
Based on such an approach, 
a significant amount of work has been done in the last four years on various aspects of solar wind properties and the turbulence. We refer the reader to Raouafi et al \cite{RaouafiEA22SSR}
for a detailed discussion of many other aspects that were not covered here. 

As the solar wind expands from inner heliosphere to the outer heliosphere, its turbulent nature changes dramatically \cite{DeforestApJ16, ChenApJS20}. Expanding from the solar corona, the wind becomes supersonic in the Alfv\'en critical region. Beyond the Alfv\'en critical region, the magnetic field loses control and velocity shears at flux tube boundaries can create flocculation, isotropizing the wind at the largest scales, and advancing the wind's turbulent evolution \cite{DeforestApJ16, ChhiberApJL18, RuffoloApJ20}. 

The amplitudes of turbulent fluctuations decrease radially outwards while the outer scale gets larger. The outer scales increases as a power-law with increasing heliocentric distance. The power-laws exponents found for outer scale vary between $0.5-1.5$ whereas theoretical models predict the exponent to be generally less than 1 \cite{KleinSW92, HorburyAA96, RuizSP14, BrunoApJL14, ChenApJS20, WuApJL20}. Intermittency increases from inner heliosphere up to 1au. A clear transition is seen at roughly 0.4au \cite{AlbertiApJ20, CuestaApJL22}. Although it might appear that the bandwidth available for the turbulent cascade might keep increasing with increasing outer scale even beyond 1au, the inner scale (for example $d_i$) expands faster than the outer scale. This results in a reduced Reynolds number with increasing heliocentric distance. The reduction in the Reynolds number in-turn implies reduced intermittency, as quantified by radially decreasing small scale kurtosis and PVI values \cite{ParasharApJL19, CuestaApJS22}. 

The fluctuations are highly Alfv\'enic and imbalanced in the inner heliosphere with outward propagating fluctuations dominating the energy budget \cite{BavassanoJGR00}. The Alfv\'enicity drops as the wind expands with the dominant species ($\m{z}^+$) decaying significantly faster than the minor species ($\m{z}^-$) \cite{GoldsteinGRL95, BreechEA08, RobertsJGR87, MatthaeusGRL04, StriblingMatt91}. This is reflected in the decreasing cross helicity with increasing heliocentric distance as well. 

Going sunwards, the enhanced fluctuation levels of velocity, magnetic field, as well as density imply that the energy input rate at the largest scales increases \cite{WuApJ22a}. The \vkh decay rate increases by two orders of magnitude going from 1au to $\sim$0.2au \cite{BandyopadhyayApJS20b}. This enhanced decay rate balances estimated heating rates for the solar wind. In the inertial range, the incompressive decay rates compare favourably with the \vkh decay rate and increase by a couple of orders of magnitude approaching the sun. The compressive component of the cascade rate could be significant in the inner heliosphere owing to the increased density fluctuations as well. It can decrease by as much as five orders of magnitude from 0.2au to 2au \cite{AndresApJ21}, likely a consequence of the solar wind's evolution towards a nearly incompressible state.

At the kinetic scale, the cascade is modified by kinetic physics as well as dissipation \cite{BowenArXiv21}. The spectral slopes in the imbalanced inner heliospheric turbulence routinely show a steep transition range with powerlaw $\sim f^{-4}$ just below proton kinetic scales \cite{DenskatJGR84, LeamonJGR99, HuangApJL21}. This potentially happens because of enhanced heating of protons and an inhibition of the cascade to smaller scales \cite{MeyrandJPP21, SquireNatureAstron22}.

The transfer of energy into internal degrees of freedom has been shown to happen via wave-particle interactions, stochastic heating as well as in intermittent locations \cite{VechAA21, MartinovicApJ19, MartinovicApJS20, QudsiApJS20}. Distribution functions in the presence of waves show signatures expected of quasilinear diffusion in the velocity space \cite{BowenArXiv21, VernieroApJ22} . Larger PVI values have been shown to associate favourably with hotter plasma, consistent with simulations as well as 1au observations \cite{QudsiApJ20}. Local measures of dissipation have also been shown to relate with enhanced intermittent heating \cite{DavidApJ22, Sioulas_2022}.

Many of the properties discussed above modify significantly in the presence of switchbacks. The turbulence appears to be more evolved inside switchbacks. The power spectra become Kolmogorov like inside SBs even when nearby NSB intervals show IK-like behaviour \cite{BourouaineApJL20}. The stochastic heating rate, the incompressive cascade rate, and PVI values etc. are enhanced in the presence of SBs \cite{MartinovicApJS20}. 

Although PSP is already enabling a significant progress in studying turbulent transfer of energy across scales in the heliosphere, many open questions remain. 
{\em 
\begin{itemize} 
\item How do we get true estimates of heating rates from single spacecraft measurements? 
\item Do such estimates modify the balance between the \vkh decay rate and true heating rates? 
\item How do local dissipation measures (e.g. LET \cite{Sioulas_2022}, $\mathcal{D}$ \cite{DavidApJ22}, field particle correlation \cite{VernieroJGR21}) match with the large scale decay rates? 
\item What are the relative contributions of various cascade terms (incompressible, compressible, hall, etc.) to the inertial range estimates of decay rates?, 
\item How does this cascaded energy partition between ions and electrons? How much energy is taken away from the cascade by ion heating and how much is left to cascade down to electron scales? 
\item How does the cascade proceed closer to the electron scales?, 
\item Is there an identifiable relationship between intermittent dissipation and wave-particle interactions? 
\item How do the conclusions to above mentioned questions vary with ambient conditions?
\end{itemize}
}

Building upon the approximately six decades of research, PSP is allowing an in-depth exploration of turbulent regimes that were not accessible up to now, both in (time/length) scales and in proximity to the sun. With the abundance of switchbacks, identification of waves in turbulence, new proton VDF features, modified turbulence properties, and observations of turbulence in the sub-Alfv\'enic solar corona, PSP is only getting started. Over the next few decades, PSP will allow greater discoveries related to the origins of the solar wind, its evolution and the consequent turbulent properties.

\section{Acknowledgements}
The authors would like to dedicate this paper to late Dr. Eugene Parker, the pioneer of this field. Dr. Parker's seminal works on plasma astrophysics not only inspired generations of plasma astrophysicists but also paved the way for modern space physics. Inspired by his seminal findings, the Parker Solar Probe is taking our understanding and view of the Solar Wind to the next level. We would like to thank the PSP instrument teams for creating state-of-the-art instruments that are enabling cutting edge science in the inner heliosphere. We also thank our collaborators who made 
possible much of the science discussed in this paper. 
TNP would like to thank the AAPPS-DPP committee for the invitation to deliver a talk on the subject and to write this review paper. 
This research is supported in part by
the NASA Parker Solar Probe Mission
under a GI grant 80NSSC21K1765 and the 
ISOIS team (Princeton SUB0000165),
by the IMAP project (Princeton
SUB0000317), by the MMS mission under a 
Theory and Modeling grant  80NSSC19K0565,
and by HSR grants
80NSSC18K1648 and 80NSSC19K0284.
\bibliography{psp}


\begin{thebibliography}{240}
\ifx \bisbn   \undefined \def \bisbn  #1{ISBN #1}\fi
\ifx \binits  \undefined \def \binits#1{#1}\fi
\ifx \bauthor  \undefined \def \bauthor#1{#1}\fi
\ifx \batitle  \undefined \def \batitle#1{#1}\fi
\ifx \bjtitle  \undefined \def \bjtitle#1{#1}\fi
\ifx \bvolume  \undefined \def \bvolume#1{\textbf{#1}}\fi
\ifx \byear  \undefined \def \byear#1{#1}\fi
\ifx \bissue  \undefined \def \bissue#1{#1}\fi
\ifx \bfpage  \undefined \def \bfpage#1{#1}\fi
\ifx \blpage  \undefined \def \blpage #1{#1}\fi
\ifx \burl  \undefined \def \burl#1{\textsf{#1}}\fi
\ifx \doiurl  \undefined \def \doiurl#1{\url{https://doi.org/#1}}\fi
\ifx \betal  \undefined \def \betal{\textit{et al.}}\fi
\ifx \binstitute  \undefined \def \binstitute#1{#1}\fi
\ifx \binstitutionaled  \undefined \def \binstitutionaled#1{#1}\fi
\ifx \bctitle  \undefined \def \bctitle#1{#1}\fi
\ifx \beditor  \undefined \def \beditor#1{#1}\fi
\ifx \bpublisher  \undefined \def \bpublisher#1{#1}\fi
\ifx \bbtitle  \undefined \def \bbtitle#1{#1}\fi
\ifx \bedition  \undefined \def \bedition#1{#1}\fi
\ifx \bseriesno  \undefined \def \bseriesno#1{#1}\fi
\ifx \blocation  \undefined \def \blocation#1{#1}\fi
\ifx \bsertitle  \undefined \def \bsertitle#1{#1}\fi
\ifx \bsnm \undefined \def \bsnm#1{#1}\fi
\ifx \bsuffix \undefined \def \bsuffix#1{#1}\fi
\ifx \bparticle \undefined \def \bparticle#1{#1}\fi
\ifx \barticle \undefined \def \barticle#1{#1}\fi
\bibcommenthead
\ifx \bconfdate \undefined \def \bconfdate #1{#1}\fi
\ifx \botherref \undefined \def \botherref #1{#1}\fi
\ifx \url \undefined \def \url#1{\textsf{#1}}\fi
\ifx \bchapter \undefined \def \bchapter#1{#1}\fi
\ifx \bbook \undefined \def \bbook#1{#1}\fi
\ifx \bcomment \undefined \def \bcomment#1{#1}\fi
\ifx \oauthor \undefined \def \oauthor#1{#1}\fi
\ifx \citeauthoryear \undefined \def \citeauthoryear#1{#1}\fi
\ifx \endbibitem  \undefined \def \endbibitem {}\fi
\ifx \bconflocation  \undefined \def \bconflocation#1{#1}\fi
\ifx \arxivurl  \undefined \def \arxivurl#1{\textsf{#1}}\fi
\csname PreBibitemsHook\endcsname

\bibitem{ParkerApJ58a}
\begin{barticle}
\bauthor{\bsnm{Parker}, \binits{E.N.}}:
\batitle{Dynamics of the interplanetary gas and magnetic fields.}
\bjtitle{The Astrophysical Journal}
\bvolume{128},
\bfpage{664}
(\byear{1958})
\end{barticle}
\endbibitem

\bibitem{NeugebauerJGR66}
\begin{barticle}
\bauthor{\bsnm{Neugebauer}, \binits{M.}},
\bauthor{\bsnm{Snyder}, \binits{C.W.}}:
\batitle{Mariner 2 observations of the solar wind: 1. average properties}.
\bjtitle{Journal of Geophysical Research}
\bvolume{71}(\bissue{19}),
\bfpage{4469}--\blpage{4484}
(\byear{1966})
\end{barticle}
\endbibitem

\bibitem{krommes2002fundamental}
\begin{barticle}
\bauthor{\bsnm{Krommes}, \binits{J.A.}}:
\batitle{Fundamental statistical descriptions of plasma turbulence in magnetic
  fields}.
\bjtitle{Physics Reports}
\bvolume{360}(\bissue{1-4}),
\bfpage{1}--\blpage{352}
(\byear{2002})
\end{barticle}
\endbibitem

\bibitem{tsytovich2016introduction}
\begin{bbook}
\bauthor{\bsnm{Tsytovich}, \binits{V.N.}}:
\bbtitle{An Introduction to the Theory of Plasma Turbulence: International
  Series of Monographs in Natural Philosophy}
vol. \bseriesno{44}.
\bpublisher{Elsevier}, \blocation{???}
(\byear{2016})
\end{bbook}
\endbibitem

\bibitem{yamada2008anatomy}
\begin{barticle}
\bauthor{\bsnm{Yamada}, \binits{T.}},
\bauthor{\bsnm{Itoh}, \binits{S.-I.}},
\bauthor{\bsnm{Maruta}, \binits{T.}},
\bauthor{\bsnm{Kasuya}, \binits{N.}},
\bauthor{\bsnm{Nagashima}, \binits{Y.}},
\bauthor{\bsnm{Shinohara}, \binits{S.}},
\bauthor{\bsnm{Terasaka}, \binits{K.}},
\bauthor{\bsnm{Yagi}, \binits{M.}},
\bauthor{\bsnm{Inagaki}, \binits{S.}},
\bauthor{\bsnm{Kawai}, \binits{Y.}}, \betal:
\batitle{Anatomy of plasma turbulence}.
\bjtitle{Nature physics}
\bvolume{4}(\bissue{9}),
\bfpage{721}--\blpage{725}
(\byear{2008})
\end{barticle}
\endbibitem

\bibitem{ColemanApJ68}
\begin{barticle}
\bauthor{\bsnm{{Coleman Jr, P~J}}}:
\batitle{{Turbulence, Viscosity, and Dissipation in the Solar-Wind Plasma}}.
\bjtitle{The Astrophysical Journal}
\bvolume{153},
\bfpage{371}
(\byear{1968}).
\doiurl{10.1086/149674}
\end{barticle}
\endbibitem

\bibitem{MatthaeusJGR82}
\begin{barticle}
\bauthor{\bsnm{Matthaeus}, \binits{W.H.}},
\bauthor{\bsnm{Goldstein}, \binits{M.L.}}:
\batitle{{Measurement of the rugged invariants of magnetohydrodynamic
  turbulence in the solar wind}}.
\bjtitle{Journal of Geophysical Research}
\bvolume{87}(\bissue{A8}),
\bfpage{6011}--\blpage{6028}
(\byear{1982})
\end{barticle}
\endbibitem

\bibitem{BrunoLRSP13}
\begin{botherref}
\oauthor{\bsnm{Bruno}, \binits{R.}},
\oauthor{\bsnm{Carbone}, \binits{V.}}:
The solar wind as a turbulence laboratory.
Living Reviews in Solar Physics
\textbf{10}(2)
(2013).
\doiurl{10.12942/lrsp-2013-2}
\end{botherref}
\endbibitem

\bibitem{cordes1985small}
\begin{barticle}
\bauthor{\bsnm{Cordes}, \binits{J.}},
\bauthor{\bsnm{Weisberg}, \binits{J.}},
\bauthor{\bsnm{Boriakoff}, \binits{V.}}:
\batitle{Small-scale electron density turbulence in the interstellar medium}.
\bjtitle{The Astrophysical Journal}
\bvolume{288},
\bfpage{221}--\blpage{247}
(\byear{1985})
\end{barticle}
\endbibitem

\bibitem{falceta2014turbulence}
\begin{barticle}
\bauthor{\bsnm{Falceta-Gon{\c{c}}alves}, \binits{D.}},
\bauthor{\bsnm{Kowal}, \binits{G.}},
\bauthor{\bsnm{Falgarone}, \binits{E.}},
\bauthor{\bsnm{Chian}, \binits{A.-L.}}:
\batitle{Turbulence in the interstellar medium}.
\bjtitle{Nonlinear Processes in Geophysics}
\bvolume{21}(\bissue{3}),
\bfpage{587}--\blpage{604}
(\byear{2014})
\end{barticle}
\endbibitem

\bibitem{armstrong1995electron}
\begin{barticle}
\bauthor{\bsnm{Armstrong}, \binits{J.}},
\bauthor{\bsnm{Rickett}, \binits{B.}},
\bauthor{\bsnm{Spangler}, \binits{S.}}:
\batitle{Electron density power spectrum in the local interstellar medium}.
\bjtitle{The Astrophysical Journal}
\bvolume{443},
\bfpage{209}--\blpage{221}
(\byear{1995})
\end{barticle}
\endbibitem

\bibitem{balbus1998instability}
\begin{barticle}
\bauthor{\bsnm{Balbus}, \binits{S.A.}},
\bauthor{\bsnm{Hawley}, \binits{J.F.}}:
\batitle{Instability, turbulence, and enhanced transport in accretion disks}.
\bjtitle{Reviews of modern physics}
\bvolume{70}(\bissue{1}),
\bfpage{1}
(\byear{1998})
\end{barticle}
\endbibitem

\bibitem{abramowicz2013foundations}
\begin{barticle}
\bauthor{\bsnm{Abramowicz}, \binits{M.A.}},
\bauthor{\bsnm{Fragile}, \binits{P.C.}}:
\batitle{Foundations of black hole accretion disk theory}.
\bjtitle{Living Reviews in Relativity}
\bvolume{16}(\bissue{1}),
\bfpage{1}--\blpage{88}
(\byear{2013})
\end{barticle}
\endbibitem

\bibitem{SchueckerAA04}
\begin{barticle}
\bauthor{\bsnm{Schuecker}, \binits{P.}},
\bauthor{\bsnm{Finoguenov}, \binits{A.}},
\bauthor{\bsnm{Miniati}, \binits{F.}},
\bauthor{\bsnm{B{\"o}hringer}, \binits{H.}},
\bauthor{\bsnm{Briel}, \binits{U.}}:
\batitle{Probing turbulence in the coma galaxy cluster}.
\bjtitle{Astronomy \& Astrophysics}
\bvolume{426}(\bissue{2}),
\bfpage{387}--\blpage{397}
(\byear{2004})
\end{barticle}
\endbibitem

\bibitem{churazov2012x}
\begin{barticle}
\bauthor{\bsnm{Churazov}, \binits{E.}},
\bauthor{\bsnm{Vikhlinin}, \binits{A.}},
\bauthor{\bsnm{Zhuravleva}, \binits{I.}},
\bauthor{\bsnm{Schekochihin}, \binits{A.}},
\bauthor{\bsnm{Parrish}, \binits{I.}},
\bauthor{\bsnm{Sunyaev}, \binits{R.}},
\bauthor{\bsnm{Forman}, \binits{W.}},
\bauthor{\bsnm{B{\"o}hringer}, \binits{H.}},
\bauthor{\bsnm{Randall}, \binits{S.}}:
\batitle{X-ray surface brightness and gas density fluctuations in the coma
  cluster}.
\bjtitle{Monthly Notices of the Royal Astronomical Society}
\bvolume{421}(\bissue{2}),
\bfpage{1123}--\blpage{1135}
(\byear{2012})
\end{barticle}
\endbibitem

\bibitem{mohapatra2020turbulence}
\begin{barticle}
\bauthor{\bsnm{Mohapatra}, \binits{R.}},
\bauthor{\bsnm{Federrath}, \binits{C.}},
\bauthor{\bsnm{Sharma}, \binits{P.}}:
\batitle{Turbulence in stratified atmospheres: implications for the
  intracluster medium}.
\bjtitle{Monthly Notices of the Royal Astronomical Society}
\bvolume{493}(\bissue{4}),
\bfpage{5838}--\blpage{5853}
(\byear{2020})
\end{barticle}
\endbibitem

\bibitem{hendrix1996magnetohydrodynamic}
\begin{barticle}
\bauthor{\bsnm{Hendrix}, \binits{D.}},
\bauthor{\bsnm{Van~Hoven}, \binits{G.}}:
\batitle{Magnetohydrodynamic turbulence and implications for solar coronal
  heating}.
\bjtitle{The Astrophysical Journal}
\bvolume{467},
\bfpage{887}
(\byear{1996})
\end{barticle}
\endbibitem

\bibitem{MattEA99-ch}
\begin{barticle}
\bauthor{\bsnm{Matthaeus}, \binits{W.H.}},
\bauthor{\bsnm{Zank}, \binits{G.P.}},
\bauthor{\bsnm{Oughton}, \binits{S.}},
\bauthor{\bsnm{Mullan}, \binits{D.}},
\bauthor{\bsnm{Dmitruk}, \binits{P.}}:
\batitle{Coronal heating by magnetohydrodynamic turbulence driven by reflected
  low-frequency waves}.
\bjtitle{The Astrophysical Journal}
\bvolume{523}(\bissue{1}),
\bfpage{93}
(\byear{1999})
\end{barticle}
\endbibitem

\bibitem{cranmer2007self}
\begin{barticle}
\bauthor{\bsnm{Cranmer}, \binits{S.R.}},
\bauthor{\bsnm{Van~Ballegooijen}, \binits{A.A.}},
\bauthor{\bsnm{Edgar}, \binits{R.J.}}:
\batitle{Self-consistent coronal heating and solar wind acceleration from
  anisotropic magnetohydrodynamic turbulence}.
\bjtitle{The Astrophysical Journal Supplement Series}
\bvolume{171}(\bissue{2}),
\bfpage{520}
(\byear{2007})
\end{barticle}
\endbibitem

\bibitem{DeforestApJ16}
\begin{barticle}
\bauthor{\bsnm{DeForest}, \binits{C.}},
\bauthor{\bsnm{Matthaeus}, \binits{W.}},
\bauthor{\bsnm{Viall}, \binits{N.}},
\bauthor{\bsnm{Cranmer}, \binits{S.}}:
\batitle{Fading coronal structure and the onset of turbulence in the young
  solar wind}.
\bjtitle{The Astrophysical Journal}
\bvolume{828}(\bissue{2}),
\bfpage{66}
(\byear{2016})
\end{barticle}
\endbibitem

\bibitem{TelloniApJ22}
\begin{barticle}
\bauthor{\bsnm{Telloni}, \binits{D.}},
\bauthor{\bsnm{Adhikari}, \binits{L.}},
\bauthor{\bsnm{Zank}, \binits{G.P.}},
\bauthor{\bsnm{Zhao}, \binits{L.}},
\bauthor{\bsnm{Sorriso-Valvo}, \binits{L.}},
\bauthor{\bsnm{Antonucci}, \binits{E.}},
\bauthor{\bsnm{Giordano}, \binits{S.}},
\bauthor{\bsnm{Mancuso}, \binits{S.}}:
\batitle{Possible evidence for shear-driven kelvin--helmholtz instability along
  the boundary of fast and slow solar wind in the corona}.
\bjtitle{The Astrophysical Journal}
\bvolume{929}(\bissue{1}),
\bfpage{98}
(\byear{2022})
\end{barticle}
\endbibitem

\bibitem{RuffoloApJ20}
\begin{barticle}
\bauthor{\bsnm{Ruffolo}, \binits{D.}},
\bauthor{\bsnm{Matthaeus}, \binits{W.H.}},
\bauthor{\bsnm{Chhiber}, \binits{R.}},
\bauthor{\bsnm{Usmanov}, \binits{A.V.}},
\bauthor{\bsnm{Yang}, \binits{Y.}},
\bauthor{\bsnm{Bandyopadhyay}, \binits{R.}},
\bauthor{\bsnm{Parashar}, \binits{T.}},
\bauthor{\bsnm{Goldstein}, \binits{M.L.}},
\bauthor{\bsnm{DeForest}, \binits{C.}},
\bauthor{\bsnm{Wan}, \binits{M.}}, \betal:
\batitle{Shear-driven transition to isotropically turbulent solar wind outside
  the alfv{\'e}n critical zone}.
\bjtitle{The Astrophysical Journal}
\bvolume{902}(\bissue{2}),
\bfpage{94}
(\byear{2020})
\end{barticle}
\endbibitem

\bibitem{RichardsonEA95}
\begin{barticle}
\bauthor{\bsnm{Richardson}, \binits{J.D.}},
\bauthor{\bsnm{Paularena}, \binits{K.I.}},
\bauthor{\bsnm{Lazarus}, \binits{A.J.}},
\bauthor{\bsnm{Belcher}, \binits{J.W.}}:
\batitle{Radial evolution of the solar wind from {I}{M}{P} 8 to {Voyager} 2}.
\bjtitle{Geophysical Research Letters}
\bvolume{22},
\bfpage{325}
(\byear{1995})
\end{barticle}
\endbibitem

\bibitem{UsmanovApJ11}
\begin{barticle}
\bauthor{\bsnm{Usmanov}, \binits{A.V.}},
\bauthor{\bsnm{Matthaeus}, \binits{W.H.}},
\bauthor{\bsnm{Breech}, \binits{B.A.}},
\bauthor{\bsnm{Goldstein}, \binits{M.L.}}:
\batitle{Solar wind modeling with turbulence transport and heating}.
\bjtitle{The Astrophysical Journal}
\bvolume{727}(\bissue{2}),
\bfpage{84}
(\byear{2011})
\end{barticle}
\endbibitem

\bibitem{GamayunovApJ12}
\begin{barticle}
\bauthor{\bsnm{Gamayunov}, \binits{K.V.}},
\bauthor{\bsnm{Zhang}, \binits{M.}},
\bauthor{\bsnm{Pogorelov}, \binits{N.V.}},
\bauthor{\bsnm{HEERIkHUISEN}, \binits{J.}},
\bauthor{\bsnm{Rassoul}, \binits{H.K.}}:
\batitle{Self-consistent model of the interstellar pickup protons, alfv{\'e}nic
  turbulence, and core solar wind in the outer heliosphere}.
\bjtitle{The Astrophysical Journal}
\bvolume{757}(\bissue{1}),
\bfpage{74}
(\byear{2012})
\end{barticle}
\endbibitem

\bibitem{SokolovApJ13}
\begin{barticle}
\bauthor{\bsnm{Sokolov}, \binits{I.V.}},
\bauthor{\bparticle{Van~der} \bsnm{Holst}, \binits{B.}},
\bauthor{\bsnm{Oran}, \binits{R.}},
\bauthor{\bsnm{Downs}, \binits{C.}},
\bauthor{\bsnm{Roussev}, \binits{I.I.}},
\bauthor{\bsnm{Jin}, \binits{M.}},
\bauthor{\bsnm{Manchester}, \binits{W.B.}},
\bauthor{\bsnm{Evans}, \binits{R.M.}},
\bauthor{\bsnm{Gombosi}, \binits{T.I.}}:
\batitle{Magnetohydrodynamic waves and coronal heating: Unifying empirical and
  mhd turbulence models}.
\bjtitle{The Astrophysical Journal}
\bvolume{764}(\bissue{1}),
\bfpage{23}
(\byear{2013})
\end{barticle}
\endbibitem

\bibitem{oran2014coronal}
\begin{botherref}
\oauthor{\bsnm{Oran}, \binits{R.}}:
Coronal heating and solar wind acceleration by alfv{\'e}n wave turbulence: a
  global computational model and observations.
PhD thesis,
The University of Michigan
(2014)
\end{botherref}
\endbibitem

\bibitem{ChhiberApJS17}
\begin{barticle}
\bauthor{\bsnm{Chhiber}, \binits{R.}},
\bauthor{\bsnm{Subedi}, \binits{P.}},
\bauthor{\bsnm{Usmanov}, \binits{A.V.}},
\bauthor{\bsnm{Matthaeus}, \binits{W.H.}},
\bauthor{\bsnm{Ruffolo}, \binits{D.}},
\bauthor{\bsnm{Goldstein}, \binits{M.L.}},
\bauthor{\bsnm{Parashar}, \binits{T.N.}}:
\batitle{Cosmic-ray diffusion coefficients throughout the inner heliosphere
  from a global solar wind simulation}.
\bjtitle{The Astrophysical Journal Supplement Series}
\bvolume{230}(\bissue{2}),
\bfpage{21}
(\byear{2017})
\end{barticle}
\endbibitem

\bibitem{vanderHolstApJ22}
\begin{barticle}
\bauthor{\bparticle{van~der} \bsnm{Holst}, \binits{B.}},
\bauthor{\bsnm{Huang}, \binits{J.}},
\bauthor{\bsnm{Sachdeva}, \binits{N.}},
\bauthor{\bsnm{Kasper}, \binits{J.}},
\bauthor{\bsnm{Manchester~IV}, \binits{W.}},
\bauthor{\bsnm{Borovikov}, \binits{D.}},
\bauthor{\bsnm{Chandran}, \binits{B.}},
\bauthor{\bsnm{Case}, \binits{A.}},
\bauthor{\bsnm{Korreck}, \binits{K.}},
\bauthor{\bsnm{Larson}, \binits{D.}}, \betal:
\batitle{Improving the alfv{\'e}n wave solar atmosphere model based on parker
  solar probe data}.
\bjtitle{The Astrophysical Journal}
\bvolume{925}(\bissue{2}),
\bfpage{146}
(\byear{2022})
\end{barticle}
\endbibitem

\bibitem{TuSSR95}
\begin{barticle}
\bauthor{\bsnm{Tu}, \binits{C.Y.}},
\bauthor{\bsnm{Marsch}, \binits{E.}}:
\batitle{{MHD structures, waves and turbulence in the solar wind: Observations
  and theories}}.
\bjtitle{Space Science Reviews}
\bvolume{73}(\bissue{1}),
\bfpage{1}--\blpage{210}
(\byear{1995})
\end{barticle}
\endbibitem

\bibitem{VerscharenLRSP19}
\begin{barticle}
\bauthor{\bsnm{Verscharen}, \binits{D.}},
\bauthor{\bsnm{Klein}, \binits{K.G.}},
\bauthor{\bsnm{Maruca}, \binits{B.A.}}:
\batitle{The multi-scale nature of the solar wind}.
\bjtitle{Living Reviews in Solar Physics}
\bvolume{16}(\bissue{1}),
\bfpage{5}
(\byear{2019})
\end{barticle}
\endbibitem

\bibitem{Hundhausen}
\begin{bbook}
\bauthor{\bsnm{Hundhausen}, \binits{A.J.}}:
\bbtitle{Coronal Expansion and the Solar Wind}.
\bpublisher{Springer},
\blocation{New York}
(\byear{1972})
\end{bbook}
\endbibitem

\bibitem{FoxSSR16}
\begin{barticle}
\bauthor{\bsnm{Fox}, \binits{N.}},
\bauthor{\bsnm{Velli}, \binits{M.}},
\bauthor{\bsnm{Bale}, \binits{S.}},
\bauthor{\bsnm{Decker}, \binits{R.}},
\bauthor{\bsnm{Driesman}, \binits{A.}},
\bauthor{\bsnm{Howard}, \binits{R.}},
\bauthor{\bsnm{Kasper}, \binits{J.C.}},
\bauthor{\bsnm{Kinnison}, \binits{J.}},
\bauthor{\bsnm{Kusterer}, \binits{M.}},
\bauthor{\bsnm{Lario}, \binits{D.}}, \betal:
\batitle{The solar probe plus mission: Humanity’s first visit to our star}.
\bjtitle{Space Science Reviews}
\bvolume{204}(\bissue{1-4}),
\bfpage{7}--\blpage{48}
(\byear{2016})
\end{barticle}
\endbibitem

\bibitem{BatchelorTHT}
\begin{bbook}
\bauthor{\bsnm{Batchelor}, \binits{G.K.}}:
\bbtitle{The Theory of Homogeneous Turbulence}.
\bpublisher{Cambridge University Press},
\blocation{Cambridge, UK}
(\byear{1970})
\end{bbook}
\endbibitem

\bibitem{TennekesLumley}
\begin{bbook}
\bauthor{\bsnm{Tennekes}, \binits{H.}},
\bauthor{\bsnm{Lumley}, \binits{J.L.}}:
\bbtitle{A First Course in Turbulence}.
\bpublisher{The MIT press}, \blocation{???}
(\byear{1972})
\end{bbook}
\endbibitem

\bibitem{WuPRL13}
\begin{barticle}
\bauthor{\bsnm{Wu}, \binits{P.}},
\bauthor{\bsnm{Wan}, \binits{M.}},
\bauthor{\bsnm{Matthaeus}, \binits{W.}},
\bauthor{\bsnm{Shay}, \binits{M.}},
\bauthor{\bsnm{Swisdak}, \binits{M.}}:
\batitle{von k{\'a}rm{\'a}n energy decay and heating of protons and electrons
  in a kinetic turbulent plasma}.
\bjtitle{Physical review letters}
\bvolume{111}(\bissue{12}),
\bfpage{121105}
(\byear{2013})
\end{barticle}
\endbibitem

\bibitem{KarimabadiPP13}
\begin{barticle}
\bauthor{\bsnm{Karimabadi}, \binits{H.}},
\bauthor{\bsnm{Roytershteyn}, \binits{V.}},
\bauthor{\bsnm{Wan}, \binits{M.}},
\bauthor{\bsnm{Matthaeus}, \binits{W.H.}},
\bauthor{\bsnm{Daughton}, \binits{W.}},
\bauthor{\bsnm{Wu}, \binits{P.}},
\bauthor{\bsnm{Shay}, \binits{M.}},
\bauthor{\bsnm{Loring}, \binits{B.}},
\bauthor{\bsnm{Borovsky}, \binits{J.}},
\bauthor{\bsnm{Leonardis}, \binits{E.}},
\bauthor{\bsnm{Chapman}, \binits{S.C.}},
\bauthor{\bsnm{Nakamura}, \binits{T.K.M.}}:
\batitle{Coherent structures, intermittent turbulence, and dissipation in
  high-temperature plasmas}.
\bjtitle{Physics of Plasmas}
\bvolume{20}(\bissue{1}),
(\byear{2013}).
\doiurl{10.1063/1.4773205}
\end{barticle}
\endbibitem

\bibitem{WanPP16}
\begin{barticle}
\bauthor{\bsnm{Wan}, \binits{M.}},
\bauthor{\bsnm{Matthaeus}, \binits{W.}},
\bauthor{\bsnm{Roytershteyn}, \binits{V.}},
\bauthor{\bsnm{Parashar}, \binits{T.}},
\bauthor{\bsnm{Wu}, \binits{P.}},
\bauthor{\bsnm{Karimabadi}, \binits{H.}}:
\batitle{Intermittency, coherent structures and dissipation in plasma
  turbulence}.
\bjtitle{Physics of Plasmas (1994-present)}
\bvolume{23}(\bissue{4}),
\bfpage{042307}
(\byear{2016})
\end{barticle}
\endbibitem

\bibitem{SchekochihinApJS09}
\begin{barticle}
\bauthor{\bsnm{Schekochihin}, \binits{A.}},
\bauthor{\bsnm{Cowley}, \binits{S.}},
\bauthor{\bsnm{Dorland}, \binits{W.}},
\bauthor{\bsnm{Hammett}, \binits{G.}},
\bauthor{\bsnm{Howes}, \binits{G.G.}},
\bauthor{\bsnm{Quataert}, \binits{E.}},
\bauthor{\bsnm{Tatsuno}, \binits{T.}}:
\batitle{Astrophysical gyrokinetics: kinetic and fluid turbulent cascades in
  magnetized weakly collisional plasmas}.
\bjtitle{The Astrophysical Journal Supplement Series}
\bvolume{182}(\bissue{1}),
\bfpage{310}
(\byear{2009})
\end{barticle}
\endbibitem

\bibitem{BoldyrevApJ13}
\begin{barticle}
\bauthor{\bsnm{Boldyrev}, \binits{S.}},
\bauthor{\bsnm{Horaites}, \binits{K.}},
\bauthor{\bsnm{Xia}, \binits{Q.}},
\bauthor{\bsnm{Perez}, \binits{J.C.}}:
\batitle{Toward a theory of astrophysical plasma turbulence at subproton
  scales}.
\bjtitle{The Astrophysical Journal}
\bvolume{777}(\bissue{1}),
\bfpage{41}
(\byear{2013})
\end{barticle}
\endbibitem

\bibitem{EyinkPRX18}
\begin{barticle}
\bauthor{\bsnm{Eyink}, \binits{G.L.}}:
\batitle{Cascades and dissipative anomalies in nearly collisionless plasma
  turbulence}.
\bjtitle{Physical Review X}
\bvolume{8}(\bissue{4}),
\bfpage{041020}
(\byear{2018})
\end{barticle}
\endbibitem

\bibitem{BiskampBook03}
\begin{bbook}
\bauthor{\bsnm{Biskamp}, \binits{D.}}:
\bbtitle{Magnetohydrodynamic Turbulence}.
\bpublisher{Cambridge University Press}, \blocation{???}
(\byear{2003})
\end{bbook}
\endbibitem

\bibitem{KarmanPRSLA38}
\begin{bchapter}
\bauthor{\bsnm{De~Karman}, \binits{T.}},
\bauthor{\bsnm{Howarth}, \binits{L.}}:
\bctitle{On the statistical theory of isotropic turbulence}.
In: \bbtitle{Proceedings of the Royal Society of London A: Mathematical,
  Physical and Engineering Sciences},
vol. \bseriesno{164},
pp. \bfpage{192}--\blpage{215}
(\byear{1938}).
\bcomment{The Royal Society}
\end{bchapter}
\endbibitem

\bibitem{HossainPFL95}
\begin{barticle}
\bauthor{\bsnm{Hossain}, \binits{M.}},
\bauthor{\bsnm{Gray}, \binits{P.C.}},
\bauthor{\bsnm{Pontius}, \binits{D.H.}},
\bauthor{\bsnm{Matthaeus}, \binits{W.H.}},
\bauthor{\bsnm{Oughton}, \binits{S.}}:
\batitle{Phenomenology for the decay of energy containing eddies in homogeneous
  mhd turbulence}.
\bjtitle{Physics of Fluids (1994-present)}
\bvolume{7}(\bissue{11}),
\bfpage{2886}--\blpage{2904}
(\byear{1995}).
\doiurl{10.1063/1.868665}
\end{barticle}
\endbibitem

\bibitem{WanJFM12}
\begin{barticle}
\bauthor{\bsnm{Wan}, \binits{M.}},
\bauthor{\bsnm{Oughton}, \binits{S.}},
\bauthor{\bsnm{Servidio}, \binits{S.}},
\bauthor{\bsnm{Matthaeus}, \binits{W.H.}}:
\batitle{von k{\'a}rm{\'a}n self-preservation hypothesis for
  magnetohydrodynamic turbulence and its consequences for universality}.
\bjtitle{Journal of Fluid Mechanics}
\bvolume{697},
\bfpage{296}--\blpage{315}
(\byear{2012})
\end{barticle}
\endbibitem

\bibitem{BandyopadhyayPRX18}
\begin{barticle}
\bauthor{\bsnm{Bandyopadhyay}, \binits{R.}},
\bauthor{\bsnm{Oughton}, \binits{S.}},
\bauthor{\bsnm{Wan}, \binits{M.}},
\bauthor{\bsnm{Matthaeus}, \binits{W.H.}},
\bauthor{\bsnm{Chhiber}, \binits{R.}},
\bauthor{\bsnm{Parashar}, \binits{T.N.}}:
\batitle{Finite dissipation in anisotropic magnetohydrodynamic turbulence}.
\bjtitle{Phys. Rev. X}
\bvolume{8},
\bfpage{041052}
(\byear{2018}).
\doiurl{10.1103/PhysRevX.8.041052}
\end{barticle}
\endbibitem

\bibitem{MccombPRE15}
\begin{barticle}
\bauthor{\bsnm{McComb}, \binits{W.}},
\bauthor{\bsnm{Berera}, \binits{A.}},
\bauthor{\bsnm{Yoffe}, \binits{S.}},
\bauthor{\bsnm{Linkmann}, \binits{M.}}:
\batitle{Energy transfer and dissipation in forced isotropic turbulence}.
\bjtitle{Physical Review E}
\bvolume{91}(\bissue{4}),
\bfpage{043013}
(\byear{2015})
\end{barticle}
\endbibitem

\bibitem{MatthaeusApJL16}
\begin{barticle}
\bauthor{\bsnm{Matthaeus}, \binits{W.H.}},
\bauthor{\bsnm{Parashar}, \binits{T.N.}},
\bauthor{\bsnm{Wan}, \binits{M.}},
\bauthor{\bsnm{Wu}, \binits{P.}}:
\batitle{Turbulence and proton–electron heating in kinetic plasma}.
\bjtitle{The Astrophysical Journal Letters}
\bvolume{827}(\bissue{1}),
\bfpage{7}
(\byear{2016})
\end{barticle}
\endbibitem

\bibitem{ParasharApJ15}
\begin{barticle}
\bauthor{\bsnm{Parashar}, \binits{T.N.}},
\bauthor{\bsnm{Matthaeus}, \binits{W.H.}},
\bauthor{\bsnm{Shay}, \binits{M.A.}},
\bauthor{\bsnm{Wan}, \binits{M.}}:
\batitle{Transition from kinetic to mhd behavior in a collisionless plasma}.
\bjtitle{The Astrophysical Journal}
\bvolume{811}(\bissue{2}),
\bfpage{112}
(\byear{2015})
\end{barticle}
\endbibitem

\bibitem{RoyApJL21}
\begin{barticle}
\bauthor{\bsnm{Roy}, \binits{S.}},
\bauthor{\bsnm{Chhiber}, \binits{R.}},
\bauthor{\bsnm{Dasso}, \binits{S.}},
\bauthor{\bsnm{Ruiz}, \binits{M.}},
\bauthor{\bsnm{Matthaeus}, \binits{W.}}:
\batitle{von karman correlation similarity of the turbulent interplanetary
  magnetic field}.
\bjtitle{The Astrophysical Journal Letters}
\bvolume{919}(\bissue{2}),
\bfpage{27}
(\byear{2021})
\end{barticle}
\endbibitem

\bibitem{Kolmogorov41a}
\begin{barticle}
\bauthor{\bsnm{{Kolmogorov}}, \binits{A.}}:
\batitle{{The Local Structure of Turbulence in Incompressible Viscous Fluid for
  Very Large Reynolds' Numbers}}.
\bjtitle{Akademiia Nauk SSSR Doklady}
\bvolume{30},
\bfpage{301}--\blpage{305}
(\byear{1941})
\end{barticle}
\endbibitem

\bibitem{KraichnanPFL65}
\begin{barticle}
\bauthor{\bsnm{Kraichnan}, \binits{R.H.}}:
\batitle{Inertial-range spectrum of hydromagnetic turbulence}.
\bjtitle{Physics of Fluids (1958-1988)}
\bvolume{8}(\bissue{7}),
\bfpage{1385}--\blpage{1387}
(\byear{1965})
\end{barticle}
\endbibitem

\bibitem{GoldreichApJ95}
\begin{barticle}
\bauthor{\bsnm{{Goldreich}}, \binits{P.}},
\bauthor{\bsnm{{Sridhar}}, \binits{S.}}:
\batitle{{Toward a theory of interstellar turbulence. 2: Strong alfvenic
  turbulence}}.
\bjtitle{The Astrophysical Journal}
\bvolume{438},
\bfpage{763}--\blpage{775}
(\byear{1995}).
\doiurl{10.1086/175121}
\end{barticle}
\endbibitem

\bibitem{VermaPP99}
\begin{barticle}
\bauthor{\bsnm{Verma}, \binits{M.K.}}:
\batitle{Mean magnetic field renormalization and kolmogorov’s energy spectrum
  in magnetohydrodynamic turbulence}.
\bjtitle{Physics of Plasmas}
\bvolume{6}(\bissue{5}),
\bfpage{1455}--\blpage{1460}
(\byear{1999})
\end{barticle}
\endbibitem

\bibitem{ZhouRMP04}
\begin{barticle}
\bauthor{\bsnm{Zhou}, \binits{Y.}},
\bauthor{\bsnm{Matthaeus}, \binits{W.H.}},
\bauthor{\bsnm{Dmitruk}, \binits{P.}}:
\batitle{Colloquium: Magnetohydrodynamic turbulence and time scales in
  astrophysical and space plasmas}.
\bjtitle{Rev. Mod. Phys.}
\bvolume{76}(\bissue{4}),
\bfpage{1015}--\blpage{1035}
(\byear{2004}).
\doiurl{10.1103/RevModPhys.76.1015}
\end{barticle}
\endbibitem

\bibitem{ChampagneJFM78}
\begin{barticle}
\bauthor{\bsnm{Champagne}, \binits{F.}}:
\batitle{The fine-scale structure of the turbulent velocity field}.
\bjtitle{Journal of Fluid Mechanics}
\bvolume{86}(\bissue{1}),
\bfpage{67}--\blpage{108}
(\byear{1978})
\end{barticle}
\endbibitem

\bibitem{ParasharPRL18}
\begin{barticle}
\bauthor{\bsnm{Parashar}, \binits{T.N.}},
\bauthor{\bsnm{Chasapis}, \binits{A.}},
\bauthor{\bsnm{Bandyopadhyay}, \binits{R.}},
\bauthor{\bsnm{Chhiber}, \binits{R.}},
\bauthor{\bsnm{Matthaeus}, \binits{W.H.}},
\bauthor{\bsnm{Maruca}, \binits{B.}},
\bauthor{\bsnm{Shay}, \binits{M.A.}},
\bauthor{\bsnm{Burch}, \binits{J.L.}},
\bauthor{\bsnm{Moore}, \binits{T.E.}},
\bauthor{\bsnm{Giles}, \binits{B.L.}},
\bauthor{\bsnm{Gershman}, \binits{D.J.}},
\bauthor{\bsnm{Pollock}, \binits{C.J.}},
\bauthor{\bsnm{Torbert}, \binits{R.B.}},
\bauthor{\bsnm{Russell}, \binits{C.T.}},
\bauthor{\bsnm{Strangeway}, \binits{R.J.}},
\bauthor{\bsnm{Roytershteyn}, \binits{V.}}:
\batitle{Kinetic range spectral features of cross helicity using the
  magnetospheric multiscale spacecraft}.
\bjtitle{Phys. Rev. Lett.}
\bvolume{121},
\bfpage{265101}
(\byear{2018}).
\doiurl{10.1103/PhysRevLett.121.265101}
\end{barticle}
\endbibitem

\bibitem{KiyaniPTRSA15}
\begin{barticle}
\bauthor{\bsnm{Kiyani}, \binits{K.H.}},
\bauthor{\bsnm{Osman}, \binits{K.T.}},
\bauthor{\bsnm{Chapman}, \binits{S.C.}}:
\batitle{Dissipation and heating in solar wind turbulence: from the macro to
  the micro and back again}.
\bjtitle{Philosophical Transactions of the Royal Society A: Mathematical,
  Physical and Engineering Sciences}
\bvolume{373}(\bissue{2041}),
\bfpage{20140155}
(\byear{2015})
{\href{https://arxiv.org/abs/https://royalsocietypublishing.org/doi/pdf/10.1098/rsta.2014.0155}{{https://royalsocietypublishing.org/doi/pdf/10.1098/rsta.2014.0155}}}.
\doiurl{10.1098/rsta.2014.0155}
\end{barticle}
\endbibitem

\bibitem{FraternaleApJ19}
\begin{barticle}
\bauthor{\bsnm{Fraternale}, \binits{F.}},
\bauthor{\bsnm{Pogorelov}, \binits{N.V.}},
\bauthor{\bsnm{Richardson}, \binits{J.D.}},
\bauthor{\bsnm{Tordella}, \binits{D.}}:
\batitle{Magnetic turbulence spectra and intermittency in the heliosheath and
  in the local interstellar medium}.
\bjtitle{The Astrophysical Journal}
\bvolume{872}(\bissue{1}),
\bfpage{40}
(\byear{2019})
\end{barticle}
\endbibitem

\bibitem{KolmogorovJFM62}
\begin{barticle}
\bauthor{\bsnm{Kolmogorov}, \binits{A.N.}}:
\batitle{A refinement of previous hypotheses concerning the local structure of
  turbulence in a viscous incompressible fluid at high reynolds number}.
\bjtitle{Journal of Fluid Mechanics}
\bvolume{13}(\bissue{1}),
\bfpage{82}--\blpage{85}
(\byear{1962})
\end{barticle}
\endbibitem

\bibitem{PolitanoPRE95}
\begin{barticle}
\bauthor{\bsnm{Politano}, \binits{H.}},
\bauthor{\bsnm{Pouquet}, \binits{A.}}:
\batitle{Model of intermittency in magnetohydrodynamic turbulence}.
\bjtitle{Physical Review E}
\bvolume{52}(\bissue{1}),
\bfpage{636}
(\byear{1995})
\end{barticle}
\endbibitem

\bibitem{VermaPR04}
\begin{barticle}
\bauthor{\bsnm{Verma}, \binits{M.K.}}:
\batitle{Statistical theory of magnetohydrodynamic turbulence: recent results}.
\bjtitle{Physics Reports}
\bvolume{401}(\bissue{5-6}),
\bfpage{229}--\blpage{380}
(\byear{2004})
\end{barticle}
\endbibitem

\bibitem{PopeBook}
\begin{bbook}
\bauthor{\bsnm{Pope}, \binits{S.B.}}:
\bbtitle{Turbulent Flows}.
\bpublisher{Cambridge university press}, \blocation{???}
(\byear{2000})
\end{bbook}
\endbibitem

\bibitem{PolitanoPRE98}
\begin{barticle}
\bauthor{\bsnm{Politano}, \binits{H.}},
\bauthor{\bsnm{Pouquet}, \binits{A.}}:
\batitle{von k{\'a}rm{\'a}n--howarth equation for magnetohydrodynamics and its
  consequences on third-order longitudinal structure and correlation
  functions}.
\bjtitle{Physical Review E}
\bvolume{57}(\bissue{1}),
\bfpage{21}
(\byear{1998})
\end{barticle}
\endbibitem

\bibitem{PodestaJFM08}
\begin{barticle}
\bauthor{\bsnm{Podesta}, \binits{J.}}:
\batitle{Laws for third-order moments in homogeneous anisotropic incompressible
  magnetohydrodynamic turbulence}.
\bjtitle{Journal of Fluid Mechanics}
\bvolume{609},
\bfpage{171}--\blpage{194}
(\byear{2008})
\end{barticle}
\endbibitem

\bibitem{AndresPRE17}
\begin{barticle}
\bauthor{\bsnm{Andr{\'e}s}, \binits{N.}},
\bauthor{\bsnm{Sahraoui}, \binits{F.}}:
\batitle{Alternative derivation of exact law for compressible and isothermal
  magnetohydrodynamics turbulence}.
\bjtitle{Physical Review E}
\bvolume{96}(\bissue{5}),
\bfpage{053205}
(\byear{2017})
\end{barticle}
\endbibitem

\bibitem{WanPP09}
\begin{barticle}
\bauthor{\bsnm{Wan}, \binits{M.}},
\bauthor{\bsnm{Servidio}, \binits{S.}},
\bauthor{\bsnm{Oughton}, \binits{S.}},
\bauthor{\bsnm{Matthaeus}, \binits{W.H.}}:
\batitle{The third-order law for increments in magnetohydrodynamic turbulence
  with constant shear}.
\bjtitle{Physics of plasmas}
\bvolume{16}(\bissue{9}),
\bfpage{090703}
(\byear{2009})
\end{barticle}
\endbibitem

\bibitem{GaltierPRE08}
\begin{barticle}
\bauthor{\bsnm{Galtier}, \binits{S.}}:
\batitle{von k{\'a}rm{\'a}n--howarth equations for hall magnetohydrodynamic
  flows}.
\bjtitle{Physical Review E}
\bvolume{77}(\bissue{1}),
\bfpage{015302}
(\byear{2008})
\end{barticle}
\endbibitem

\bibitem{FerrandJPP21}
\begin{botherref}
\oauthor{\bsnm{Ferrand}, \binits{R.}},
\oauthor{\bsnm{Galtier}, \binits{S.}},
\oauthor{\bsnm{Sahraoui}, \binits{F.}}:
A compact exact law for compressible isothermal hall magnetohydrodynamic
  turbulence.
Journal of Plasma Physics
\textbf{87}(2)
(2021)
\end{botherref}
\endbibitem

\bibitem{GaltierJGR08}
\begin{botherref}
\oauthor{\bsnm{Galtier}, \binits{S.}}:
Exact scaling laws for 3d electron mhd turbulence.
Journal of Geophysical Research: Space Physics
\textbf{113}(A1)
(2008)
\end{botherref}
\endbibitem

\bibitem{Sorisso-ValvoPRL07}
\begin{barticle}
\bauthor{\bsnm{Sorriso-Valvo}, \binits{L.}},
\bauthor{\bsnm{Marino}, \binits{R.}},
\bauthor{\bsnm{Carbone}, \binits{V.}},
\bauthor{\bsnm{Noullez}, \binits{A.}},
\bauthor{\bsnm{Lepreti}, \binits{F.}},
\bauthor{\bsnm{Veltri}, \binits{P.}},
\bauthor{\bsnm{Bruno}, \binits{R.}},
\bauthor{\bsnm{Bavassano}, \binits{B.}},
\bauthor{\bsnm{Pietropaolo}, \binits{E.}}:
\batitle{Observation of inertial energy cascade in interplanetary space
  plasma}.
\bjtitle{Physical review letters}
\bvolume{99}(\bissue{11}),
\bfpage{115001}
(\byear{2007})
\end{barticle}
\endbibitem

\bibitem{MarinoApJL08}
\begin{barticle}
\bauthor{\bsnm{Marino}, \binits{R.}},
\bauthor{\bsnm{Sorriso-Valvo}, \binits{L.}},
\bauthor{\bsnm{Carbone}, \binits{V.}},
\bauthor{\bsnm{Noullez}, \binits{A.}},
\bauthor{\bsnm{Bruno}, \binits{R.}},
\bauthor{\bsnm{Bavassano}, \binits{B.}}:
\batitle{Heating the solar wind by a magnetohydrodynamic turbulent energy
  cascade}.
\bjtitle{The Astrophysical Journal}
\bvolume{677}(\bissue{1}),
\bfpage{71}
(\byear{2008})
\end{barticle}
\endbibitem

\bibitem{OsmanEA11-3rd}
\begin{barticle}
\bauthor{\bsnm{Osman}, \binits{K.T.}},
\bauthor{\bsnm{Wan}, \binits{M.}},
\bauthor{\bsnm{Matthaeus}, \binits{W.H.}},
\bauthor{\bsnm{Weygand}, \binits{J.M.}},
\bauthor{\bsnm{Dasso}, \binits{S.}}:
\batitle{Anisotropic third-moment estimates of the energy cascade in solar wind
  turbulence using multispacecraft data}.
\bjtitle{Physical Review Letters}
\bvolume{107},
\bfpage{165001}
(\byear{2011}).
\doiurl{10.1103/PhysRevLett.107.165001}
\end{barticle}
\endbibitem

\bibitem{VerdiniApJ15}
\begin{barticle}
\bauthor{\bsnm{Verdini}, \binits{A.}},
\bauthor{\bsnm{Grappin}, \binits{R.}},
\bauthor{\bsnm{Hellinger}, \binits{P.}},
\bauthor{\bsnm{Landi}, \binits{S.}},
\bauthor{\bsnm{M{\"u}ller}, \binits{W.C.}}:
\batitle{Anisotropy of third-order structure functions in mhd turbulence}.
\bjtitle{The Astrophysical Journal}
\bvolume{804}(\bissue{2}),
\bfpage{119}
(\byear{2015})
\end{barticle}
\endbibitem

\bibitem{HellingerApJL18}
\begin{barticle}
\bauthor{\bsnm{Hellinger}, \binits{P.}},
\bauthor{\bsnm{Verdini}, \binits{A.}},
\bauthor{\bsnm{Landi}, \binits{S.}},
\bauthor{\bsnm{Franci}, \binits{L.}},
\bauthor{\bsnm{Matteini}, \binits{L.}}:
\batitle{von k{\'{a}}rm{\'{a}}n{\textendash}howarth equation for hall
  magnetohydrodynamics: Hybrid simulations}.
\bjtitle{The Astrophysical Journal}
\bvolume{857}(\bissue{2}),
\bfpage{19}
(\byear{2018}).
\doiurl{10.3847/2041-8213/aabc06}
\end{barticle}
\endbibitem

\bibitem{BandyopadhyayPRL20-PP}
\begin{barticle}
\bauthor{\bsnm{Bandyopadhyay}, \binits{R.}},
\bauthor{\bsnm{Sorriso-Valvo}, \binits{L.}},
\bauthor{\bsnm{Chasapis}, \binits{A.}},
\bauthor{\bsnm{Hellinger}, \binits{P.}},
\bauthor{\bsnm{Matthaeus}, \binits{W.H.}},
\bauthor{\bsnm{Verdini}, \binits{A.}},
\bauthor{\bsnm{Landi}, \binits{S.}},
\bauthor{\bsnm{Franci}, \binits{L.}},
\bauthor{\bsnm{Matteini}, \binits{L.}},
\bauthor{\bsnm{Giles}, \binits{B.L.}}, \betal:
\batitle{In situ observation of hall magnetohydrodynamic cascade in space
  plasma}.
\bjtitle{Physical Review Letters}
\bvolume{124}(\bissue{22}),
\bfpage{225101}
(\byear{2020})
\end{barticle}
\endbibitem

\bibitem{AluiePRL11}
\begin{barticle}
\bauthor{\bsnm{Aluie}, \binits{H.}}:
\batitle{Compressible turbulence: the cascade and its locality}.
\bjtitle{Physical review letters}
\bvolume{106}(\bissue{17}),
\bfpage{174502}
(\byear{2011})
\end{barticle}
\endbibitem

\bibitem{BianPRL19}
\begin{barticle}
\bauthor{\bsnm{Bian}, \binits{X.}},
\bauthor{\bsnm{Aluie}, \binits{H.}}:
\batitle{Decoupled cascades of kinetic and magnetic energy in
  magnetohydrodynamic turbulence}.
\bjtitle{Physical review letters}
\bvolume{122}(\bissue{13}),
\bfpage{135101}
(\byear{2019})
\end{barticle}
\endbibitem

\bibitem{SquirePRL17}
\begin{barticle}
\bauthor{\bsnm{Squire}, \binits{J.}},
\bauthor{\bsnm{Kunz}, \binits{M.W.}},
\bauthor{\bsnm{Quataert}, \binits{E.}},
\bauthor{\bsnm{Schekochihin}, \binits{A.}}:
\batitle{Kinetic simulations of the interruption of large-amplitude
  shear-alfv{\'e}n waves in a high-$\beta$ plasma}.
\bjtitle{Physical review letters}
\bvolume{119}(\bissue{15}),
\bfpage{155101}
(\byear{2017})
\end{barticle}
\endbibitem

\bibitem{FranciApJL17}
\begin{barticle}
\bauthor{\bsnm{Franci}, \binits{L.}},
\bauthor{\bsnm{Cerri}, \binits{S.S.}},
\bauthor{\bsnm{Califano}, \binits{F.}},
\bauthor{\bsnm{Landi}, \binits{S.}},
\bauthor{\bsnm{Papini}, \binits{E.}},
\bauthor{\bsnm{Verdini}, \binits{A.}},
\bauthor{\bsnm{Matteini}, \binits{L.}},
\bauthor{\bsnm{Jenko}, \binits{F.}},
\bauthor{\bsnm{Hellinger}, \binits{P.}}:
\batitle{Magnetic reconnection as a driver for a sub-ion-scale cascade in
  plasma turbulence}.
\bjtitle{The Astrophysical Journal Letters}
\bvolume{850}(\bissue{1}),
\bfpage{16}
(\byear{2017})
\end{barticle}
\endbibitem

\bibitem{KunzJPP20}
\begin{botherref}
\oauthor{\bsnm{Kunz}, \binits{M.}},
\oauthor{\bsnm{Squire}, \binits{J.}},
\oauthor{\bsnm{Schekochihin}, \binits{A.}},
\oauthor{\bsnm{Quataert}, \binits{E.}}:
Self-sustaining sound in collisionless, high-$\beta$ plasma.
Journal of Plasma Physics
\textbf{86}(6)
(2020)
\end{botherref}
\endbibitem

\bibitem{YangPP17}
\begin{barticle}
\bauthor{\bsnm{Yang}, \binits{Y.}},
\bauthor{\bsnm{Matthaeus}, \binits{W.H.}},
\bauthor{\bsnm{Parashar}, \binits{T.N.}},
\bauthor{\bsnm{Haggerty}, \binits{C.C.}},
\bauthor{\bsnm{Roytershteyn}, \binits{V.}},
\bauthor{\bsnm{Daughton}, \binits{W.}},
\bauthor{\bsnm{Wan}, \binits{M.}},
\bauthor{\bsnm{Shi}, \binits{Y.}},
\bauthor{\bsnm{Chen}, \binits{S.}}:
\batitle{Energy transfer, pressure tensor, and heating of kinetic plasma}.
\bjtitle{Physics of Plasmas}
\bvolume{24}(\bissue{7}),
\bfpage{072306}
(\byear{2017})
{\href{https://arxiv.org/abs/http://dx.doi.org/10.1063/1.4990421}{{http://dx.doi.org/10.1063/1.4990421}}}.
\doiurl{10.1063/1.4990421}
\end{barticle}
\endbibitem

\bibitem{CamporealePRL18}
\begin{barticle}
\bauthor{\bsnm{Camporeale}, \binits{E.}},
\bauthor{\bsnm{Sorriso-Valvo}, \binits{L.}},
\bauthor{\bsnm{Califano}, \binits{F.}},
\bauthor{\bsnm{Retin{\`o}}, \binits{A.}}:
\batitle{Coherent structures and spectral energy transfer in turbulent plasma:
  a space-filter approach}.
\bjtitle{Physical review letters}
\bvolume{120}(\bissue{12}),
\bfpage{125101}
(\byear{2018})
\end{barticle}
\endbibitem

\bibitem{CerriPP20}
\begin{barticle}
\bauthor{\bsnm{Cerri}, \binits{S.}},
\bauthor{\bsnm{Camporeale}, \binits{E.}}:
\batitle{Space-filter techniques for quasi-neutral hybrid-kinetic models}.
\bjtitle{Physics of Plasmas}
\bvolume{27}(\bissue{8}),
\bfpage{082102}
(\byear{2020})
\end{barticle}
\endbibitem

\bibitem{YangApJ22}
\begin{barticle}
\bauthor{\bsnm{Yang}, \binits{Y.}},
\bauthor{\bsnm{Matthaeus}, \binits{W.H.}},
\bauthor{\bsnm{Roy}, \binits{S.}},
\bauthor{\bsnm{Roytershteyn}, \binits{V.}},
\bauthor{\bsnm{Parashar}, \binits{T.}},
\bauthor{\bsnm{Bandyopadhyay}, \binits{R.}},
\bauthor{\bsnm{Wan}, \binits{M.}}:
\batitle{Pressure-strain interaction as the energy dissipation estimate in
  collisionless plasma}.
\bjtitle{Astrophys. J.}
\bvolume{929},
\bfpage{142}
(\byear{2022}).
\doiurl{10.3847/1538-4357/ac5d3e}
\end{barticle}
\endbibitem

\bibitem{HellingerArXiv22}
\begin{botherref}
\oauthor{\bsnm{Hellinger}, \binits{P.}},
\oauthor{\bsnm{Montagud-Camps}, \binits{V.}},
\oauthor{\bsnm{Franci}, \binits{L.}},
\oauthor{\bsnm{Matteini}, \binits{L.}},
\oauthor{\bsnm{Papini}, \binits{E.}},
\oauthor{\bsnm{Verdini}, \binits{A.}},
\oauthor{\bsnm{Landi}, \binits{S.}}:
Ion-scale transition of plasma turbulence: Pressure-strain effect.
arXiv preprint arXiv:2203.12322
(2022)
\end{botherref}
\endbibitem

\bibitem{HollwegPRL71}
\begin{barticle}
\bauthor{\bsnm{Hollweg}, \binits{J.V.}}:
\batitle{Nonlinear landau damping of alfv{\'e}n waves}.
\bjtitle{Physical Review Letters}
\bvolume{27}(\bissue{20}),
\bfpage{1349}
(\byear{1971})
\end{barticle}
\endbibitem

\bibitem{ChenNature19}
\begin{barticle}
\bauthor{\bsnm{Chen}, \binits{C.}},
\bauthor{\bsnm{Klein}, \binits{K.}},
\bauthor{\bsnm{Howes}, \binits{G.G.}}:
\batitle{Evidence for electron landau damping in space plasma turbulence}.
\bjtitle{Nature communications}
\bvolume{10}(\bissue{1}),
\bfpage{1}--\blpage{8}
(\byear{2019})
\end{barticle}
\endbibitem

\bibitem{HollwegJGR02}
\begin{barticle}
\bauthor{\bsnm{Hollweg}, \binits{J.V.}},
\bauthor{\bsnm{Isenberg}, \binits{P.A.}}:
\batitle{Generation of the fast solar wind: A review with emphasis on the
  resonant cyclotron interaction}.
\bjtitle{Journal of Geophysical Research: Space Physics}
\bvolume{107}(\bissue{A7}),
\bfpage{12}
(\byear{2002})
\end{barticle}
\endbibitem

\bibitem{KasperPRL13}
\begin{barticle}
\bauthor{\bsnm{Kasper}, \binits{J.C.}},
\bauthor{\bsnm{Maruca}, \binits{B.A.}},
\bauthor{\bsnm{Stevens}, \binits{M.L.}},
\bauthor{\bsnm{Zaslavsky}, \binits{A.}}:
\batitle{Sensitive test for ion-cyclotron resonant heating in the solar wind}.
\bjtitle{Physical review letters}
\bvolume{110}(\bissue{9}),
\bfpage{091102}
(\byear{2013})
\end{barticle}
\endbibitem

\bibitem{DawsonNF65}
\begin{barticle}
\bauthor{\bsnm{Dawson}, \binits{J.M.}},
\bauthor{\bsnm{Uman}, \binits{M.F.}}:
\batitle{Heating a plasma by means of magnetic pumping}.
\bjtitle{Nuclear Fusion}
\bvolume{5}(\bissue{3}),
\bfpage{242}
(\byear{1965})
\end{barticle}
\endbibitem

\bibitem{LichkoApJL17}
\begin{barticle}
\bauthor{\bsnm{Lichko}, \binits{E.}},
\bauthor{\bsnm{Egedal}, \binits{J.}},
\bauthor{\bsnm{Daughton}, \binits{W.}},
\bauthor{\bsnm{Kasper}, \binits{J.}}:
\batitle{Magnetic pumping as a source of particle heating and power-law
  distributions in the solar wind}.
\bjtitle{The Astrophysical Journal Letters}
\bvolume{850}(\bissue{2}),
\bfpage{28}
(\byear{2017})
\end{barticle}
\endbibitem

\bibitem{ChandranApJ10-1}
\begin{barticle}
\bauthor{\bsnm{Chandran}, \binits{B.D.G.}},
\bauthor{\bsnm{Li}, \binits{B.}},
\bauthor{\bsnm{Rogers}, \binits{B.N.}},
\bauthor{\bsnm{Quataert}, \binits{E.}},
\bauthor{\bsnm{Germaschewski}, \binits{K.}}:
\batitle{Perpendicular ion heating by low-frequency alfv\'en-wave turbulence in
  the solar wind}.
\bjtitle{The Astrophysical Journal}
\bvolume{720}(\bissue{1}),
\bfpage{503}
(\byear{2010})
\end{barticle}
\endbibitem

\bibitem{XiaApJ13}
\begin{barticle}
\bauthor{\bsnm{Xia}, \binits{Q.}},
\bauthor{\bsnm{Perez}, \binits{J.C.}},
\bauthor{\bsnm{Chandran}, \binits{B.D.}},
\bauthor{\bsnm{Quataert}, \binits{E.}}:
\batitle{Perpendicular ion heating by reduced magnetohydrodynamic turbulence}.
\bjtitle{The Astrophysical Journal}
\bvolume{776}(\bissue{2}),
\bfpage{90}
(\byear{2013})
\end{barticle}
\endbibitem

\bibitem{MalletJPP19}
\begin{botherref}
\oauthor{\bsnm{Mallet}, \binits{A.}},
\oauthor{\bsnm{Klein}, \binits{K.G.}},
\oauthor{\bsnm{Chandran}, \binits{B.D.}},
\oauthor{\bsnm{Gro{\v{s}}elj}, \binits{D.}},
\oauthor{\bsnm{Hoppock}, \binits{I.W.}},
\oauthor{\bsnm{Bowen}, \binits{T.A.}},
\oauthor{\bsnm{Salem}, \binits{C.S.}},
\oauthor{\bsnm{Bale}, \binits{S.D.}}:
Interplay between intermittency and dissipation in collisionless plasma
  turbulence.
Journal of Plasma Physics
\textbf{85}(3)
(2019)
\end{botherref}
\endbibitem

\bibitem{CerriApJ21}
\begin{barticle}
\bauthor{\bsnm{Cerri}, \binits{S.S.}},
\bauthor{\bsnm{Arzamasskiy}, \binits{L.}},
\bauthor{\bsnm{Kunz}, \binits{M.W.}}:
\batitle{On stochastic heating and its phase-space signatures in low-beta
  kinetic turbulence}.
\bjtitle{The Astrophysical Journal}
\bvolume{916}(\bissue{2}),
\bfpage{120}
(\byear{2021})
\end{barticle}
\endbibitem

\bibitem{MartinovicApJ21}
\begin{barticle}
\bauthor{\bsnm{Martinovi{\'c}}, \binits{M.M.}},
\bauthor{\bsnm{Klein}, \binits{K.G.}},
\bauthor{\bsnm{Huang}, \binits{J.}},
\bauthor{\bsnm{Chandran}, \binits{B.D.}},
\bauthor{\bsnm{Kasper}, \binits{J.C.}},
\bauthor{\bsnm{Lichko}, \binits{E.}},
\bauthor{\bsnm{Bowen}, \binits{T.}},
\bauthor{\bsnm{Chen}, \binits{C.H.}},
\bauthor{\bsnm{Matteini}, \binits{L.}},
\bauthor{\bsnm{Stevens}, \binits{M.}}, \betal:
\batitle{Multiscale solar wind turbulence properties inside and near
  switchbacks measured by the parker solar probe}.
\bjtitle{The Astrophysical Journal}
\bvolume{912}(\bissue{1}),
\bfpage{28}
(\byear{2021})
\end{barticle}
\endbibitem

\bibitem{TenbargeApJ13}
\begin{barticle}
\bauthor{\bsnm{TenBarge}, \binits{J.}},
\bauthor{\bsnm{Howes}, \binits{G.}},
\bauthor{\bsnm{Dorland}, \binits{W.}}:
\batitle{Collisionless damping at electron scales in solar wind turbulence}.
\bjtitle{The Astrophysical Journal}
\bvolume{774}(\bissue{2}),
\bfpage{139}
(\byear{2013})
\end{barticle}
\endbibitem

\bibitem{HughesGRL14}
\begin{barticle}
\bauthor{\bsnm{Hughes}, \binits{R.S.}},
\bauthor{\bsnm{Gary}, \binits{S.P.}},
\bauthor{\bsnm{Wang}, \binits{J.}}:
\batitle{Electron and ion heating by whistler turbulence: Three-dimensional
  particle-in-cell simulations}.
\bjtitle{Geophysical Research Letters}
\bvolume{41}(\bissue{24}),
\bfpage{8681}--\blpage{8687}
(\byear{2014})
\end{barticle}
\endbibitem

\bibitem{RoyApJL22}
\begin{botherref}
\oauthor{\bsnm{Roy}, \binits{S.}},
\oauthor{\bsnm{Bandyopadhyay}, \binits{R.}},
\oauthor{\bsnm{Yang}, \binits{Y.}},
\oauthor{\bsnm{Matthaeus}, \binits{W.H.}},
\oauthor{\bsnm{Adhikari}, \binits{S.}},
\oauthor{\bsnm{Parashar}, \binits{T.N.}},
\oauthor{\bsnm{Chasapis}, \binits{A.}},
\oauthor{\bsnm{Li}, \binits{H.}},
\oauthor{\bsnm{Gershman}, \binits{D.J.}},
\oauthor{\bsnm{Giles}, \binits{B.L.}},
\oauthor{\bsnm{Burch}, \binits{J.L.}}:
Turbulent energy transfer and proton-electron heating in collisionless plasmas.
Physical Review X
(Under Review)
\end{botherref}
\endbibitem

\bibitem{YangPRE17}
\begin{barticle}
\bauthor{\bsnm{Yang}, \binits{Y.}},
\bauthor{\bsnm{Matthaeus}, \binits{W.H.}},
\bauthor{\bsnm{Parashar}, \binits{T.N.}},
\bauthor{\bsnm{Wu}, \binits{P.}},
\bauthor{\bsnm{Wan}, \binits{M.}},
\bauthor{\bsnm{Shi}, \binits{Y.}},
\bauthor{\bsnm{Chen}, \binits{S.}},
\bauthor{\bsnm{Roytershteyn}, \binits{V.}},
\bauthor{\bsnm{Daughton}, \binits{W.}}:
\batitle{Energy transfer channels and turbulence cascade in vlasov-maxwell
  turbulence}.
\bjtitle{Phys. Rev. E}
\bvolume{95},
\bfpage{061201}
(\byear{2017}).
\doiurl{10.1103/PhysRevE.95.061201}
\end{barticle}
\endbibitem

\bibitem{DelSartoPRE16}
\begin{barticle}
\bauthor{\bsnm{Del~Sarto}, \binits{D.}},
\bauthor{\bsnm{Pegoraro}, \binits{F.}},
\bauthor{\bsnm{Califano}, \binits{F.}}:
\batitle{Pressure anisotropy and small spatial scales induced by velocity
  shear}.
\bjtitle{Physical Review E}
\bvolume{93}(\bissue{5}),
\bfpage{053203}
(\byear{2016})
\end{barticle}
\endbibitem

\bibitem{HuangBook}
\begin{bbook}
\bauthor{\bsnm{Huang}, \binits{K.}}:
\bbtitle{Statistical Mechanics}.
\bpublisher{John Wiley \& Sons}, \blocation{???}
(\byear{2008})
\end{bbook}
\endbibitem

\bibitem{ServidioPRL12}
\begin{barticle}
\bauthor{\bsnm{Servidio}, \binits{S.}},
\bauthor{\bsnm{Valentini}, \binits{F.}},
\bauthor{\bsnm{Califano}, \binits{F.}},
\bauthor{\bsnm{Veltri}, \binits{P.}}:
\batitle{Local kinetic effects in two-dimensional plasma turbulence}.
\bjtitle{Physical review letters}
\bvolume{108}(\bissue{4}),
\bfpage{045001}
(\byear{2012})
\end{barticle}
\endbibitem

\bibitem{ServidioApJL14}
\begin{barticle}
\bauthor{\bsnm{Servidio}, \binits{S.}},
\bauthor{\bsnm{Osman}, \binits{K.}},
\bauthor{\bsnm{Valentini}, \binits{F.}},
\bauthor{\bsnm{Perrone}, \binits{D.}},
\bauthor{\bsnm{Califano}, \binits{F.}},
\bauthor{\bsnm{Chapman}, \binits{S.}},
\bauthor{\bsnm{Matthaeus}, \binits{W.}},
\bauthor{\bsnm{Veltri}, \binits{P.}}:
\batitle{Proton kinetic effects in vlasov and solar wind turbulence}.
\bjtitle{The Astrophysical Journal Letters}
\bvolume{781}(\bissue{2}),
\bfpage{27}
(\byear{2014})
\end{barticle}
\endbibitem

\bibitem{FranciAIP15}
\begin{bchapter}
\bauthor{\bsnm{Franci}, \binits{L.}},
\bauthor{\bsnm{Hellinger}, \binits{P.}},
\bauthor{\bsnm{Matteini}, \binits{L.}},
\bauthor{\bsnm{Verdini}, \binits{A.}},
\bauthor{\bsnm{Landi}, \binits{S.}}:
\bctitle{Two-dimensional hybrid simulations of kinetic plasma turbulence:
  Current and vorticity vs proton temperature}.
In: \bbtitle{AIP Conference Proceedings},
vol. \bseriesno{1720},
p. \bfpage{040003}
(\byear{2016}).
\bcomment{AIP Publishing LLC}
\end{bchapter}
\endbibitem

\bibitem{ParasharApJ16b}
\begin{barticle}
\bauthor{\bsnm{Parashar}, \binits{T.N.}},
\bauthor{\bsnm{Matthaeus}, \binits{W.H.}}:
\batitle{Propinquity of current and vortex structures: Effects on collisionless
  plasma heating}.
\bjtitle{The Astrophysical Journal}
\bvolume{832}(\bissue{1}),
\bfpage{57}
(\byear{2016})
\end{barticle}
\endbibitem

\bibitem{MatthaeusGRL82}
\begin{barticle}
\bauthor{\bsnm{Matthaeus}, \binits{W.H.}}:
\batitle{Reconnection in two dimensions: Localization of vorticity and current
  near magnetic x-points}.
\bjtitle{Geophysical Research Letters}
\bvolume{9}(\bissue{6}),
\bfpage{660}--\blpage{663}
(\byear{1982})
\end{barticle}
\endbibitem

\bibitem{SitnovGRL18}
\begin{barticle}
\bauthor{\bsnm{Sitnov}, \binits{M.}},
\bauthor{\bsnm{Merkin}, \binits{V.}},
\bauthor{\bsnm{Roytershteyn}, \binits{V.}},
\bauthor{\bsnm{Swisdak}, \binits{M.}}:
\batitle{Kinetic dissipation around a dipolarization front}.
\bjtitle{Geophysical Research Letters}
\bvolume{45}(\bissue{10}),
\bfpage{4639}--\blpage{4647}
(\byear{2018})
\end{barticle}
\endbibitem

\bibitem{MatthaeusApJ20}
\begin{barticle}
\bauthor{\bsnm{Matthaeus}, \binits{W.H.}},
\bauthor{\bsnm{Yang}, \binits{Y.}},
\bauthor{\bsnm{Wan}, \binits{M.}},
\bauthor{\bsnm{Parashar}, \binits{T.N.}},
\bauthor{\bsnm{Bandyopadhyay}, \binits{R.}},
\bauthor{\bsnm{Chasapis}, \binits{A.}},
\bauthor{\bsnm{Pezzi}, \binits{O.}},
\bauthor{\bsnm{Valentini}, \binits{F.}}:
\batitle{Pathways to dissipation in weakly collisional plasmas}.
\bjtitle{The Astrophysical Journal}
\bvolume{891}(\bissue{1}),
\bfpage{101}
(\byear{2020}).
\doiurl{10.3847/1538-4357/ab6d6a}
\end{barticle}
\endbibitem

\bibitem{BandyopadhyayPRL20}
\begin{barticle}
\bauthor{\bsnm{Bandyopadhyay}, \binits{R.}},
\bauthor{\bsnm{Matthaeus}, \binits{W.H.}},
\bauthor{\bsnm{Parashar}, \binits{T.N.}},
\bauthor{\bsnm{Yang}, \binits{Y.}},
\bauthor{\bsnm{Chasapis}, \binits{A.}},
\bauthor{\bsnm{Giles}, \binits{B.L.}},
\bauthor{\bsnm{Gershman}, \binits{D.J.}},
\bauthor{\bsnm{Pollock}, \binits{C.J.}},
\bauthor{\bsnm{Russell}, \binits{C.T.}},
\bauthor{\bsnm{Strangeway}, \binits{R.J.}}, \betal:
\batitle{Statistics of kinetic dissipation in the earth’s magnetosheath: Mms
  observations}.
\bjtitle{Physical Review Letters}
\bvolume{124}(\bissue{25}),
\bfpage{255101}
(\byear{2020})
\end{barticle}
\endbibitem

\bibitem{GrecoSSR18}
\begin{barticle}
\bauthor{\bsnm{Greco}, \binits{A.}},
\bauthor{\bsnm{Matthaeus}, \binits{W.}},
\bauthor{\bsnm{Perri}, \binits{S.}},
\bauthor{\bsnm{Osman}, \binits{K.}},
\bauthor{\bsnm{Servidio}, \binits{S.}},
\bauthor{\bsnm{Wan}, \binits{M.}},
\bauthor{\bsnm{Dmitruk}, \binits{P.}}:
\batitle{Partial variance of increments method in solar wind observations and
  plasma simulations}.
\bjtitle{Space Science Reviews}
\bvolume{214}(\bissue{1}),
\bfpage{1}
(\byear{2018})
\end{barticle}
\endbibitem

\bibitem{Sorisso-ValvoSP18}
\begin{barticle}
\bauthor{\bsnm{Sorriso-Valvo}, \binits{L.}},
\bauthor{\bsnm{Carbone}, \binits{F.}},
\bauthor{\bsnm{Perri}, \binits{S.}},
\bauthor{\bsnm{Greco}, \binits{A.}},
\bauthor{\bsnm{Marino}, \binits{R.}},
\bauthor{\bsnm{Bruno}, \binits{R.}}:
\batitle{On the statistical properties of turbulent energy transfer rate in the
  inner heliosphere}.
\bjtitle{Solar Physics}
\bvolume{293}(\bissue{1}),
\bfpage{1}--\blpage{16}
(\byear{2018})
\end{barticle}
\endbibitem

\bibitem{OsmanApJL11}
\begin{barticle}
\bauthor{\bsnm{Osman}, \binits{K.T.}},
\bauthor{\bsnm{Matthaeus}, \binits{W.H.}},
\bauthor{\bsnm{Greco}, \binits{A.}},
\bauthor{\bsnm{Servidio}, \binits{S.}}:
\batitle{Evidence for inhomogeneous heating in the solar wind}.
\bjtitle{The Astrophysical Journal Letters}
\bvolume{727}(\bissue{1}),
\bfpage{11}
(\byear{2011})
\end{barticle}
\endbibitem

\bibitem{TesseinApJL13}
\begin{barticle}
\bauthor{\bsnm{Tessein}, \binits{J.}},
\bauthor{\bsnm{Matthaeus}, \binits{W.}},
\bauthor{\bsnm{Wan}, \binits{M.}},
\bauthor{\bsnm{Osman}, \binits{K.}},
\bauthor{\bsnm{Ruffolo}, \binits{D.}},
\bauthor{\bsnm{Giacalone}, \binits{J.}}:
\batitle{Association of suprathermal particles with coherent structures and
  shocks}.
\bjtitle{The Astrophysical Journal Letters}
\bvolume{776}(\bissue{1}),
\bfpage{8}
(\byear{2013})
\end{barticle}
\endbibitem

\bibitem{SquireNatureAstron22}
\begin{botherref}
\oauthor{\bsnm{Squire}, \binits{J.}},
\oauthor{\bsnm{Meyrand}, \binits{R.}},
\oauthor{\bsnm{Kunz}, \binits{M.W.}},
\oauthor{\bsnm{Arzamasskiy}, \binits{L.}},
\oauthor{\bsnm{Schekochihin}, \binits{A.A.}},
\oauthor{\bsnm{Quataert}, \binits{E.}}:
High-frequency heating of the solar wind triggered by low-frequency turbulence.
Nature Astronomy,
1--9
(2022)
\end{botherref}
\endbibitem

\bibitem{AlexandrovaPRE21}
\begin{barticle}
\bauthor{\bsnm{Alexandrova}, \binits{O.}},
\bauthor{\bsnm{Jagarlamudi}, \binits{V.K.}},
\bauthor{\bsnm{Hellinger}, \binits{P.}},
\bauthor{\bsnm{Maksimovic}, \binits{M.}},
\bauthor{\bsnm{Shprits}, \binits{Y.}},
\bauthor{\bsnm{Mangeney}, \binits{A.}}:
\batitle{Spectrum of kinetic plasma turbulence at 0.3--0.9 astronomical units
  from the sun}.
\bjtitle{Physical Review E}
\bvolume{103}(\bissue{6}),
\bfpage{063202}
(\byear{2021})
\end{barticle}
\endbibitem

\bibitem{AlexandrovaApJ12}
\begin{barticle}
\bauthor{\bsnm{Alexandrova}, \binits{O.}},
\bauthor{\bsnm{Lacombe}, \binits{C.}},
\bauthor{\bsnm{Mangeney}, \binits{A.}},
\bauthor{\bsnm{Grappin}, \binits{R.}},
\bauthor{\bsnm{Maksimovic}, \binits{M.}}:
\batitle{Solar wind turbulent spectrum at plasma kinetic scales}.
\bjtitle{The Astrophysical Journal}
\bvolume{760}(\bissue{2}),
\bfpage{121}
(\byear{2012})
\end{barticle}
\endbibitem

\bibitem{ArroArXiv21}
\begin{botherref}
\oauthor{\bsnm{Arr{\`o}}, \binits{G.}},
\oauthor{\bsnm{Califano}, \binits{F.}},
\oauthor{\bsnm{Lapenta}, \binits{G.}}:
Spectral properties and energy cascade at kinetic scales in collisionless
  plasma turbulence.
arXiv preprint arXiv:2112.12753
(2021)
\end{botherref}
\endbibitem

\bibitem{HolzerJGR66}
\begin{barticle}
\bauthor{\bsnm{Holzer}, \binits{R.E.}},
\bauthor{\bsnm{McLeod}, \binits{M.G.}},
\bauthor{\bsnm{Smith}, \binits{E.J.}}:
\batitle{Preliminary results from the ogo 1 search coil magnetometer: Boundary
  positions and magnetic noise spectra}.
\bjtitle{Journal of Geophysical Research}
\bvolume{71}(\bissue{5}),
\bfpage{1481}--\blpage{1486}
(\byear{1966})
\end{barticle}
\endbibitem

\bibitem{Coleman66}
\begin{barticle}
\bauthor{\bsnm{Coleman}, \binits{P.J.}}:
\batitle{Hydromagnetic waves in the interplanetary plasma}.
\bjtitle{Phys. Rev. Lett.}
\bvolume{17},
\bfpage{207}--\blpage{211}
(\byear{1966})
\end{barticle}
\endbibitem

\bibitem{SturrockPRL66}
\begin{barticle}
\bauthor{\bsnm{Sturrock}, \binits{P.}},
\bauthor{\bsnm{Hartle}, \binits{R.}}:
\batitle{Two-fluid model of the solar wind}.
\bjtitle{Physical Review Letters}
\bvolume{16}(\bissue{14}),
\bfpage{628}
(\byear{1966})
\end{barticle}
\endbibitem

\bibitem{Chandra49-apj}
\begin{barticle}
\bauthor{\bsnm{Chandrasekhar}, \binits{S.}}:
\batitle{Turbulence---{A} physical theory of astrophysical interest}.
\bjtitle{Astrophys. J.}
\bvolume{110},
\bfpage{329}--\blpage{339}
(\byear{1949})
\end{barticle}
\endbibitem

\bibitem{BelcherJGR74}
\begin{barticle}
\bauthor{\bsnm{Belcher}, \binits{J.W.}},
\bauthor{\bsnm{Burchsted}, \binits{R.}}:
\batitle{Energy densities of alfv{\'e}n waves between 0.7 and 1.6 au}.
\bjtitle{Journal of Geophysical Research}
\bvolume{79}(\bissue{31}),
\bfpage{4765}--\blpage{4768}
(\byear{1974})
\end{barticle}
\endbibitem

\bibitem{RobertsEA90}
\begin{barticle}
\bauthor{\bsnm{Roberts}, \binits{D.A.}},
\bauthor{\bsnm{Goldstein}, \binits{M.L.}},
\bauthor{\bsnm{Klein}, \binits{L.W.}}:
\batitle{The amplitudes of interplanetary fluctuations: {Stream} structure,
  heliocentric distance, and frequency dependence}.
\bjtitle{J. Geophys. Res.}
\bvolume{95},
\bfpage{4203}--\blpage{4216}
(\byear{1990})
\end{barticle}
\endbibitem

\bibitem{VermaRoberts93}
\begin{barticle}
\bauthor{\bsnm{Verma}, \binits{M.K.}},
\bauthor{\bsnm{Roberts}, \binits{D.A.}}:
\batitle{The radial evolution of the amplitudes of ``dissipationless''
  turbulent solar wind fluctuations}.
\bjtitle{J. Geophys. Res.}
\bvolume{98},
\bfpage{5625}
(\byear{1993})
\end{barticle}
\endbibitem

\bibitem{ZankJGR96}
\begin{barticle}
\bauthor{\bsnm{Zank}, \binits{G.}},
\bauthor{\bsnm{Matthaeus}, \binits{W.}},
\bauthor{\bsnm{Smith}, \binits{C.}}:
\batitle{Evolution of turbulent magnetic fluctuation power with heliospheric
  distance}.
\bjtitle{Journal of Geophysical Research: Space Physics}
\bvolume{101}(\bissue{A8}),
\bfpage{17093}--\blpage{17107}
(\byear{1996})
\end{barticle}
\endbibitem

\bibitem{BavassanoJGR82}
\begin{barticle}
\bauthor{\bsnm{Bavassano}, \binits{B.}},
\bauthor{\bsnm{Dobrowolny}, \binits{M.}},
\bauthor{\bsnm{Mariani}, \binits{F.}},
\bauthor{\bsnm{Ness}, \binits{N.}}:
\batitle{Radial evolution of power spectra of interplanetary alfv{\'e}nic
  turbulence}.
\bjtitle{Journal of Geophysical Research: Space Physics}
\bvolume{87}(\bissue{A5}),
\bfpage{3617}--\blpage{3622}
(\byear{1982})
\end{barticle}
\endbibitem

\bibitem{HorburyJGR01}
\begin{barticle}
\bauthor{\bsnm{Horbury}, \binits{T.}},
\bauthor{\bsnm{Balogh}, \binits{A.}}:
\batitle{Evolution of magnetic field fluctuations in high-speed solar wind
  streams: Ulysses and helios observations}.
\bjtitle{Journal of Geophysical Research: Space Physics}
\bvolume{106}(\bissue{A8}),
\bfpage{15929}--\blpage{15940}
(\byear{2001})
\end{barticle}
\endbibitem

\bibitem{BrunoEMP09}
\begin{barticle}
\bauthor{\bsnm{Bruno}, \binits{R.}},
\bauthor{\bsnm{Carbone}, \binits{V.}},
\bauthor{\bsnm{V{\"o}r{\"o}s}, \binits{Z.}},
\bauthor{\bsnm{D’Amicis}, \binits{R.}},
\bauthor{\bsnm{Bavassano}, \binits{B.}},
\bauthor{\bsnm{Cattaneo}, \binits{M.}},
\bauthor{\bsnm{Mura}, \binits{A.}},
\bauthor{\bsnm{Milillo}, \binits{A.}},
\bauthor{\bsnm{Orsini}, \binits{S.}},
\bauthor{\bsnm{Veltri}, \binits{P.}}, \betal:
\batitle{Coordinated study on solar wind turbulence during the venus-express,
  ace and ulysses alignment of august 2007}.
\bjtitle{Earth, Moon, and Planets}
\bvolume{104}(\bissue{1}),
\bfpage{101}--\blpage{104}
(\byear{2009})
\end{barticle}
\endbibitem

\bibitem{Hollweg86}
\begin{barticle}
\bauthor{\bsnm{Hollweg}, \binits{J.V.}}:
\batitle{Transition region, corona, and solar wind in coronal holes}.
\bjtitle{J. Geophys. Res.}
\bvolume{91},
\bfpage{4111}
(\byear{1986})
\end{barticle}
\endbibitem

\bibitem{HollwegJohnson88}
\begin{barticle}
\bauthor{\bsnm{Hollweg}, \binits{J.V.}},
\bauthor{\bsnm{Johnson}, \binits{W.}}:
\batitle{Transition region, corona, and solar wind in coronal holes: {Some} two
  fluid models}.
\bjtitle{J. Geophys. Res.}
\bvolume{93},
\bfpage{9547}
(\byear{1988})
\end{barticle}
\endbibitem

\bibitem{TuJGR84}
\begin{barticle}
\bauthor{\bsnm{Tu}, \binits{C.-Y.}},
\bauthor{\bsnm{Pu}, \binits{Z.-Y.}},
\bauthor{\bsnm{Wei}, \binits{F.-S.}}:
\batitle{The power spectrum of interplanetary alfv{\'e}nic fluctuations:
  Derivation of the governing equation and its solution}.
\bjtitle{Journal of Geophysical Research: Space Physics}
\bvolume{89}(\bissue{A11}),
\bfpage{9695}--\blpage{9702}
(\byear{1984})
\end{barticle}
\endbibitem

\bibitem{TuJGR88}
\begin{barticle}
\bauthor{\bsnm{Tu}, \binits{C.-y.}}:
\batitle{The damping of interplanetary alfv{\'e}nic fluctuations and the
  heating of the solar wind}.
\bjtitle{Journal of Geophysical Research: Space Physics}
\bvolume{93}(\bissue{A1}),
\bfpage{7}--\blpage{20}
(\byear{1988})
\end{barticle}
\endbibitem

\bibitem{Hollweg74}
\begin{barticle}
\bauthor{\bsnm{Hollweg}, \binits{J.V.}}:
\batitle{Transverse alfv\'en waves in the solar wind: {Arbitrary} {$\bf k,
  v_{0}, B_{0}$} and $\delta {\bf b}$}.
\bjtitle{J. Geophys. Res.}
\bvolume{79},
\bfpage{1539}
(\byear{1974})
\end{barticle}
\endbibitem

\bibitem{MarschTu89}
\begin{barticle}
\bauthor{\bsnm{Marsch}, \binits{E.}},
\bauthor{\bsnm{Tu}, \binits{C.-Y.}}:
\batitle{Dynamics of correlation functions with {Els\"asser} variables for
  inhomogeneous mhd turbulence}.
\bjtitle{Journal of Plasma Physics}
\bvolume{41},
\bfpage{479}--\blpage{491}
(\byear{1989})
\end{barticle}
\endbibitem

\bibitem{ZhouMatt89}
\begin{barticle}
\bauthor{\bsnm{Zhou}, \binits{Y.}},
\bauthor{\bsnm{Matthaeus}, \binits{W.H.}}:
\batitle{Non-{WKB} evolution of solar wind fluctuations: {A} turbulence
  modeling approach}.
\bjtitle{Geophysical Research Letters}
\bvolume{16},
\bfpage{755}
(\byear{1989})
\end{barticle}
\endbibitem

\bibitem{MatthaeusJGR94}
\begin{barticle}
\bauthor{\bsnm{Matthaeus}, \binits{W.H.}},
\bauthor{\bsnm{Oughton}, \binits{S.}},
\bauthor{\bsnm{Pontius~Jr}, \binits{D.H.}},
\bauthor{\bsnm{Zhou}, \binits{Y.}}:
\batitle{Evolution of energy-containing turbulent eddies in the solar wind}.
\bjtitle{Journal of Geophysical Research: Space Physics}
\bvolume{99}(\bissue{A10}),
\bfpage{19267}--\blpage{19287}
(\byear{1994})
\end{barticle}
\endbibitem

\bibitem{SmithJGR01}
\begin{barticle}
\bauthor{\bsnm{Smith}, \binits{C.W.}},
\bauthor{\bsnm{Matthaeus}, \binits{W.H.}},
\bauthor{\bsnm{Zank}, \binits{G.P.}},
\bauthor{\bsnm{Ness}, \binits{N.F.}},
\bauthor{\bsnm{Oughton}, \binits{S.}},
\bauthor{\bsnm{Richardson}, \binits{J.D.}}:
\batitle{Heating of the low-latitude solar wind by dissipation of turbulent
  magnetic fluctuations}.
\bjtitle{Journal of Geophysical Research: Space Physics}
\bvolume{106}(\bissue{A5}),
\bfpage{8253}--\blpage{8272}
(\byear{2001})
\end{barticle}
\endbibitem

\bibitem{RuizSP14}
\begin{barticle}
\bauthor{\bsnm{Ruiz}, \binits{M.E.}},
\bauthor{\bsnm{Dasso}, \binits{S.}},
\bauthor{\bsnm{Matthaeus}, \binits{W.}},
\bauthor{\bsnm{Weygand}, \binits{J.}}:
\batitle{Characterization of the turbulent magnetic integral length in the
  solar wind: from 0.3 to 5 astronomical units}.
\bjtitle{Solar Physics}
\bvolume{289}(\bissue{10}),
\bfpage{3917}--\blpage{3933}
(\byear{2014})
\end{barticle}
\endbibitem

\bibitem{RobertsJGR10}
\begin{botherref}
\oauthor{\bsnm{Roberts}, \binits{D.A.}}:
Evolution of the spectrum of solar wind velocity fluctuations from 0.3 to 5 au.
Journal of Geophysical Research: Space Physics
\textbf{115}(A12)
(2010)
\end{botherref}
\endbibitem

\bibitem{MattEA99-swh}
\begin{barticle}
\bauthor{\bsnm{Matthaeus}, \binits{W.H.}},
\bauthor{\bsnm{Zank}, \binits{G.P.}},
\bauthor{\bsnm{Smith}, \binits{C.W.}},
\bauthor{\bsnm{Oughton}, \binits{S.}}:
\batitle{Turbulence, spatial transport, and heating of the solar wind}.
\bjtitle{Physical Review Letters}
\bvolume{82},
\bfpage{3444}--\blpage{3447}
(\byear{1999}).
\doiurl{10.1103/PhysRevLett.82.3444}
\end{barticle}
\endbibitem

\bibitem{GoldsteinGRL95}
\begin{barticle}
\bauthor{\bsnm{Goldstein}, \binits{B.}},
\bauthor{\bsnm{Smith}, \binits{E.}},
\bauthor{\bsnm{Balogh}, \binits{A.}},
\bauthor{\bsnm{Horbury}, \binits{T.}},
\bauthor{\bsnm{Goldstein}, \binits{M.}},
\bauthor{\bsnm{Roberts}, \binits{D.}}:
\batitle{Properties of magnetohydrodynamic turbulence in the solar wind as
  observed by ulysses at high heliographic latitudes}.
\bjtitle{Geophysical Research Letters}
\bvolume{22}(\bissue{23}),
\bfpage{3393}--\blpage{3396}
(\byear{1995})
\end{barticle}
\endbibitem

\bibitem{BreechEA08}
\begin{botherref}
\oauthor{\bsnm{Breech}, \binits{B.}},
\oauthor{\bsnm{Matthaeus}, \binits{W.H.}},
\oauthor{\bsnm{Minnie}, \binits{J.}},
\oauthor{\bsnm{Bieber}, \binits{J.W.}},
\oauthor{\bsnm{Oughton}, \binits{S.}},
\oauthor{\bsnm{Smith}, \binits{C.W.}},
\oauthor{\bsnm{Isenberg}, \binits{P.A.}}:
Turbulence transport throughout the heliosphere.
Journal of Geophysical Research
\textbf{113}
(2008).
\doiurl{10.1029/2007JA012711}
\end{botherref}
\endbibitem

\bibitem{RobertsJGR87}
\begin{barticle}
\bauthor{\bsnm{Roberts}, \binits{D.}},
\bauthor{\bsnm{Klein}, \binits{L.}},
\bauthor{\bsnm{Goldstein}, \binits{M.}},
\bauthor{\bsnm{Matthaeus}, \binits{W.}}:
\batitle{The nature and evolution of magnetohydrodynamic fluctuations in the
  solar wind: Voyager observations}.
\bjtitle{Journal of Geophysical Research: Space Physics}
\bvolume{92}(\bissue{A10}),
\bfpage{11021}--\blpage{11040}
(\byear{1987})
\end{barticle}
\endbibitem

\bibitem{MatthaeusGRL04}
\begin{botherref}
\oauthor{\bsnm{Matthaeus}, \binits{W.H.}},
\oauthor{\bsnm{Minnie}, \binits{J.}},
\oauthor{\bsnm{Breech}, \binits{B.}},
\oauthor{\bsnm{Parhi}, \binits{S.}},
\oauthor{\bsnm{Bieber}, \binits{J.}},
\oauthor{\bsnm{Oughton}, \binits{S.}}:
Transport of cross helicity and radial evolution of alfv{\'e}nicity in the
  solar wind.
Geophysical research letters
\textbf{31}(12)
(2004)
\end{botherref}
\endbibitem

\bibitem{BavassanoJGR00}
\begin{barticle}
\bauthor{\bsnm{Bavassano}, \binits{B.}},
\bauthor{\bsnm{Pietropaolo}, \binits{E.}},
\bauthor{\bsnm{Bruno}, \binits{R.}}:
\batitle{On the evolution of outward and inward alfv{\'e}nic fluctuations in
  the polar wind}.
\bjtitle{Journal of Geophysical Research: Space Physics}
\bvolume{105}(\bissue{A7}),
\bfpage{15959}--\blpage{15964}
(\byear{2000})
\end{barticle}
\endbibitem

\bibitem{StriblingMatt91}
\begin{barticle}
\bauthor{\bsnm{Stribling}, \binits{T.}},
\bauthor{\bsnm{Matthaeus}, \binits{W.H.}}:
\batitle{Relaxation processes in a low order three-dimensional
  magnetohydrodynamics model}.
\bjtitle{Physics of Fluids B}
\bvolume{boldVol{3}},
\bfpage{1848}
(\byear{1991})
\end{barticle}
\endbibitem

\bibitem{MullerGrappin04}
\begin{barticle}
\bauthor{\bsnm{Müller}, \binits{W.-C.}},
\bauthor{\bsnm{Grappin}, \binits{R.}}:
\batitle{The residual energy in freely decaying magnetohydrodynamic
  turbulence}.
\bjtitle{Plasma Physics and Controlled Fusion}
\bvolume{46}(\bissue{12B}),
\bfpage{91}--\blpage{96}
(\byear{2004}).
\doiurl{10.1088/0741-3335/46/12b/008}
\end{barticle}
\endbibitem

\bibitem{BoldyrevEA11}
\begin{barticle}
\bauthor{\bsnm{Boldyrev}, \binits{S.}},
\bauthor{\bsnm{Perez}, \binits{J.C.}},
\bauthor{\bsnm{Borovsky}, \binits{J.E.}},
\bauthor{\bsnm{Podesta}, \binits{J.J.}}:
\batitle{Spectral scaling laws in magnetohydrodynamic turbulence simulations
  and in the solar wind}.
\bjtitle{The Astrophysical Journal}
\bvolume{741},
\bfpage{19}
(\byear{2011}).
\doiurl{10.1088/2041-8205/741/1/L19}
\end{barticle}
\endbibitem

\bibitem{GrappinEA16}
\begin{barticle}
\bauthor{\bsnm{{Grappin}}, \binits{R.}},
\bauthor{\bsnm{{M{\"u}ller}}, \binits{W.-C.}},
\bauthor{\bsnm{{Verdini}}, \binits{A.}}:
\batitle{{Alfv{\'e}n-dynamo balance and magnetic excess in magnetohydrodynamic
  turbulence}}.
\bjtitle{Astron. Astrophys.}
\bvolume{589},
\bfpage{131}
(\byear{2016})
{\href{https://arxiv.org/abs/1603.03559}{{arXiv:1603.03559}}}
{[astro-ph.SR]}.
\doiurl{10.1051/0004-6361/201628097}
\end{barticle}
\endbibitem

\bibitem{CarbonePRL09}
\begin{barticle}
\bauthor{\bsnm{Carbone}, \binits{V.}},
\bauthor{\bsnm{Marino}, \binits{R.}},
\bauthor{\bsnm{Sorriso-Valvo}, \binits{L.}},
\bauthor{\bsnm{Noullez}, \binits{A.}},
\bauthor{\bsnm{Bruno}, \binits{R.}}:
\batitle{Scaling laws of turbulence and heating of fast solar wind: the role of
  density fluctuations}.
\bjtitle{Physical review letters}
\bvolume{103}(\bissue{6}),
\bfpage{061102}
(\byear{2009})
\end{barticle}
\endbibitem

\bibitem{MarschJGR82}
\begin{barticle}
\bauthor{\bsnm{Marsch}, \binits{E.}},
\bauthor{\bsnm{M{\"u}hlh{\"a}user}, \binits{K.-H.}},
\bauthor{\bsnm{Schwenn}, \binits{R.}},
\bauthor{\bsnm{Rosenbauer}, \binits{H.}},
\bauthor{\bsnm{Pilipp}, \binits{W.}},
\bauthor{\bsnm{Neubauer}, \binits{F.}}:
\batitle{Solar wind protons: Three-dimensional velocity distributions and
  derived plasma parameters measured between 0.3 and 1 au}.
\bjtitle{Journal of Geophysical Research: Space Physics}
\bvolume{87}(\bissue{A1}),
\bfpage{52}--\blpage{72}
(\byear{1982})
\end{barticle}
\endbibitem

\bibitem{WangJGR01}
\begin{barticle}
\bauthor{\bsnm{Wang}, \binits{C.}},
\bauthor{\bsnm{Richardson}, \binits{J.}}:
\batitle{{Energy partition between solar wind protons and pickup ions in the
  distant heliosphere: A three-fluid approach}}.
\bjtitle{Journal of Geophysical Research}
\bvolume{106},
\bfpage{29401}--\blpage{29408}
(\byear{2001})
\end{barticle}
\endbibitem

\bibitem{MatteiniGRL07}
\begin{botherref}
\oauthor{\bsnm{Matteini}, \binits{L.}},
\oauthor{\bsnm{Landi}, \binits{S.}},
\oauthor{\bsnm{Hellinger}, \binits{P.}},
\oauthor{\bsnm{Pantellini}, \binits{F.}},
\oauthor{\bsnm{Maksimovic}, \binits{M.}},
\oauthor{\bsnm{Velli}, \binits{M.}},
\oauthor{\bsnm{Goldstein}, \binits{B.E.}},
\oauthor{\bsnm{Marsch}, \binits{E.}}:
Evolution of the solar wind proton temperature anisotropy from 0.3 to 2.5 au.
Geophysical Research Letters
\textbf{34}(20)
(2007)
\end{botherref}
\endbibitem

\bibitem{HellingerJGR13}
\begin{botherref}
\oauthor{\bsnm{Hellinger}, \binits{P.}},
\oauthor{\bsnm{Tr{\'a}vn{\'\i}{\v{c}}ek}, \binits{P.M.}},
\oauthor{\bsnm{{\v{S}}tver{\'a}k}, \binits{{\v{S}}.}},
\oauthor{\bsnm{Matteini}, \binits{L.}},
\oauthor{\bsnm{Velli}, \binits{M.}}:
Proton thermal energetics in the solar wind: Helios reloaded.
Journal of Geophysical Research: Space Physics
(2013)
\end{botherref}
\endbibitem

\bibitem{VasquezJGR07}
\begin{botherref}
\oauthor{\bsnm{Vasquez}, \binits{B.J.}},
\oauthor{\bsnm{Smith}, \binits{C.W.}},
\oauthor{\bsnm{Hamilton}, \binits{K.}},
\oauthor{\bsnm{MacBride}, \binits{B.T.}},
\oauthor{\bsnm{Leamon}, \binits{R.J.}}:
Evaluation of the turbulent energy cascade rates from the upper inertial range
  in the solar wind at 1 au.
Journal of Geophysical Research: Space Physics
\textbf{112}(A7)
(2007)
\end{botherref}
\endbibitem

\bibitem{MacBrideEA08}
\begin{barticle}
\bauthor{\bsnm{Mac{Bride}}, \binits{B.T.}},
\bauthor{\bsnm{Smith}, \binits{C.W.}},
\bauthor{\bsnm{Forman}, \binits{M.A.}}:
\batitle{The turbulent cascade at 1\,{A}{U}: {Energy} transfer and the
  third-order scaling for {M}{H}{D}}.
\bjtitle{The Astrophysical Journal}
\bvolume{679},
\bfpage{1644}--\blpage{1660}
(\byear{2008}).
\doiurl{10.1086/529575}
\end{barticle}
\endbibitem

\bibitem{MacBrideEA-sw11}
\begin{bchapter}
\bauthor{\bsnm{Mac{Bride}}, \binits{B.T.}},
\bauthor{\bsnm{Forman}, \binits{M.A.}},
\bauthor{\bsnm{Smith}, \binits{C.W.}}:
\bctitle{Turbulence and third moment of fluctuations: {Kolmogorov's} $4/5$ law
  and its mhd analogues in the solar wind}.
In: \beditor{\bsnm{Fleck}, \binits{B.}},
\beditor{\bsnm{Zurbuchen}, \binits{T.H.}},
\beditor{\bsnm{Lacoste}, \binits{H.}} (eds.)
\bbtitle{Proc. Solar Wind 11 -- Soho 16 ``Connecting Sun and Heliosphere''},
vol. \bseriesno{{S}{P}-592},
pp. \bfpage{613}--\blpage{616}.
\bpublisher{ESA},
\blocation{Noordwijk, The Netherlands}
(\byear{2005})
\end{bchapter}
\endbibitem

\bibitem{MarinoApJ12}
\begin{barticle}
\bauthor{\bsnm{Marino}, \binits{R.}},
\bauthor{\bsnm{Sorriso-Valvo}, \binits{L.}},
\bauthor{\bsnm{D’Amicis}, \binits{R.}},
\bauthor{\bsnm{Carbone}, \binits{V.}},
\bauthor{\bsnm{Bruno}, \binits{R.}},
\bauthor{\bsnm{Veltri}, \binits{P.}}:
\batitle{On the occurrence of the third-order scaling in high latitude solar
  wind}.
\bjtitle{The Astrophysical Journal}
\bvolume{750}(\bissue{1}),
\bfpage{41}
(\byear{2012})
\end{barticle}
\endbibitem

\bibitem{VermaJGR95}
\begin{barticle}
\bauthor{\bsnm{Verma}, \binits{M.}},
\bauthor{\bsnm{Roberts}, \binits{D.}},
\bauthor{\bsnm{Goldstein}, \binits{M.}}:
\batitle{Turbulent heating and temperature evolution in the solar wind plasma}.
\bjtitle{Journal of Geophysical Research: Space Physics}
\bvolume{100}(\bissue{A10}),
\bfpage{19839}--\blpage{19850}
(\byear{1995})
\end{barticle}
\endbibitem

\bibitem{BourouaineApJ13}
\begin{barticle}
\bauthor{\bsnm{Bourouaine}, \binits{S.}},
\bauthor{\bsnm{Chandran}, \binits{B.D.}}:
\batitle{Observational test of stochastic heating in low-$\beta$
  fast-solar-wind streams}.
\bjtitle{The Astrophysical Journal}
\bvolume{774}(\bissue{2}),
\bfpage{96}
(\byear{2013})
\end{barticle}
\endbibitem

\bibitem{MartinovicApJ19}
\begin{barticle}
\bauthor{\bsnm{Martinovi{\'c}}, \binits{M.M.}},
\bauthor{\bsnm{Klein}, \binits{K.G.}},
\bauthor{\bsnm{Bourouaine}, \binits{S.}}:
\batitle{Radial evolution of stochastic heating in low-$\beta$ solar wind}.
\bjtitle{The Astrophysical Journal}
\bvolume{879}(\bissue{1}),
\bfpage{43}
(\byear{2019})
\end{barticle}
\endbibitem

\bibitem{ParasharPP11}
\begin{barticle}
\bauthor{\bsnm{Parashar}, \binits{T.N.}},
\bauthor{\bsnm{Servidio}, \binits{S.}},
\bauthor{\bsnm{Breech}, \binits{B.}},
\bauthor{\bsnm{Shay}, \binits{M.A.}},
\bauthor{\bsnm{Matthaeus}, \binits{W.H.}}:
\batitle{Effect of driving frequency on excitation of turbulence in a kinetic
  plasma}.
\bjtitle{Physics of Plasmas}
\bvolume{18}(\bissue{9}),
\bfpage{092302}
(\byear{2011})
\end{barticle}
\endbibitem

\bibitem{OsmanPRL12a}
\begin{barticle}
\bauthor{\bsnm{Osman}, \binits{K.}},
\bauthor{\bsnm{Matthaeus}, \binits{W.}},
\bauthor{\bsnm{Wan}, \binits{M.}},
\bauthor{\bsnm{Rappazzo}, \binits{A.}}:
\batitle{Intermittency and local heating in the solar wind}.
\bjtitle{Physical review letters}
\bvolume{108}(\bissue{26}),
\bfpage{261102}
(\byear{2012})
\end{barticle}
\endbibitem

\bibitem{BrunoJGR03}
\begin{botherref}
\oauthor{\bsnm{Bruno}, \binits{R.}},
\oauthor{\bsnm{Carbone}, \binits{V.}},
\oauthor{\bsnm{Sorriso-Valvo}, \binits{L.}},
\oauthor{\bsnm{Bavassano}, \binits{B.}}:
Radial evolution of solar wind intermittency in the inner heliosphere.
Journal of Geophysical Research: Space Physics
\textbf{108}(A3)
(2003)
\end{botherref}
\endbibitem

\bibitem{ParasharApJL19}
\begin{barticle}
\bauthor{\bsnm{Parashar}, \binits{T.N.}},
\bauthor{\bsnm{Cuesta}, \binits{M.}},
\bauthor{\bsnm{Matthaeus}, \binits{W.H.}}:
\batitle{Reynolds number and intermittency in the expanding solar wind:
  Predictions based on voyager observations}.
\bjtitle{The Astrophysical Journal Letters}
\bvolume{884}(\bissue{2}),
\bfpage{57}
(\byear{2019}).
\doiurl{10.3847/2041-8213/ab4a82}
\end{barticle}
\endbibitem

\bibitem{CuestaApJS22}
\begin{barticle}
\bauthor{\bsnm{Cuesta}, \binits{M.E.}},
\bauthor{\bsnm{Parashar}, \binits{T.N.}},
\bauthor{\bsnm{Chhiber}, \binits{R.}},
\bauthor{\bsnm{Matthaeus}, \binits{W.H.}}:
\batitle{Intermittency in the expanding solar wind: Observations from parker
  solar probe (0.16 au), helios 1 (0.3--1 au), and voyager 1 (1--10 au)}.
\bjtitle{The Astrophysical Journal Supplement Series}
\bvolume{259}(\bissue{1}),
\bfpage{23}
(\byear{2022})
\end{barticle}
\endbibitem

\bibitem{BaleSSR16}
\begin{barticle}
\bauthor{\bsnm{Bale}, \binits{S.}},
\bauthor{\bsnm{Goetz}, \binits{K.}},
\bauthor{\bsnm{Harvey}, \binits{P.}},
\bauthor{\bsnm{Turin}, \binits{P.}},
\bauthor{\bsnm{Bonnell}, \binits{J.}},
\bauthor{\bparticle{Dudok~de} \bsnm{Wit}, \binits{T.}},
\bauthor{\bsnm{Ergun}, \binits{R.}},
\bauthor{\bsnm{MacDowall}, \binits{R.}},
\bauthor{\bsnm{Pulupa}, \binits{M.}},
\bauthor{\bsnm{Andr{\'e}}, \binits{M.}}, \betal:
\batitle{The fields instrument suite for solar probe plus}.
\bjtitle{Space science reviews}
\bvolume{204}(\bissue{1}),
\bfpage{49}--\blpage{82}
(\byear{2016})
\end{barticle}
\endbibitem

\bibitem{McComasSSR16}
\begin{barticle}
\bauthor{\bsnm{McComas}, \binits{D.}},
\bauthor{\bsnm{Alexander}, \binits{N.}},
\bauthor{\bsnm{Angold}, \binits{N.}},
\bauthor{\bsnm{Bale}, \binits{S.}},
\bauthor{\bsnm{Beebe}, \binits{C.}},
\bauthor{\bsnm{Birdwell}, \binits{B.}},
\bauthor{\bsnm{Boyle}, \binits{M.}},
\bauthor{\bsnm{Burgum}, \binits{J.}},
\bauthor{\bsnm{Burnham}, \binits{J.}},
\bauthor{\bsnm{Christian}, \binits{E.}}, \betal:
\batitle{Integrated science investigation of the sun (isis): Design of the
  energetic particle investigation}.
\bjtitle{Space Science Reviews}
\bvolume{204}(\bissue{1}),
\bfpage{187}--\blpage{256}
(\byear{2016})
\end{barticle}
\endbibitem

\bibitem{KasperSSR16}
\begin{barticle}
\bauthor{\bsnm{Kasper}, \binits{J.C.}},
\bauthor{\bsnm{Abiad}, \binits{R.}},
\bauthor{\bsnm{Austin}, \binits{G.}},
\bauthor{\bsnm{Balat-Pichelin}, \binits{M.}},
\bauthor{\bsnm{Bale}, \binits{S.D.}},
\bauthor{\bsnm{Belcher}, \binits{J.W.}},
\bauthor{\bsnm{Berg}, \binits{P.}},
\bauthor{\bsnm{Bergner}, \binits{H.}},
\bauthor{\bsnm{Berthomier}, \binits{M.}},
\bauthor{\bsnm{Bookbinder}, \binits{J.}}, \betal:
\batitle{Solar wind electrons alphas and protons (sweap) investigation: Design
  of the solar wind and coronal plasma instrument suite for solar probe plus}.
\bjtitle{Space Science Reviews}
\bvolume{204}(\bissue{1}),
\bfpage{131}--\blpage{186}
(\byear{2016})
\end{barticle}
\endbibitem

\bibitem{VourlidasSSR16}
\begin{barticle}
\bauthor{\bsnm{Vourlidas}, \binits{A.}},
\bauthor{\bsnm{Howard}, \binits{R.A.}},
\bauthor{\bsnm{Plunkett}, \binits{S.P.}},
\bauthor{\bsnm{Korendyke}, \binits{C.M.}},
\bauthor{\bsnm{Thernisien}, \binits{A.F.}},
\bauthor{\bsnm{Wang}, \binits{D.}},
\bauthor{\bsnm{Rich}, \binits{N.}},
\bauthor{\bsnm{Carter}, \binits{M.T.}},
\bauthor{\bsnm{Chua}, \binits{D.H.}},
\bauthor{\bsnm{Socker}, \binits{D.G.}}, \betal:
\batitle{The wide-field imager for solar probe plus (wispr)}.
\bjtitle{Space Science Reviews}
\bvolume{204}(\bissue{1}),
\bfpage{83}--\blpage{130}
(\byear{2016})
\end{barticle}
\endbibitem

\bibitem{RaouafiEA22SSR}
\begin{botherref}
\oauthor{\bparticle{et} \bsnm{al.}, \binits{N.R.}}:
Parker solar probe: Three years of solar cycle minimum discoveries.
Space Science Rev.
(2022)
\end{botherref}
\endbibitem

\bibitem{BaleNature19}
\begin{barticle}
\bauthor{\bsnm{Bale}, \binits{S.}},
\bauthor{\bsnm{Badman}, \binits{S.}},
\bauthor{\bsnm{Bonnell}, \binits{J.}},
\bauthor{\bsnm{Bowen}, \binits{T.}},
\bauthor{\bsnm{Burgess}, \binits{D.}},
\bauthor{\bsnm{Case}, \binits{A.}},
\bauthor{\bsnm{Cattell}, \binits{C.}},
\bauthor{\bsnm{Chandran}, \binits{B.}},
\bauthor{\bsnm{Chaston}, \binits{C.}},
\bauthor{\bsnm{Chen}, \binits{C.}}, \betal:
\batitle{Highly structured slow solar wind emerging from an equatorial coronal
  hole}.
\bjtitle{Nature}
\bvolume{576}(\bissue{7786}),
\bfpage{237}--\blpage{242}
(\byear{2019})
\end{barticle}
\endbibitem

\bibitem{BaloghGRL99}
\begin{barticle}
\bauthor{\bsnm{Balogh}, \binits{A.}},
\bauthor{\bsnm{Forsyth}, \binits{R.}},
\bauthor{\bsnm{Lucek}, \binits{E.}},
\bauthor{\bsnm{Horbury}, \binits{T.}},
\bauthor{\bsnm{Smith}, \binits{E.}}:
\batitle{Heliospheric magnetic field polarity inversions at high heliographic
  latitudes}.
\bjtitle{Geophysical research letters}
\bvolume{26}(\bissue{6}),
\bfpage{631}--\blpage{634}
(\byear{1999})
\end{barticle}
\endbibitem

\bibitem{MatteiniGRL14}
\begin{barticle}
\bauthor{\bsnm{Matteini}, \binits{L.}},
\bauthor{\bsnm{Horbury}, \binits{T.S.}},
\bauthor{\bsnm{Neugebauer}, \binits{M.}},
\bauthor{\bsnm{Goldstein}, \binits{B.E.}}:
\batitle{Dependence of solar wind speed on the local magnetic field
  orientation: Role of alfv{\'e}nic fluctuations}.
\bjtitle{Geophysical Research Letters}
\bvolume{41}(\bissue{2}),
\bfpage{259}--\blpage{265}
(\byear{2014})
\end{barticle}
\endbibitem

\bibitem{BorovskyJGR16}
\begin{barticle}
\bauthor{\bsnm{Borovsky}, \binits{J.E.}}:
\batitle{The plasma structure of coronal hole solar wind: Origins and
  evolution}.
\bjtitle{Journal of Geophysical Research: Space Physics}
\bvolume{121}(\bissue{6}),
\bfpage{5055}--\blpage{5087}
(\byear{2016})
\end{barticle}
\endbibitem

\bibitem{deWitApJS20}
\begin{barticle}
\bauthor{\bparticle{de} \bsnm{Wit}, \binits{T.D.}},
\bauthor{\bsnm{Krasnoselskikh}, \binits{V.V.}},
\bauthor{\bsnm{Bale}, \binits{S.D.}},
\bauthor{\bsnm{Bonnell}, \binits{J.W.}},
\bauthor{\bsnm{Bowen}, \binits{T.A.}},
\bauthor{\bsnm{Chen}, \binits{C.H.}},
\bauthor{\bsnm{Froment}, \binits{C.}},
\bauthor{\bsnm{Goetz}, \binits{K.}},
\bauthor{\bsnm{Harvey}, \binits{P.R.}},
\bauthor{\bsnm{Jagarlamudi}, \binits{V.K.}}, \betal:
\batitle{Switchbacks in the near-sun magnetic field: long memory and impact on
  the turbulence cascade}.
\bjtitle{The Astrophysical Journal Supplement Series}
\bvolume{246}(\bissue{2}),
\bfpage{39}
(\byear{2020})
\end{barticle}
\endbibitem

\bibitem{FiskApJL20}
\begin{barticle}
\bauthor{\bsnm{Fisk}, \binits{L.}},
\bauthor{\bsnm{Kasper}, \binits{J.}}:
\batitle{Global circulation of the open magnetic flux of the sun}.
\bjtitle{The Astrophysical Journal Letters}
\bvolume{894}(\bissue{1}),
\bfpage{4}
(\byear{2020})
\end{barticle}
\endbibitem

\bibitem{SquireApJL20}
\begin{barticle}
\bauthor{\bsnm{Squire}, \binits{J.}},
\bauthor{\bsnm{Chandran}, \binits{B.D.}},
\bauthor{\bsnm{Meyrand}, \binits{R.}}:
\batitle{In-situ switchback formation in the expanding solar wind}.
\bjtitle{The Astrophysical Journal Letters}
\bvolume{891}(\bissue{1}),
\bfpage{2}
(\byear{2020})
\end{barticle}
\endbibitem

\bibitem{ChhiberApJL18}
\begin{barticle}
\bauthor{\bsnm{Chhiber}, \binits{R.}},
\bauthor{\bsnm{Usmanov}, \binits{A.V.}},
\bauthor{\bsnm{DeForest}, \binits{C.E.}},
\bauthor{\bsnm{Matthaeus}, \binits{W.H.}},
\bauthor{\bsnm{Parashar}, \binits{T.N.}},
\bauthor{\bsnm{Goldstein}, \binits{M.L.}}:
\batitle{Weakened magnetization and onset of large-scale turbulence in the
  young solar wind—comparisons of remote sensing observations with
  simulation}.
\bjtitle{The Astrophysical Journal Letters}
\bvolume{856}(\bissue{2}),
\bfpage{39}
(\byear{2018})
\end{barticle}
\endbibitem

\bibitem{BandyopadhyayApJL22}
\begin{barticle}
\bauthor{\bsnm{Bandyopadhyay}, \binits{R.}},
\bauthor{\bsnm{Matthaeus}, \binits{W.}},
\bauthor{\bsnm{McComas}, \binits{D.}},
\bauthor{\bsnm{Chhiber}, \binits{R.}},
\bauthor{\bsnm{Usmanov}, \binits{A.}},
\bauthor{\bsnm{Huang}, \binits{J.}},
\bauthor{\bsnm{Livi}, \binits{R.}},
\bauthor{\bsnm{Larson}, \binits{D.}},
\bauthor{\bsnm{Kasper}, \binits{J.}},
\bauthor{\bsnm{Case}, \binits{A.}}, \betal:
\batitle{Sub-alfv{\'e}nic solar wind observed by the parker solar probe:
  Characterization of turbulence, anisotropy, intermittency, and switchback}.
\bjtitle{The Astrophysical Journal Letters}
\bvolume{926}(\bissue{1}),
\bfpage{1}
(\byear{2022})
\end{barticle}
\endbibitem

\bibitem{KasperPRL21}
\begin{barticle}
\bauthor{\bsnm{Kasper}, \binits{J.}},
\bauthor{\bsnm{Klein}, \binits{K.}},
\bauthor{\bsnm{Lichko}, \binits{E.}},
\bauthor{\bsnm{Huang}, \binits{J.}},
\bauthor{\bsnm{Chen}, \binits{C.}},
\bauthor{\bsnm{Badman}, \binits{S.}},
\bauthor{\bsnm{Bonnell}, \binits{J.}},
\bauthor{\bsnm{Whittlesey}, \binits{P.}},
\bauthor{\bsnm{Livi}, \binits{R.}},
\bauthor{\bsnm{Larson}, \binits{D.}}, \betal:
\batitle{Parker solar probe enters the magnetically dominated solar corona}.
\bjtitle{Physical review letters}
\bvolume{127}(\bissue{25}),
\bfpage{255101}
(\byear{2021})
\end{barticle}
\endbibitem

\bibitem{BourouaineApJL20}
\begin{barticle}
\bauthor{\bsnm{Bourouaine}, \binits{S.}},
\bauthor{\bsnm{Perez}, \binits{J.C.}},
\bauthor{\bsnm{Klein}, \binits{K.G.}},
\bauthor{\bsnm{Chen}, \binits{C.H.}},
\bauthor{\bsnm{Martinovi{\'c}}, \binits{M.}},
\bauthor{\bsnm{Bale}, \binits{S.D.}},
\bauthor{\bsnm{Kasper}, \binits{J.C.}},
\bauthor{\bsnm{Raouafi}, \binits{N.E.}}:
\batitle{Turbulence characteristics of switchback and nonswitchback intervals
  observed by parker solar probe}.
\bjtitle{The Astrophysical Journal Letters}
\bvolume{904}(\bissue{2}),
\bfpage{30}
(\byear{2020})
\end{barticle}
\endbibitem

\bibitem{MartinovicApJS20}
\begin{barticle}
\bauthor{\bsnm{Martinovi{\'c}}, \binits{M.M.}},
\bauthor{\bsnm{Klein}, \binits{K.G.}},
\bauthor{\bsnm{Kasper}, \binits{J.C.}},
\bauthor{\bsnm{Case}, \binits{A.W.}},
\bauthor{\bsnm{Korreck}, \binits{K.E.}},
\bauthor{\bsnm{Larson}, \binits{D.}},
\bauthor{\bsnm{Livi}, \binits{R.}},
\bauthor{\bsnm{Stevens}, \binits{M.}},
\bauthor{\bsnm{Whittlesey}, \binits{P.}},
\bauthor{\bsnm{Chandran}, \binits{B.D.}}, \betal:
\batitle{The enhancement of proton stochastic heating in the near-sun solar
  wind}.
\bjtitle{The Astrophysical Journal Supplement Series}
\bvolume{246}(\bissue{2}),
\bfpage{30}
(\byear{2020})
\end{barticle}
\endbibitem

\bibitem{TeneraniApJS20}
\begin{barticle}
\bauthor{\bsnm{Tenerani}, \binits{A.}},
\bauthor{\bsnm{Velli}, \binits{M.}},
\bauthor{\bsnm{Matteini}, \binits{L.}},
\bauthor{\bsnm{R{\'e}ville}, \binits{V.}},
\bauthor{\bsnm{Shi}, \binits{C.}},
\bauthor{\bsnm{Bale}, \binits{S.D.}},
\bauthor{\bsnm{Kasper}, \binits{J.C.}},
\bauthor{\bsnm{Bonnell}, \binits{J.W.}},
\bauthor{\bsnm{Case}, \binits{A.W.}},
\bauthor{\bsnm{De~Wit}, \binits{T.D.}}, \betal:
\batitle{Magnetic field kinks and folds in the solar wind}.
\bjtitle{The Astrophysical Journal Supplement Series}
\bvolume{246}(\bissue{2}),
\bfpage{32}
(\byear{2020})
\end{barticle}
\endbibitem

\bibitem{McManusApJS20}
\begin{barticle}
\bauthor{\bsnm{McManus}, \binits{M.D.}},
\bauthor{\bsnm{Bowen}, \binits{T.A.}},
\bauthor{\bsnm{Mallet}, \binits{A.}},
\bauthor{\bsnm{Chen}, \binits{C.H.}},
\bauthor{\bsnm{Chandran}, \binits{B.D.}},
\bauthor{\bsnm{Bale}, \binits{S.D.}},
\bauthor{\bsnm{Larson}, \binits{D.E.}},
\bauthor{\bparticle{de} \bsnm{Wit}, \binits{T.D.}},
\bauthor{\bsnm{Kasper}, \binits{J.C.}},
\bauthor{\bsnm{Stevens}, \binits{M.}}, \betal:
\batitle{Cross helicity reversals in magnetic switchbacks}.
\bjtitle{The Astrophysical Journal Supplement Series}
\bvolume{246}(\bissue{2}),
\bfpage{67}
(\byear{2020})
\end{barticle}
\endbibitem

\bibitem{ChenApJS20}
\begin{barticle}
\bauthor{\bsnm{Chen}, \binits{C.}},
\bauthor{\bsnm{Bale}, \binits{S.}},
\bauthor{\bsnm{Bonnell}, \binits{J.}},
\bauthor{\bsnm{Borovikov}, \binits{D.}},
\bauthor{\bsnm{Bowen}, \binits{T.}},
\bauthor{\bsnm{Burgess}, \binits{D.}},
\bauthor{\bsnm{Case}, \binits{A.}},
\bauthor{\bsnm{Chandran}, \binits{B.}},
\bauthor{\bsnm{De~Wit}, \binits{T.D.}},
\bauthor{\bsnm{Goetz}, \binits{K.}}, \betal:
\batitle{The evolution and role of solar wind turbulence in the inner
  heliosphere}.
\bjtitle{The Astrophysical Journal Supplement Series}
\bvolume{246}(\bissue{2}),
\bfpage{53}
(\byear{2020})
\end{barticle}
\endbibitem

\bibitem{AlbertiApJ20}
\begin{barticle}
\bauthor{\bsnm{Alberti}, \binits{T.}},
\bauthor{\bsnm{Laurenza}, \binits{M.}},
\bauthor{\bsnm{Consolini}, \binits{G.}},
\bauthor{\bsnm{Milillo}, \binits{A.}},
\bauthor{\bsnm{Marcucci}, \binits{M.F.}},
\bauthor{\bsnm{Carbone}, \binits{V.}},
\bauthor{\bsnm{Bale}, \binits{S.D.}}:
\batitle{On the scaling properties of magnetic-field fluctuations through the
  inner heliosphere}.
\bjtitle{The Astrophysical Journal}
\bvolume{902}(\bissue{1}),
\bfpage{84}
(\byear{2020})
\end{barticle}
\endbibitem

\bibitem{ParasharApJS20}
\begin{barticle}
\bauthor{\bsnm{Parashar}, \binits{T.N.}},
\bauthor{\bsnm{Goldstein}, \binits{M.L.}},
\bauthor{\bsnm{Maruca}, \binits{B.A.}},
\bauthor{\bsnm{Matthaeus}, \binits{W.H.}},
\bauthor{\bsnm{Ruffolo}, \binits{D.}},
\bauthor{\bsnm{Bandyopadhyay}, \binits{R.}},
\bauthor{\bsnm{Chhiber}, \binits{R.}},
\bauthor{\bsnm{Chasapis}, \binits{A.}},
\bauthor{\bsnm{Qudsi}, \binits{R.}},
\bauthor{\bsnm{Vech}, \binits{D.}},
\bauthor{\bsnm{Roberts}, \binits{D.A.}},
\bauthor{\bsnm{Bale}, \binits{S.D.}},
\bauthor{\bsnm{Bonnell}, \binits{J.W.}},
\bauthor{\bparticle{de} \bsnm{Wit}, \binits{T.D.}},
\bauthor{\bsnm{Goetz}, \binits{K.}},
\bauthor{\bsnm{Harvey}, \binits{P.R.}},
\bauthor{\bsnm{MacDowall}, \binits{R.J.}},
\bauthor{\bsnm{Malaspina}, \binits{D.}},
\bauthor{\bsnm{Pulupa}, \binits{M.}},
\bauthor{\bsnm{Kasper}, \binits{J.C.}},
\bauthor{\bsnm{Korreck}, \binits{K.E.}},
\bauthor{\bsnm{Case}, \binits{A.W.}},
\bauthor{\bsnm{Stevens}, \binits{M.}},
\bauthor{\bsnm{Whittlesey}, \binits{P.}},
\bauthor{\bsnm{Larson}, \binits{D.}},
\bauthor{\bsnm{Livi}, \binits{R.}},
\bauthor{\bsnm{Velli}, \binits{M.}},
\bauthor{\bsnm{Raouafi}, \binits{N.}}:
\batitle{Measures of scale-dependent alfv{\'{e}}nicity in the first {PSP} solar
  encounter}.
\bjtitle{The Astrophysical Journal Supplement Series}
\bvolume{246}(\bissue{2}),
\bfpage{58}
(\byear{2020}).
\doiurl{10.3847/1538-4365/ab64e6}
\end{barticle}
\endbibitem

\bibitem{BrunoApJL14}
\begin{barticle}
\bauthor{\bsnm{Bruno}, \binits{R.}},
\bauthor{\bsnm{Trenchi}, \binits{L.}}:
\batitle{Radial dependence of the frequency break between fluid and kinetic
  scales in the solar wind fluctuations}.
\bjtitle{The Astrophysical Journal Letters}
\bvolume{787}(\bissue{2}),
\bfpage{24}
(\byear{2014})
\end{barticle}
\endbibitem

\bibitem{WuApJL20}
\begin{barticle}
\bauthor{\bsnm{Wu}, \binits{H.}},
\bauthor{\bsnm{Tu}, \binits{C.}},
\bauthor{\bsnm{Wang}, \binits{X.}},
\bauthor{\bsnm{He}, \binits{J.}},
\bauthor{\bsnm{Yang}, \binits{L.}}:
\batitle{Energy supply for heating the slow solar wind observed by parker solar
  probe between 0.17 and 0.7 au}.
\bjtitle{The Astrophysical Journal Letters}
\bvolume{904}(\bissue{1}),
\bfpage{8}
(\byear{2020})
\end{barticle}
\endbibitem

\bibitem{HorburyAA96}
\begin{barticle}
\bauthor{\bsnm{Horbury}, \binits{T.}},
\bauthor{\bsnm{Balogh}, \binits{A.}},
\bauthor{\bsnm{Forsyth}, \binits{R.}},
\bauthor{\bsnm{Smith}, \binits{E.}}:
\batitle{The rate of turbulent evolution over the sun's poles.}
\bjtitle{Astronomy and Astrophysics}
\bvolume{316},
\bfpage{333}--\blpage{341}
(\byear{1996})
\end{barticle}
\endbibitem

\bibitem{KleinSW92}
\begin{bchapter}
\bauthor{\bsnm{Klein}, \binits{L.}},
\bauthor{\bsnm{Matthaeus}, \binits{W.}},
\bauthor{\bsnm{Roberts}, \binits{D.}},
\bauthor{\bsnm{Goldstein}, \binits{M.}}:
\bctitle{Evolution of spatial and temporal correlations in the solar wind:
  Observations and interpretation}.
In: \bbtitle{Solar Wind Seven},
pp. \bfpage{197}--\blpage{200}.
\bpublisher{Elsevier}, \blocation{???}
(\byear{1992})
\end{bchapter}
\endbibitem

\bibitem{MatthaeusJPP96}
\begin{barticle}
\bauthor{\bsnm{Matthaeus}, \binits{W.H.}},
\bauthor{\bsnm{Zank}, \binits{G.P.}},
\bauthor{\bsnm{Oughton}, \binits{S.}}:
\batitle{Phenomenology of hydromagnetic turbulence in a uniformly expanding
  medium}.
\bjtitle{Journal of plasma physics}
\bvolume{56}(\bissue{3}),
\bfpage{659}--\blpage{675}
(\byear{1996})
\end{barticle}
\endbibitem

\bibitem{DuanApJS20}
\begin{barticle}
\bauthor{\bsnm{Duan}, \binits{D.}},
\bauthor{\bsnm{Bowen}, \binits{T.A.}},
\bauthor{\bsnm{Chen}, \binits{C.H.}},
\bauthor{\bsnm{Mallet}, \binits{A.}},
\bauthor{\bsnm{He}, \binits{J.}},
\bauthor{\bsnm{Bale}, \binits{S.D.}},
\bauthor{\bsnm{Vech}, \binits{D.}},
\bauthor{\bsnm{Kasper}, \binits{J.}},
\bauthor{\bsnm{Pulupa}, \binits{M.}},
\bauthor{\bsnm{Bonnell}, \binits{J.W.}}, \betal:
\batitle{The radial dependence of proton-scale magnetic spectral break in slow
  solar wind during psp encounter 2}.
\bjtitle{The Astrophysical Journal Supplement Series}
\bvolume{246}(\bissue{2}),
\bfpage{55}
(\byear{2020})
\end{barticle}
\endbibitem

\bibitem{ShiAA21}
\begin{barticle}
\bauthor{\bsnm{Shi}, \binits{C.}},
\bauthor{\bsnm{Velli}, \binits{M.}},
\bauthor{\bsnm{Panasenco}, \binits{O.}},
\bauthor{\bsnm{Tenerani}, \binits{A.}},
\bauthor{\bsnm{R{\'e}ville}, \binits{V.}},
\bauthor{\bsnm{Bale}, \binits{S.D.}},
\bauthor{\bsnm{Kasper}, \binits{J.}},
\bauthor{\bsnm{Korreck}, \binits{K.}},
\bauthor{\bsnm{Bonnell}, \binits{J.}},
\bauthor{\bparticle{de} \bsnm{Wit}, \binits{T.D.}}, \betal:
\batitle{Alfv{\'e}nic versus non-alfv{\'e}nic turbulence in the inner
  heliosphere as observed by parker solar probe}.
\bjtitle{Astronomy \& Astrophysics}
\bvolume{650},
\bfpage{21}
(\byear{2021})
\end{barticle}
\endbibitem

\bibitem{RobertsJGR92}
\begin{barticle}
\bauthor{\bsnm{Roberts}, \binits{D.A.}},
\bauthor{\bsnm{Goldstein}, \binits{M.L.}},
\bauthor{\bsnm{Matthaeus}, \binits{W.H.}},
\bauthor{\bsnm{Ghosh}, \binits{S.}}:
\batitle{Velocity shear generation of solar wind turbulence}.
\bjtitle{Journal of Geophysical Research: Space Physics}
\bvolume{97}(\bissue{A11}),
\bfpage{17115}--\blpage{17130}
(\byear{1992})
\end{barticle}
\endbibitem

\bibitem{HuangApJL21}
\begin{barticle}
\bauthor{\bsnm{Huang}, \binits{S.}},
\bauthor{\bsnm{Sahraoui}, \binits{F.}},
\bauthor{\bsnm{Andr{\'e}s}, \binits{N.}},
\bauthor{\bsnm{Hadid}, \binits{L.}},
\bauthor{\bsnm{Yuan}, \binits{Z.}},
\bauthor{\bsnm{He}, \binits{J.}},
\bauthor{\bsnm{Zhao}, \binits{J.}},
\bauthor{\bsnm{Galtier}, \binits{S.}},
\bauthor{\bsnm{Zhang}, \binits{J.}},
\bauthor{\bsnm{Deng}, \binits{X.}}, \betal:
\batitle{The ion transition range of solar wind turbulence in the inner
  heliosphere: Parker solar probe observations}.
\bjtitle{The Astrophysical journal letters}
\bvolume{909}(\bissue{1}),
\bfpage{7}
(\byear{2021})
\end{barticle}
\endbibitem

\bibitem{BowenJGR20}
\begin{barticle}
\bauthor{\bsnm{Bowen}, \binits{T.A.}},
\bauthor{\bsnm{Bale}, \binits{S.D.}},
\bauthor{\bsnm{Bonnell}, \binits{J.W.}},
\bauthor{\bparticle{Dudok~de} \bsnm{Wit}, \binits{T.}},
\bauthor{\bsnm{Goetz}, \binits{K.}},
\bauthor{\bsnm{Goodrich}, \binits{K.}},
\bauthor{\bsnm{Gruesbeck}, \binits{J.}},
\bauthor{\bsnm{Harvey}, \binits{P.R.}},
\bauthor{\bsnm{Jannet}, \binits{G.}},
\bauthor{\bsnm{Koval}, \binits{A.}}, \betal:
\batitle{A merged search-coil and fluxgate magnetometer data product for parker
  solar probe fields}.
\bjtitle{Journal of Geophysical Research: Space Physics}
\bvolume{125}(\bissue{5}),
\bfpage{2020}--\blpage{027813}
(\byear{2020})
\end{barticle}
\endbibitem

\bibitem{DenskatJGR84}
\begin{barticle}
\bauthor{\bsnm{Denskat}, \binits{K.}},
\bauthor{\bsnm{Beinroth}, \binits{H.}},
\bauthor{\bsnm{Neubauer}, \binits{F.}}, \betal:
\batitle{Interplanetary magnetic field power spectra with frequencies from 2.4
  x 10\^{}-5 hz to 470 hz from helios-observations during solar minimum
  conditions}.
\bjtitle{Journal of geophysics}
\bvolume{54}(\bissue{1}),
\bfpage{60}--\blpage{67}
(\byear{1984})
\end{barticle}
\endbibitem

\bibitem{LeamonJGR99}
\begin{barticle}
\bauthor{\bsnm{Leamon}, \binits{R.J.}},
\bauthor{\bsnm{Smith}, \binits{C.W.}},
\bauthor{\bsnm{Ness}, \binits{N.F.}},
\bauthor{\bsnm{Wong}, \binits{H.K.}}:
\batitle{{Dissipation range dynamics: Kinetic Alfv{\'e}n waves and the
  importance of $\beta$ e}}.
\bjtitle{Journal of Geophysical Research}
\bvolume{104}(\bissue{A10}),
\bfpage{22331}
(\byear{1999})
\end{barticle}
\endbibitem

\bibitem{MeyrandJPP21}
\begin{botherref}
\oauthor{\bsnm{Meyrand}, \binits{R.}},
\oauthor{\bsnm{Squire}, \binits{J.}},
\oauthor{\bsnm{Schekochihin}, \binits{A.A.}},
\oauthor{\bsnm{Dorland}, \binits{W.}}:
On the violation of the zeroth law of turbulence in space plasmas.
Journal of Plasma Physics
\textbf{87}(3)
(2021)
\end{botherref}
\endbibitem

\bibitem{ZhuApJL20}
\begin{barticle}
\bauthor{\bsnm{Zhu}, \binits{X.}},
\bauthor{\bsnm{He}, \binits{J.}},
\bauthor{\bsnm{Verscharen}, \binits{D.}},
\bauthor{\bsnm{Duan}, \binits{D.}},
\bauthor{\bsnm{Bale}, \binits{S.D.}}:
\batitle{Wave composition, propagation, and polarization of magnetohydrodynamic
  turbulence within 0.3 au as observed by parker solar probe}.
\bjtitle{The Astrophysical Journal Letters}
\bvolume{901}(\bissue{1}),
\bfpage{3}
(\byear{2020})
\end{barticle}
\endbibitem

\bibitem{DuanApJL21}
\begin{barticle}
\bauthor{\bsnm{Duan}, \binits{D.}},
\bauthor{\bsnm{He}, \binits{J.}},
\bauthor{\bsnm{Bowen}, \binits{T.A.}},
\bauthor{\bsnm{Woodham}, \binits{L.D.}},
\bauthor{\bsnm{Wang}, \binits{T.}},
\bauthor{\bsnm{Chen}, \binits{C.H.}},
\bauthor{\bsnm{Mallet}, \binits{A.}},
\bauthor{\bsnm{Bale}, \binits{S.D.}}:
\batitle{Anisotropy of solar wind turbulence in the inner heliosphere at
  kinetic scales: Psp observations}.
\bjtitle{The Astrophysical Journal Letters}
\bvolume{915}(\bissue{1}),
\bfpage{8}
(\byear{2021})
\end{barticle}
\endbibitem

\bibitem{ZhangApJL22}
\begin{barticle}
\bauthor{\bsnm{Zhang}, \binits{J.}},
\bauthor{\bsnm{Huang}, \binits{S.}},
\bauthor{\bsnm{He}, \binits{J.}},
\bauthor{\bsnm{Wang}, \binits{T.}},
\bauthor{\bsnm{Yuan}, \binits{Z.}},
\bauthor{\bsnm{Deng}, \binits{X.}},
\bauthor{\bsnm{Jiang}, \binits{K.}},
\bauthor{\bsnm{Wei}, \binits{Y.}},
\bauthor{\bsnm{Xu}, \binits{S.}},
\bauthor{\bsnm{Xiong}, \binits{Q.}}, \betal:
\batitle{Three-dimensional anisotropy and scaling properties of solar wind
  turbulence at kinetic scales in the inner heliosphere: Parker solar probe
  observations}.
\bjtitle{The Astrophysical Journal Letters}
\bvolume{924}(\bissue{2}),
\bfpage{21}
(\byear{2022})
\end{barticle}
\endbibitem

\bibitem{BandyopadhyayApJS20b}
\begin{barticle}
\bauthor{\bsnm{Bandyopadhyay}, \binits{R.}},
\bauthor{\bsnm{Goldstein}, \binits{M.L.}},
\bauthor{\bsnm{Maruca}, \binits{B.A.}},
\bauthor{\bsnm{Matthaeus}, \binits{W.H.}},
\bauthor{\bsnm{Parashar}, \binits{T.N.}},
\bauthor{\bsnm{Ruffolo}, \binits{D.}},
\bauthor{\bsnm{Chhiber}, \binits{R.}},
\bauthor{\bsnm{Usmanov}, \binits{A.}},
\bauthor{\bsnm{Chasapis}, \binits{A.}},
\bauthor{\bsnm{Qudsi}, \binits{R.}},
\bauthor{\bsnm{Bale}, \binits{S.D.}},
\bauthor{\bsnm{Bonnell}, \binits{J.W.}},
\bauthor{\bparticle{de} \bsnm{Wit}, \binits{T.D.}},
\bauthor{\bsnm{Goetz}, \binits{K.}},
\bauthor{\bsnm{Harvey}, \binits{P.R.}},
\bauthor{\bsnm{MacDowall}, \binits{R.J.}},
\bauthor{\bsnm{Malaspina}, \binits{D.M.}},
\bauthor{\bsnm{Pulupa}, \binits{M.}},
\bauthor{\bsnm{Kasper}, \binits{J.C.}},
\bauthor{\bsnm{Korreck}, \binits{K.E.}},
\bauthor{\bsnm{Case}, \binits{A.W.}},
\bauthor{\bsnm{Stevens}, \binits{M.}},
\bauthor{\bsnm{Whittlesey}, \binits{P.}},
\bauthor{\bsnm{Larson}, \binits{D.}},
\bauthor{\bsnm{Livi}, \binits{R.}},
\bauthor{\bsnm{Klein}, \binits{K.G.}},
\bauthor{\bsnm{Velli}, \binits{M.}},
\bauthor{\bsnm{Raouafi}, \binits{N.}}:
\batitle{Enhanced energy transfer rate in solar wind turbulence observed near
  the sun from parker solar probe}.
\bjtitle{The Astrophysical Journal Supplement Series}
\bvolume{246}(\bissue{2}),
\bfpage{48}
(\byear{2020}).
\doiurl{10.3847/1538-4365/ab5dae}
\end{barticle}
\endbibitem

\bibitem{ChhiberApJS19}
\begin{barticle}
\bauthor{\bsnm{Chhiber}, \binits{R.}},
\bauthor{\bsnm{Usmanov}, \binits{A.V.}},
\bauthor{\bsnm{Matthaeus}, \binits{W.H.}},
\bauthor{\bsnm{Parashar}, \binits{T.N.}},
\bauthor{\bsnm{Goldstein}, \binits{M.L.}}:
\batitle{Contextual predictions for parker solar probe. {II}. turbulence
  properties and taylor hypothesis}.
\bjtitle{The Astrophysical Journal Supplement Series}
\bvolume{242}(\bissue{1}),
\bfpage{12}
(\byear{2019}).
\doiurl{10.3847/1538-4365/ab16d7}
\end{barticle}
\endbibitem

\bibitem{CoburnPTRSA15}
\begin{barticle}
\bauthor{\bsnm{Coburn}, \binits{J.T.}},
\bauthor{\bsnm{Forman}, \binits{M.A.}},
\bauthor{\bsnm{Smith}, \binits{C.W.}},
\bauthor{\bsnm{Vasquez}, \binits{B.J.}},
\bauthor{\bsnm{Stawarz}, \binits{J.E.}}:
\batitle{Third-moment descriptions of the interplanetary turbulent cascade,
  intermittency and back transfer}.
\bjtitle{Philosophical Transactions of the Royal Society A: Mathematical,
  Physical and Engineering Sciences}
\bvolume{373}(\bissue{2041}),
\bfpage{20140150}
(\byear{2015})
\end{barticle}
\endbibitem

\bibitem{RajaApJ21}
\begin{barticle}
\bauthor{\bsnm{Raja}, \binits{K.S.}},
\bauthor{\bsnm{Subramanian}, \binits{P.}},
\bauthor{\bsnm{Ingale}, \binits{M.}},
\bauthor{\bsnm{Ramesh}, \binits{R.}},
\bauthor{\bsnm{Maksimovic}, \binits{M.}}:
\batitle{Turbulent proton heating rate in the solar wind from 5--45 r$\odot$}.
\bjtitle{The Astrophysical Journal}
\bvolume{914}(\bissue{2}),
\bfpage{137}
(\byear{2021})
\end{barticle}
\endbibitem

\bibitem{WuApJ22a}
\begin{barticle}
\bauthor{\bsnm{Wu}, \binits{H.}},
\bauthor{\bsnm{Tu}, \binits{C.}},
\bauthor{\bsnm{He}, \binits{J.}},
\bauthor{\bsnm{Wang}, \binits{X.}},
\bauthor{\bsnm{Yang}, \binits{L.}}:
\batitle{Consistency of von karman decay rate with the energy supply rate and
  heating rate observed by parker solar probe}.
\bjtitle{The Astrophysical Journal}
\bvolume{926}(\bissue{2}),
\bfpage{116}
(\byear{2022})
\end{barticle}
\endbibitem

\bibitem{ShayPP18}
\begin{barticle}
\bauthor{\bsnm{Shay}, \binits{M.}},
\bauthor{\bsnm{Haggerty}, \binits{C.}},
\bauthor{\bsnm{Matthaeus}, \binits{W.}},
\bauthor{\bsnm{Parashar}, \binits{T.}},
\bauthor{\bsnm{Wan}, \binits{M.}},
\bauthor{\bsnm{Wu}, \binits{P.}}:
\batitle{Turbulent heating due to magnetic reconnection}.
\bjtitle{Physics of Plasmas}
\bvolume{25}(\bissue{1}),
\bfpage{012304}
(\byear{2018})
\end{barticle}
\endbibitem

\bibitem{StawarzApJ09}
\begin{barticle}
\bauthor{\bsnm{Stawarz}, \binits{J.E.}},
\bauthor{\bsnm{Smith}, \binits{C.W.}},
\bauthor{\bsnm{Vasquez}, \binits{B.J.}},
\bauthor{\bsnm{Forman}, \binits{M.A.}},
\bauthor{\bsnm{MacBride}, \binits{B.T.}}:
\batitle{The turbulent cascade and proton heating in the solar wind at 1 au}.
\bjtitle{The Astrophysical Journal}
\bvolume{697}(\bissue{2}),
\bfpage{1119}
(\byear{2009})
\end{barticle}
\endbibitem

\bibitem{ZhaoApJL22}
\begin{barticle}
\bauthor{\bsnm{Zhao}, \binits{L.-L.}},
\bauthor{\bsnm{Zank}, \binits{G.}},
\bauthor{\bsnm{Telloni}, \binits{D.}},
\bauthor{\bsnm{Stevens}, \binits{M.}},
\bauthor{\bsnm{Kasper}, \binits{J.}},
\bauthor{\bsnm{Bale}, \binits{S.}}:
\batitle{The turbulent properties of the sub-alfv{\'e}nic solar wind measured
  by the parker solar probe}.
\bjtitle{The Astrophysical Journal Letters}
\bvolume{928}(\bissue{2}),
\bfpage{15}
(\byear{2022})
\end{barticle}
\endbibitem

\bibitem{AndresArXiv21}
\begin{botherref}
\oauthor{\bsnm{Andr{\'e}s}, \binits{N.}},
\oauthor{\bsnm{Sahraoui}, \binits{F.}},
\oauthor{\bsnm{Hadid}, \binits{L.}},
\oauthor{\bsnm{Galtier}, \binits{S.}}, et al.:
About the incompressible energy cascade rate in anisotropic solar wind
  turbulence.
arXiv preprint arXiv:2112.13748
(2021)
\end{botherref}
\endbibitem

\bibitem{AndresApJ21}
\begin{barticle}
\bauthor{\bsnm{Andr{\'e}s}, \binits{N.}},
\bauthor{\bsnm{Sahraoui}, \binits{F.}},
\bauthor{\bsnm{Hadid}, \binits{L.}},
\bauthor{\bsnm{Huang}, \binits{S.}},
\bauthor{\bsnm{Romanelli}, \binits{N.}},
\bauthor{\bsnm{Galtier}, \binits{S.}},
\bauthor{\bsnm{Dibraccio}, \binits{G.}},
\bauthor{\bsnm{Halekas}, \binits{J.}}:
\batitle{The evolution of compressible solar wind turbulence in the inner
  heliosphere: Psp, themis, and maven observations}.
\bjtitle{The Astrophysical Journal}
\bvolume{919}(\bissue{1}),
\bfpage{19}
(\byear{2021})
\end{barticle}
\endbibitem

\bibitem{MatthaeusJGR90}
\begin{barticle}
\bauthor{\bsnm{Matthaeus}, \binits{W.H.}},
\bauthor{\bsnm{Goldstein}, \binits{M.L.}},
\bauthor{\bsnm{Roberts}, \binits{D.A.}}:
\batitle{Evidence for the presence of quasi-two-dimensional nearly
  incompressible fluctuations in the solar wind}.
\bjtitle{Journal of Geophysical Research: Space Physics}
\bvolume{95}(\bissue{A12}),
\bfpage{20673}--\blpage{20683}
(\byear{1990})
\end{barticle}
\endbibitem

\bibitem{AdhikariApJ17}
\begin{barticle}
\bauthor{\bsnm{Adhikari}, \binits{L.}},
\bauthor{\bsnm{Zank}, \binits{G.}},
\bauthor{\bsnm{Hunana}, \binits{P.}},
\bauthor{\bsnm{Shiota}, \binits{D.}},
\bauthor{\bsnm{Bruno}, \binits{R.}},
\bauthor{\bsnm{Hu}, \binits{Q.}},
\bauthor{\bsnm{Telloni}, \binits{D.}}:
\batitle{Ii. transport of nearly incompressible magnetohydrodynamic turbulence
  from 1 to 75 au}.
\bjtitle{The Astrophysical Journal}
\bvolume{841}(\bissue{2}),
\bfpage{85}
(\byear{2017})
\end{barticle}
\endbibitem

\bibitem{HernandezApJL21}
\begin{barticle}
\bauthor{\bsnm{Hern{\'a}ndez}, \binits{C.S.}},
\bauthor{\bsnm{Sorriso-Valvo}, \binits{L.}},
\bauthor{\bsnm{Bandyopadhyay}, \binits{R.}},
\bauthor{\bsnm{Chasapis}, \binits{A.}},
\bauthor{\bsnm{V{\'a}sconez}, \binits{C.L.}},
\bauthor{\bsnm{Marino}, \binits{R.}},
\bauthor{\bsnm{Pezzi}, \binits{O.}}:
\batitle{Impact of switchbacks on turbulent cascade and energy transfer rate in
  the inner heliosphere}.
\bjtitle{The Astrophysical Journal Letters}
\bvolume{922}(\bissue{1}),
\bfpage{11}
(\byear{2021})
\end{barticle}
\endbibitem

\bibitem{DavidApJ22}
\begin{barticle}
\bauthor{\bsnm{David}, \binits{V.}},
\bauthor{\bsnm{Galtier}, \binits{S.}},
\bauthor{\bsnm{Sahraoui}, \binits{F.}},
\bauthor{\bsnm{Hadid}, \binits{L.}}:
\batitle{Energy transfer, discontinuities, and heating in the inner heliosphere
  measured with a weak and local formulation of the politano--pouquet law}.
\bjtitle{The Astrophysical Journal}
\bvolume{927}(\bissue{2}),
\bfpage{200}
(\byear{2022})
\end{barticle}
\endbibitem

\bibitem{BowenApJS20}
\begin{barticle}
\bauthor{\bsnm{Bowen}, \binits{T.A.}},
\bauthor{\bsnm{Mallet}, \binits{A.}},
\bauthor{\bsnm{Huang}, \binits{J.}},
\bauthor{\bsnm{Klein}, \binits{K.G.}},
\bauthor{\bsnm{Malaspina}, \binits{D.M.}},
\bauthor{\bsnm{Stevens}, \binits{M.}},
\bauthor{\bsnm{Bale}, \binits{S.D.}},
\bauthor{\bsnm{Bonnell}, \binits{J.W.}},
\bauthor{\bsnm{Case}, \binits{A.W.}},
\bauthor{\bsnm{Chandran}, \binits{B.D.}}, \betal:
\batitle{Ion-scale electromagnetic waves in the inner heliosphere}.
\bjtitle{The Astrophysical Journal Supplement Series}
\bvolume{246}(\bissue{2}),
\bfpage{66}
(\byear{2020})
\end{barticle}
\endbibitem

\bibitem{BowenApJ20}
\begin{barticle}
\bauthor{\bsnm{Bowen}, \binits{T.A.}},
\bauthor{\bsnm{Bale}, \binits{S.D.}},
\bauthor{\bsnm{Bonnell}, \binits{J.}},
\bauthor{\bsnm{Larson}, \binits{D.}},
\bauthor{\bsnm{Mallet}, \binits{A.}},
\bauthor{\bsnm{McManus}, \binits{M.D.}},
\bauthor{\bsnm{Mozer}, \binits{F.S.}},
\bauthor{\bsnm{Pulupa}, \binits{M.}},
\bauthor{\bsnm{Vasko}, \binits{I.Y.}},
\bauthor{\bsnm{Verniero}, \binits{J.}}, \betal:
\batitle{The electromagnetic signature of outward propagating ion-scale waves}.
\bjtitle{The Astrophysical Journal}
\bvolume{899}(\bissue{1}),
\bfpage{74}
(\byear{2020})
\end{barticle}
\endbibitem

\bibitem{VernieroApJS20}
\begin{barticle}
\bauthor{\bsnm{Verniero}, \binits{J.}},
\bauthor{\bsnm{Larson}, \binits{D.}},
\bauthor{\bsnm{Livi}, \binits{R.}},
\bauthor{\bsnm{Rahmati}, \binits{A.}},
\bauthor{\bsnm{McManus}, \binits{M.}},
\bauthor{\bsnm{Pyakurel}, \binits{P.S.}},
\bauthor{\bsnm{Klein}, \binits{K.}},
\bauthor{\bsnm{Bowen}, \binits{T.}},
\bauthor{\bsnm{Bonnell}, \binits{J.}},
\bauthor{\bsnm{Alterman}, \binits{B.}}, \betal:
\batitle{Parker solar probe observations of proton beams simultaneous with
  ion-scale waves}.
\bjtitle{The Astrophysical Journal Supplement Series}
\bvolume{248}(\bissue{1}),
\bfpage{5}
(\byear{2020})
\end{barticle}
\endbibitem

\bibitem{MalaspinaApJS20}
\begin{barticle}
\bauthor{\bsnm{Malaspina}, \binits{D.M.}},
\bauthor{\bsnm{Halekas}, \binits{J.}},
\bauthor{\bsnm{Ber{\v{c}}i{\v{c}}}, \binits{L.}},
\bauthor{\bsnm{Larson}, \binits{D.}},
\bauthor{\bsnm{Whittlesey}, \binits{P.}},
\bauthor{\bsnm{Bale}, \binits{S.D.}},
\bauthor{\bsnm{Bonnell}, \binits{J.W.}},
\bauthor{\bparticle{de} \bsnm{Wit}, \binits{T.D.}},
\bauthor{\bsnm{Ergun}, \binits{R.E.}},
\bauthor{\bsnm{Howes}, \binits{G.}}, \betal:
\batitle{Plasma waves near the electron cyclotron frequency in the near-sun
  solar wind}.
\bjtitle{The Astrophysical Journal Supplement Series}
\bvolume{246}(\bissue{2}),
\bfpage{21}
(\byear{2020})
\end{barticle}
\endbibitem

\bibitem{VechAA21}
\begin{barticle}
\bauthor{\bsnm{Vech}, \binits{D.}},
\bauthor{\bsnm{Martinovi{\'c}}, \binits{M.}},
\bauthor{\bsnm{Klein}, \binits{K.G.}},
\bauthor{\bsnm{Malaspina}, \binits{D.M.}},
\bauthor{\bsnm{Bowen}, \binits{T.A.}},
\bauthor{\bsnm{Verniero}, \binits{J.L.}},
\bauthor{\bsnm{Paulson}, \binits{K.}},
\bauthor{\bsnm{De~Wit}, \binits{T.D.}},
\bauthor{\bsnm{Kasper}, \binits{J.C.}},
\bauthor{\bsnm{Huang}, \binits{J.}}, \betal:
\batitle{Wave-particle energy transfer directly observed in an ion cyclotron
  wave}.
\bjtitle{Astronomy \& Astrophysics}
\bvolume{650},
\bfpage{10}
(\byear{2021})
\end{barticle}
\endbibitem

\bibitem{ZhaoApJ21}
\begin{barticle}
\bauthor{\bsnm{Zhao}, \binits{L.-L.}},
\bauthor{\bsnm{Zank}, \binits{G.}},
\bauthor{\bsnm{He}, \binits{J.}},
\bauthor{\bsnm{Telloni}, \binits{D.}},
\bauthor{\bsnm{Adhikari}, \binits{L.}},
\bauthor{\bsnm{Nakanotani}, \binits{M.}},
\bauthor{\bsnm{Kasper}, \binits{J.}},
\bauthor{\bsnm{Bale}, \binits{S.}}:
\batitle{Mhd and ion kinetic waves in field-aligned flows observed by parker
  solar probe}.
\bjtitle{The Astrophysical Journal}
\bvolume{922}(\bissue{2}),
\bfpage{188}
(\byear{2021})
\end{barticle}
\endbibitem

\bibitem{CattellApJL21}
\begin{barticle}
\bauthor{\bsnm{Cattell}, \binits{C.}},
\bauthor{\bsnm{Breneman}, \binits{A.}},
\bauthor{\bsnm{Dombeck}, \binits{J.}},
\bauthor{\bsnm{Short}, \binits{B.}},
\bauthor{\bsnm{Wygant}, \binits{J.}},
\bauthor{\bsnm{Halekas}, \binits{J.}},
\bauthor{\bsnm{Case}, \binits{T.}},
\bauthor{\bsnm{Kasper}, \binits{J.}},
\bauthor{\bsnm{Larson}, \binits{D.}},
\bauthor{\bsnm{Stevens}, \binits{M.}}, \betal:
\batitle{Parker solar probe evidence for scattering of electrons in the young
  solar wind by narrowband whistler-mode waves}.
\bjtitle{The Astrophysical journal letters}
\bvolume{911}(\bissue{2}),
\bfpage{29}
(\byear{2021})
\end{barticle}
\endbibitem

\bibitem{VernieroApJ22}
\begin{barticle}
\bauthor{\bsnm{Verniero}, \binits{J.}},
\bauthor{\bsnm{Chandran}, \binits{B.}},
\bauthor{\bsnm{Larson}, \binits{D.}},
\bauthor{\bsnm{Paulson}, \binits{K.}},
\bauthor{\bsnm{Alterman}, \binits{B.}},
\bauthor{\bsnm{Badman}, \binits{S.}},
\bauthor{\bsnm{Bale}, \binits{S.}},
\bauthor{\bsnm{Bonnell}, \binits{J.}},
\bauthor{\bsnm{Bowen}, \binits{T.}},
\bauthor{\bparticle{de} \bsnm{Wit}, \binits{T.D.}}, \betal:
\batitle{Strong perpendicular velocity-space diffusion in proton beams observed
  by parker solar probe}.
\bjtitle{The Astrophysical Journal}
\bvolume{924}(\bissue{2}),
\bfpage{112}
(\byear{2022})
\end{barticle}
\endbibitem

\bibitem{BowenArXiv21}
\begin{botherref}
\oauthor{\bsnm{Bowen}, \binits{T.A.}},
\oauthor{\bsnm{Squire}, \binits{J.}},
\oauthor{\bsnm{Bale}, \binits{S.D.}},
\oauthor{\bsnm{Chandran}, \binits{B.}},
\oauthor{\bsnm{Duan}, \binits{D.}},
\oauthor{\bsnm{Klein}, \binits{K.G.}},
\oauthor{\bsnm{Larson}, \binits{D.}},
\oauthor{\bsnm{Mallet}, \binits{A.}},
\oauthor{\bsnm{McManus}, \binits{M.D.}},
\oauthor{\bsnm{Meyrand}, \binits{R.}}, et al.:
The in situ signature of cyclotron resonant heating.
arXiv preprint arXiv:2111.05400
(2021)
\end{botherref}
\endbibitem

\bibitem{BowenPRL20}
\begin{barticle}
\bauthor{\bsnm{Bowen}, \binits{T.A.}},
\bauthor{\bsnm{Mallet}, \binits{A.}},
\bauthor{\bsnm{Bale}, \binits{S.D.}},
\bauthor{\bsnm{Bonnell}, \binits{J.}},
\bauthor{\bsnm{Case}, \binits{A.W.}},
\bauthor{\bsnm{Chandran}, \binits{B.D.}},
\bauthor{\bsnm{Chasapis}, \binits{A.}},
\bauthor{\bsnm{Chen}, \binits{C.H.}},
\bauthor{\bsnm{Duan}, \binits{D.}},
\bauthor{\bparticle{de} \bsnm{Wit}, \binits{T.D.}}, \betal:
\batitle{Constraining ion-scale heating and spectral energy transfer in
  observations of plasma turbulence}.
\bjtitle{Physical Review Letters}
\bvolume{125}(\bissue{2}),
\bfpage{025102}
(\byear{2020})
\end{barticle}
\endbibitem

\bibitem{QudsiApJ20}
\begin{barticle}
\bauthor{\bsnm{Qudsi}, \binits{R.A.}},
\bauthor{\bsnm{Bandyopadhyay}, \binits{R.}},
\bauthor{\bsnm{Maruca}, \binits{B.A.}},
\bauthor{\bsnm{Parashar}, \binits{T.N.}},
\bauthor{\bsnm{Matthaeus}, \binits{W.H.}},
\bauthor{\bsnm{Chasapis}, \binits{A.}},
\bauthor{\bsnm{Gary}, \binits{S.P.}},
\bauthor{\bsnm{Giles}, \binits{B.L.}},
\bauthor{\bsnm{Gershman}, \binits{D.J.}},
\bauthor{\bsnm{Pollock}, \binits{C.J.}},
\bauthor{\bsnm{Strangeway}, \binits{R.J.}},
\bauthor{\bsnm{Torbert}, \binits{R.B.}},
\bauthor{\bsnm{Moore}, \binits{T.E.}},
\bauthor{\bsnm{Burch}, \binits{J.L.}}:
\batitle{Intermittency and ion temperature{\textendash}anisotropy
  instabilities: Simulation and magnetosheath observation}.
\bjtitle{The Astrophysical Journal}
\bvolume{895}(\bissue{2}),
\bfpage{83}
(\byear{2020}).
\doiurl{10.3847/1538-4357/ab89ad}
\end{barticle}
\endbibitem

\bibitem{ChhiberApJS20}
\begin{barticle}
\bauthor{\bsnm{Chhiber}, \binits{R.}},
\bauthor{\bsnm{Goldstein}, \binits{M.L.}},
\bauthor{\bsnm{Maruca}, \binits{B.A.}},
\bauthor{\bsnm{Chasapis}, \binits{A.}},
\bauthor{\bsnm{Matthaeus}, \binits{W.H.}},
\bauthor{\bsnm{Ruffolo}, \binits{D.}},
\bauthor{\bsnm{Bandyopadhyay}, \binits{R.}},
\bauthor{\bsnm{Parashar}, \binits{T.N.}},
\bauthor{\bsnm{Qudsi}, \binits{R.}},
\bauthor{\bparticle{de} \bsnm{Wit}, \binits{T.D.}},
\bauthor{\bsnm{Bale}, \binits{S.D.}},
\bauthor{\bsnm{Bonnell}, \binits{J.W.}},
\bauthor{\bsnm{Goetz}, \binits{K.}},
\bauthor{\bsnm{Harvey}, \binits{P.R.}},
\bauthor{\bsnm{MacDowall}, \binits{R.J.}},
\bauthor{\bsnm{Malaspina}, \binits{D.}},
\bauthor{\bsnm{Pulupa}, \binits{M.}},
\bauthor{\bsnm{Kasper}, \binits{J.C.}},
\bauthor{\bsnm{Korreck}, \binits{K.E.}},
\bauthor{\bsnm{Case}, \binits{A.W.}},
\bauthor{\bsnm{Stevens}, \binits{M.}},
\bauthor{\bsnm{Whittlesey}, \binits{P.}},
\bauthor{\bsnm{Larson}, \binits{D.}},
\bauthor{\bsnm{Livi}, \binits{R.}},
\bauthor{\bsnm{Velli}, \binits{M.}},
\bauthor{\bsnm{Raouafi}, \binits{N.}}:
\batitle{Clustering of intermittent magnetic and flow structures near parker
  solar probe's first perihelion{\textemdash}a partial-variance-of-increments
  analysis}.
\bjtitle{The Astrophysical Journal Supplement Series}
\bvolume{246}(\bissue{2}),
\bfpage{31}
(\byear{2020}).
\doiurl{10.3847/1538-4365/ab53d2}
\end{barticle}
\endbibitem

\bibitem{QudsiApJS20}
\begin{barticle}
\bauthor{\bsnm{Qudsi}, \binits{R.A.}},
\bauthor{\bsnm{Maruca}, \binits{B.A.}},
\bauthor{\bsnm{Matthaeus}, \binits{W.H.}},
\bauthor{\bsnm{Parashar}, \binits{T.N.}},
\bauthor{\bsnm{Bandyopadhyay}, \binits{R.}},
\bauthor{\bsnm{Chhiber}, \binits{R.}},
\bauthor{\bsnm{Chasapis}, \binits{A.}},
\bauthor{\bsnm{Goldstein}, \binits{M.L.}},
\bauthor{\bsnm{Bale}, \binits{S.D.}},
\bauthor{\bsnm{Bonnell}, \binits{J.W.}},
\bauthor{\bparticle{de} \bsnm{Wit}, \binits{T.D.}},
\bauthor{\bsnm{Goetz}, \binits{K.}},
\bauthor{\bsnm{Harvey}, \binits{P.R.}},
\bauthor{\bsnm{MacDowall}, \binits{R.J.}},
\bauthor{\bsnm{Malaspina}, \binits{D.}},
\bauthor{\bsnm{Pulupa}, \binits{M.}},
\bauthor{\bsnm{Kasper}, \binits{J.C.}},
\bauthor{\bsnm{Korreck}, \binits{K.E.}},
\bauthor{\bsnm{Case}, \binits{A.W.}},
\bauthor{\bsnm{Stevens}, \binits{M.}},
\bauthor{\bsnm{Whittlesey}, \binits{P.}},
\bauthor{\bsnm{Larson}, \binits{D.}},
\bauthor{\bsnm{Livi}, \binits{R.}},
\bauthor{\bsnm{Velli}, \binits{M.}},
\bauthor{\bsnm{Raouafi}, \binits{N.}}:
\batitle{Observations of heating along intermittent structures in the inner
  heliosphere from {PSP} data}.
\bjtitle{The Astrophysical Journal Supplement Series}
\bvolume{246}(\bissue{2}),
\bfpage{46}
(\byear{2020}).
\doiurl{10.3847/1538-4365/ab5c19}
\end{barticle}
\endbibitem

\bibitem{Sioulas_2022}
\begin{barticle}
\bauthor{\bsnm{Sioulas}, \binits{N.}},
\bauthor{\bsnm{Velli}, \binits{M.}},
\bauthor{\bsnm{Chhiber}, \binits{R.}},
\bauthor{\bsnm{Vlahos}, \binits{L.}},
\bauthor{\bsnm{Matthaeus}, \binits{W.H.}},
\bauthor{\bsnm{Bandyopadhyay}, \binits{R.}},
\bauthor{\bsnm{Cuesta}, \binits{M.E.}},
\bauthor{\bsnm{Shi}, \binits{C.}},
\bauthor{\bsnm{Bowen}, \binits{T.A.}},
\bauthor{\bsnm{Qudsi}, \binits{R.A.}},
\bauthor{\bsnm{Stevens}, \binits{M.L.}},
\bauthor{\bsnm{Bale}, \binits{S.D.}}:
\batitle{Statistical analysis of intermittency and its association with proton
  heating in the near-sun environment}.
\bjtitle{The Astrophysical Journal}
\bvolume{927}(\bissue{2}),
\bfpage{140}
(\byear{2022}).
\doiurl{10.3847/1538-4357/ac4fc1}
\end{barticle}
\endbibitem

\bibitem{SorrisoJPP18}
\begin{botherref}
\oauthor{\bsnm{Sorriso-Valvo}, \binits{L.}},
\oauthor{\bsnm{Perrone}, \binits{D.}},
\oauthor{\bsnm{Pezzi}, \binits{O.}},
\oauthor{\bsnm{Valentini}, \binits{F.}},
\oauthor{\bsnm{Servidio}, \binits{S.}},
\oauthor{\bsnm{Zouganelis}, \binits{I.}},
\oauthor{\bsnm{Veltri}, \binits{P.}}:
Local energy transfer rate and kinetic processes: the fate of turbulent energy
  in two-dimensional hybrid vlasov--maxwell numerical simulations.
Journal of Plasma Physics
\textbf{84}(2)
(2018)
\end{botherref}
\endbibitem

\bibitem{ChhiberApJL21}
\begin{barticle}
\bauthor{\bsnm{Chhiber}, \binits{R.}},
\bauthor{\bsnm{Matthaeus}, \binits{W.H.}},
\bauthor{\bsnm{Bowen}, \binits{T.A.}},
\bauthor{\bsnm{Bale}, \binits{S.D.}}:
\batitle{Subproton-scale intermittency in near-sun solar wind turbulence
  observed by the parker solar probe}.
\bjtitle{The Astrophysical Journal Letters}
\bvolume{911}(\bissue{1}),
\bfpage{7}
(\byear{2021})
\end{barticle}
\endbibitem

\bibitem{MarschLRSP06}
\begin{botherref}
\oauthor{\bsnm{Marsch}, \binits{E.}}:
Kinetic physics of the solar corona and solar wind.
Living Reviews in Solar Physics
\textbf{3}(1)
(2006)
\end{botherref}
\endbibitem

\bibitem{Verma2019}
\begin{bbook}
\bauthor{\bsnm{Verma}, \binits{M.K.}}:
\bbtitle{Energy Transfers in Fluid Flows: Multiscale and Spectral
  Perspectives}.
\bpublisher{Cambridge University Press}, \blocation{???}
(\byear{2019})
\end{bbook}
\endbibitem

\bibitem{CuestaApJL22}
\begin{botherref}
\oauthor{\bsnm{Cuesta}, \binits{M.E.}},
\oauthor{\bsnm{Chhiber}, \binits{R.}},
\oauthor{\bsnm{Roy}, \binits{S.}},
\oauthor{\bsnm{Goodwill}, \binits{J.}},
\oauthor{\bsnm{Pecora}, \binits{F.}},
\oauthor{\bsnm{Jarosik}, \binits{J.}},
\oauthor{\bsnm{Matthaeus}, \binits{W.H.}},
\oauthor{\bsnm{Parashar}, \binits{T.N.}},
\oauthor{\bsnm{Bandyopadhyay}, \binits{R.}}:
Isotropization and evolution of energy-containing eddies in solar wind
  turbulence: Parker solar probe, helios 1, ace, wind, and voyager 1.
The Astrophysical Journal Letter
(2022)
\end{botherref}
\endbibitem

\bibitem{VernieroJGR21}
\begin{barticle}
\bauthor{\bsnm{Verniero}, \binits{J.}},
\bauthor{\bsnm{Howes}, \binits{G.}},
\bauthor{\bsnm{Stewart}, \binits{D.}},
\bauthor{\bsnm{Klein}, \binits{K.}}:
\batitle{Determining threshold instrumental resolutions for resolving the
  velocity-space signature of ion landau damping}.
\bjtitle{Journal of Geophysical Research: Space Physics}
\bvolume{126}(\bissue{5}),
\bfpage{2020}--\blpage{028361}
(\byear{2021})
\end{barticle}
\endbibitem

\end{thebibliography}
\end{document}